\documentclass[aps,reprint,superscriptaddress,longbibliography,bibnotes]{revtex4-1}

\usepackage{graphicx}
\usepackage{dcolumn}
\usepackage{bm, amssymb, color}
\usepackage{hyperref}
\usepackage{natbib}
\usepackage{booktabs,tabulary}
\usepackage{multirow}
\usepackage{amsmath,soul}
\usepackage[export]{adjustbox}
\usepackage[caption=false, subrefformat=parens, labelformat=parens]{subfig}
\usepackage{array}
\usepackage{cases}

\newcommand{\Z}{\mathbb{Z}}
\newcommand{\R}{\mathbb{R}}
\renewcommand{\d}[1]{\mathinner{d#1}}
\newcommand{\fn}[2]{\mathinner{#1\mathopen{\left(#2\right)}}}
\newcommand{\vect}[1]{\bm{#1}}
\newcommand{\E}[1]{\left\langle#1\right\rangle}
\newcommand{\abs}[1]{\left\vert #1 \right\vert}
\newcommand{\ceil}[1]{\lceil #1 \rceil}

\newcommand{\vv}[1]{\fn{\sigma_V ^2}{#1}}
\newcommand{\spD}[1]{\fn{\tilde{\chi}_{_V}}{#1}}
\newcommand{\order}[1]{\mathinner{\mathcal{O}\mathopen{\left(#1\right)}}}
\newcommand{\pd}[3]{\mathop{\left( \frac{\partial #1}{\partial #2} \right)_{#3}}}

\newcommand{\tens}[1]{\mathbf{#1}}

\newcommand{\subcap}[1]{\protect\subref{#1}}

\usepackage[colorinlistoftodos,shadow,textwidth=18mm]{todonotes}

\newcolumntype{L}[1]{>{\raggedright\let\newline\\\arraybackslash\hspace{0pt}}m{#1}}
\newcolumntype{C}[1]{>{\centering\let\newline\\\arraybackslash\hspace{0pt}}m{#1}}
\newcolumntype{R}[1]{>{\raggedleft\let\newline\\\arraybackslash\hspace{0pt}}m{#1}}
\setstcolor{green}

\graphicspath{{./figs/}}  

\begin{document}
\title{Methodology to Construct Large Realizations of Perfectly Hyperuniform Disordered Packings }

\author{Jaeuk Kim}
\affiliation{Department of Physics, Princeton University, Princeton, New Jersey 08544, USA}
\author{Salvatore Torquato}%
\email{torquato@princeton.edu}
\homepage{http://chemlabs.princeton.edu}
\affiliation{Department of Physics, Princeton University, Princeton, New Jersey 08544, USA}
\affiliation{Department of Chemistry, Princeton University, Princeton, New Jersey 08544, USA}
\affiliation{Princeton Institute for the Science and Technology of Materials, Princeton University, Princeton, New Jersey 08544, USA}
\affiliation{Program in Applied and Computational Mathematics, Princeton University, Princeton, New Jersey 08544, USA}

\date{\today}
\begin{abstract}
Disordered hyperuniform packings (or dispersions) are unusual amorphous two-phase materials that are endowed with exotic physical properties.
Such hyperuniform systems are characterized by an anomalous suppression of volume-fraction fluctuations at infinitely long-wavelengths, compared to ordinary disordered materials.
While there has been growing interest in such singular states of amorphous matter, a major obstacle has been an inability to produce large samples that are perfectly hyperuniform due to practical limitations of conventional numerical and experimental methods.
To overcome these limitations, we introduce a general theoretical methodology to construct perfectly hyperuniform packings in $d$-dimensional Euclidean space $\R^d$.
Specifically, beginning with an initial general tessellation of space by disjoint cells that meets a ``bounded-cell" condition, hard particles are placed inside each cell such that the local-cell particle packing fractions are identical to the global packing fraction.
We prove that the constructed packings with a polydispersity in size in $\mathbb{R}^d$ are perfectly hyperuniform in the infinite-sample-size limit and the hyperuniformity of such packings is independent of particle shapes, positions, and numbers per cell.
 We use this theoretical formulation to devise an efficient and tunable algorithm to generate extremely large realizations of such packings. We employ two distinct initial tessellations: Voronoi as well as sphere tessellations.
 Beginning with Voronoi tessellations, we show that our algorithm can remarkably convert extremely large nonhyperuniform packings into hyperuniform ones in $\R^2$ and $\R^3$.
Implementing our theoretical methodology on sphere tessellations, we establish the hyperuniformity of the classical Hashin-Shtrikman multiscale coated-spheres structures, which are known to be two-phase media microstructures that possess \textit{optimal} effective transport and elastic properties.
A consequence of our work is a rigorous demonstration that packings that have identical tessellations can either be nonhyperuniform or hyperuniform by simply tuning local characteristics.   
It is noteworthy that our computationally designed hyperuniform two-phase systems can easily be fabricated via state-of-the-art methods, such as 2D photolithographic and 3D printing technologies. 
In addition, the tunability of our methodology offers a route for the  discovery of novel disordered hyperuniform two-phase materials. 
 
\end{abstract}

\maketitle

\setstcolor{green}
\setul{0}{0.4ex}
\section{Introduction} 

A hyperuniform state of matter is characterized by an anomalous suppression of density or volume-fraction fluctuations at infinitely long wavelengths relative to those in typical disordered systems, such as liquids and structural glasses \cite{Zachary2009a, Torquato2003_hyper, Torquato2018_review}. 
Such hyperuniform states encompass all perfect crystals, many quasicrystals, as well as some exotic disordered systems.
Disordered hyperuniform states of matter have been the subject of intense interest across a variety of fields, including physics \cite{Scardicchio2009, Harrison1970, Pietronero2002, Lomba2017, Jancovici1981, Jiao2011, Feynman1956, Hexner2015,Weijs2015, Chremos2018}, materials science \cite{Man2013,Florescu2013, Leseur2016,  Scheffold2017, Thien2016, Chen2017,Zhang2016}, chemistry \cite{Rundman1967, Takeji1986, Ma2017}, biology \cite{Noh2010, Jiao2014_chickenEyes, Mayer2015, Kwon2017}, and mathematics \cite{Dyson1962, Montgomery1973, Torquato2018_2}.
The notion of hyperuniformity was first defined in the context of point-particle systems \cite{Torquato2003_hyper} and then extended to \textit{two-phase heterogeneous systems} \cite{Zachary2009a} and random scalar/vector fields \cite{Torquato2016_gen}.
General two-phase systems abound in natural and artificial materials, including colloidal suspensions, particulate composites, and concrete \cite{Brinker_sol-gel, Weeks2017, Torquato_RHM, Patel2016, Neville_concrete, Sahimi_HM1, Milton_TheoComposites, Mickel2013}. 
Packings (or dispersions), which are of central concern in this paper, comprise a class of two-phase systems in which nonoverlapping particles are spatially distributed throughout a connected ``void" (matrix) phase. 
 
 \begin{figure}[ht]
 \begin{center}
 \includegraphics[width = 0.4\textwidth]{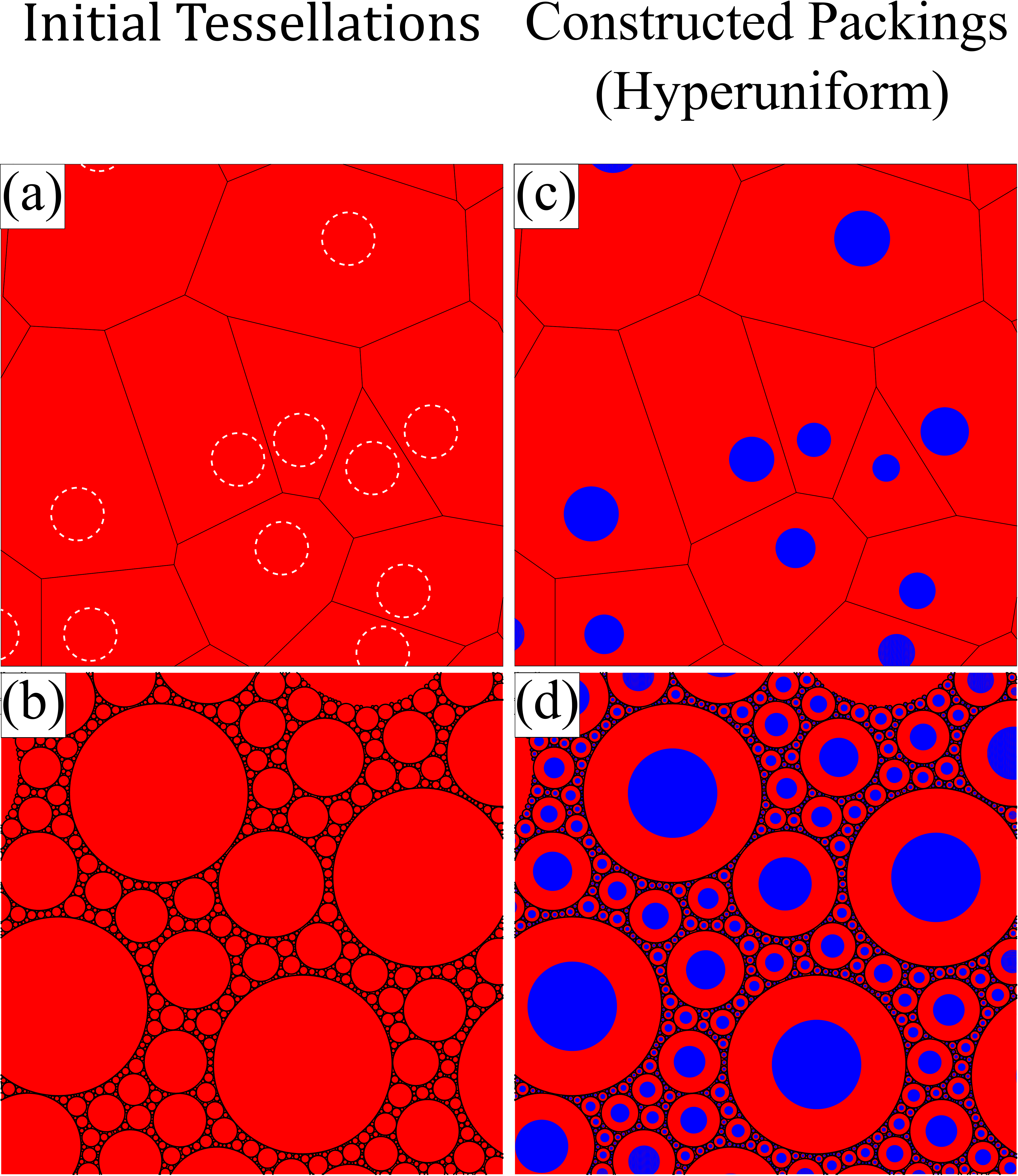}
 \end{center}
\caption{(Color online) (a, b) Portions of initial disordered tessellations: (a) Voronoi tessellation (black lines) of a nonhyperuniform packing (Sec. \ref{sec:byVoronoiTessellations}) and (b) a multiscale-disk tessellation (Sec. \ref{sec:Coated-sphere}).
A progenitor disk packing in (a) is illustrated by white dashed circles.
The tessellation-based procedure can be applied into higher dimensions, but we illustrate two-dimensional cases here for simplicity. 
(c, d) Portions of disordered hyperuniform packings (dispersions) constructed from the initial tessellations (a, b) via the tessellation-based procedure, i.e., local-cell packing fraction $\phi$ of a particle (blue disks) within each cell is identical to the global packing fraction. 
Importantly, (d) multiscale coated-disks model corresponds to optimal Hashin-Shtrikman structures. 
\label{fig:schematics} }
\end{figure}

A \textit{hyperuniform two-phase system} in $d$-dimensional Euclidean space $\R^d$ is one in which the local volume-fraction variance $\vv{R}$ inside a spherical observation window of radius $R$ decays faster than $R^{-d}$ in the large-$R$ limit \cite{Zachary2009a, Torquato2018_review}:
\begin{equation}\label{def:HU_direct}
\lim_{R\to \infty} \fn{v_1}{R}\vv{R} = 0.
\end{equation}
Equivalently, its associated spectral density $\spD{\vect{k}}$ vanishes as the wavenumber $\abs{\vect{k}}$ tends to zero  \cite{Zachary2009a, Torquato2018_review}.
For \textit{disordered} hyperuniform two-phase systems, $\spD{\vect{k}}$ typically exhibits the power-law scaling:
\begin{equation}\label{def:HU_spectral}
 \spD{\vect{k}} \sim \abs{\vect{k}}^\alpha~~~(\alpha>0).
\end{equation}
 
This exponent $\alpha$ is directly related to three distinct classes of hyperuniformity that are categorized based on the large-$R$ scalings of $\vv{R}$ \cite{Torquato2018_review}:
\begin{equation}
\vv{R} \sim 
\begin{cases}
R^{-(d+1)}, & \alpha >1~~(\mathrm{class~I})\\
R^{-(d+1)}\ln{R}, &\alpha =1~~(\mathrm{class~II})\\
R^{-d-\alpha}, &\alpha <1~~(\mathrm{class~III})
\end{cases}\label{def:HU_classes},
\end{equation}
where classes I and III represent the strongest and weakest forms of hyperuniformity, respectively.
Class I systems include all crystals, some quasicrystals, and stealthy hyperuniform systems in which $\spD{k} = 0 $ for $0< k < K$ \cite{Batten2008, Zhang2015, Zhang2015_2, Torquato2015_stealthy}.  

Disordered hyperuniform two-phase systems \cite{Zachary2009a, Zhang2016} are attracting considerable attention due to their unusual physical properties, such as complete isotropic photonic and phononic bandgaps \cite{Florescu2009, Scheffold2017, Gkantzounis2017,Lopez2018}, nearly optimal transport properties \cite{Zhang2016, Chen2017, Thien2016}, superior metamaterial designs \cite{Wu2017}, dense but transparent materials \cite{Leseur2016}, and low-density materials of blackbody-like absorption \cite{Bigourdan2019}. 
Similar to ordinary two-phase systems, their effective properties are also tunable by engineering the phase properties and volume fractions as well as the spatial arrangements \cite{Florescu2009, Scheffold2017, Rechtsman2008, Chen2017, Torquato2018_5, Zhang2016, Wu2017}.
An important class of two-phase systems are the Hashin-Shtrikman structures \cite{Hashin1962, Torquato_RHM, Torquato2004, Milton1986a, Milton_TheoComposites} that are optimal for effective elastic moduli \cite{Hashin1962, Torquato_RHM, Milton_TheoComposites}, thermal (electrical) conductivity \cite{Milton1986a}, trappiong constant \cite{Torquato2004}, and fluid permeability \cite{Torquato2004} for given phase volumes and phase properties.
Remarkably, certain disordered hyperuniform systems possess nearly optimal transport and elastic properties \cite{Zhang2016, Chen2017, Torquato2018_5, Torquato2018_3}.

Theoretical \cite{Scardicchio2009, Hexner2015, Tjhung2015,Lomba2017, Duyu2018_2}, numerical \cite{Zachary2011_3, TJ_algorithm, LS_algorithm,LS_algorithm2, Atkinson2016, Uche2004, Zhang2016, Chen2017, DiStasio2018, Wang2018}, and experimental methods \cite{Weijs2015, Kurita2011, Ricouvier2017, Zito2015} have been developed to generate disordered hyperuniform packings (dispersions). 
In practice, however, these methods are system-size limited due to computational cost or imperfections. 
For instance, the collective-coordinate optimization technique \cite{Uche2004, Zhang2016, Chen2017,DiStasio2018} and  equilibrium plasma \cite{Jancovici1981, Lomba2017, Duyu2018_2} can achieve perfect hyperuniformity, but their long-range interactions lead the computation cost to grow rapidly with system size. 
Random organization models \cite{Hexner2015, Tjhung2015} yield disordered hyperuniform packings at critical absorbing states but are limited in producing perfect hyperuniformity because of critical slowing-down phenomena \cite{Atkinson2016, Torquato2018_review}.
Determinantal point processes are perfectly hyperuniform in the thermodynamic limit, but the current numerical algorithm hardly can generate a realization of more than 100 particles due to accumulated numerical error \cite{Scardicchio2009}.
Stealthy designs via the superposition procedure \cite{DiStasio2018} provides an efficient means to construct exactly stealthy hyperuniform digitized two-phase systems.
However, this construction scheme requires one to
prepare many different small systems as building blocks,
which is computationally demanding as the system size
increases. 

Since hyperuniformity is a global property of an infinitely large system, limited sample sizes often make it difficult to ascertain whether such systems are truly hyperuniform or effectively hyperuniform. 
Furthermore, disordered hyperuniform systems often can include imperfections, such as point vacancies, stochastic displacements \cite{Gabrielli2004}, thermally excited phonon modes \cite{Jaeuk2018}, and rattlers \cite{Atkinson2016} in maximally random jammed (MRJ) packings.
Such imperfections can either destroy or degrade the hyperuniformity (even if by a small amount) of otherwise perfectly hyperuniform systems \cite{Jaeuk2018, Torquato2018_review}. 
Hence, there is a great need to devise exact and efficient procedures to construct extremely large realizations of disordered hyperuniform two-phase systems.
Such capability could then be combined with state-of-the-art photolithographic and 3D printing techniques to fabricate large disordered hyperuniform systems.

In this paper, we introduce a tessellation-based procedure that enables one to generate disordered packings in $\R^d$ that are perfectly hyperuniform in the infinite-sample-size limit. 
Based on this theoretical methodology, we formulate an efficient algorithm to generate extremely large realizations of hyperuniform packings.
Specifically, one first tessellates the space into certain disjoint cells that meet a \textit{bounded-cell} condition, i.e., the maximal length of each cell should be much shorter than side length of the entire tessellation.  
This mild restriction allows one to employ a wide class of tessellations, as discussed below and in Sec. \ref{sec:byVoronoiTessellations}.  
Then, one places hard particles in each cell such that the local packing fractions associated with the cells become identical to the global packing fraction $\phi$; see Fig. \ref{fig:schematics}.
Note that for ``periodic" tessellations of equal cell volumes (e.g., Voronoi tessellations of Bravais lattices and Weaire-Phelan foam \cite{Weaire1994, ConwayTorquato2006}), our procedure yields periodic packings of identical particles that are also stealthy hyperuniform \footnote{Generally speaking, application of our procedure to any periodic tessellation will result in a periodic packing that is stealthy hyperuniform.}.
However, our major concerns in this paper are disordered tessellations whose cells have a variability in sizes and shapes.
Thus, the application of our procedure to such tessellations generates ``disordered" hyperuniform packings in which the particles have a polydispersity in size.
We present a detailed theoretical analysis of the small-wavenumber scalings of the spectral density $\spD{k}$ for the constructed packings in the case of arbitrary particle shapes.
We thereby prove that whenever the initial tessellations meet the bounded-cell condition, these systems are strongly hyperuniform (class I) for any particle shape in the infinite-sample-size limit.
In this limit, the system size tends to be infinitely large with other intensive parameters (e.g., number density and packing fractions) held fixed.
This procedure is a packing protocol to generate packings of polydisperse particles that is uniquely different from previously known methods \cite{Torquato2018_packing}.

As a proof-of-concept, we verify our theoretical results by numerically constructing packings from certain initial tessellations, and by ascertaining their hyperuniformity from the spectral densities.
Our procedure allows for the use of any initial tessellation, including Voronoi tessellations \cite{Torquato_RHM} and their generalizations \cite{Gellatly1982, Hiroshi1985}, sphere tessellations, disordered isoradial graphs \cite{Deuschel2012}, dissected tessellations \cite{Gabbrielli2012}, Delaunay triangulations, and ``Delaunay-centroidal'' tessellations \cite{Florescu2009, Torquato2018_3}, as long as it meets the bounded-cell condition.
For concreteness and simplicity, two types of initial tessellations are considered in this work: Voronoi tessellations (Sec. \ref{sec:byVoronoiTessellations}) and sphere tessellations, which are necessarily multiscale divisions of space (Sec. \ref{sec:Coated-sphere}).  
Employing Voronoi tessellations of general disordered point patterns in $\R^2$ and $\R^3$, we demonstrate that our methodology enables a remarkable mapping that converts extremely large nonhyperuniform packings (as large as $10^8$ particles) into hyperuniform ones.
To carry out our simulations, we employ various statistically homogeneous progenitor systems. 
Based on the same idea, we establish the hyperuniformity of the aforementioned optimal Hashin-Shtrikman multiscale coated-spheres structures \cite{Hashin1962, Torquato_RHM, Torquato2004, Milton1986a} [see Fig. \ref{fig:schematics}(d)].
Here, we provide a detailed derivation of $\fn{\tilde{\chi}_{_V}^{(m)}}{k}$ in the $m$th stage and numerical simulations for two distinct types of cell-volume distributions. 
It is noteworthy that our methodology only involves calculating cell volumes, which is exactly performed and easy to parallelize. 
Furthermore, we demonstrate that large samples of many of our designs can be readily fabricated via modern photolithographic and 3D printing techniques \cite{Wong2012, Tumbleston2015, Shirazi2015, Zhao2018} (Sec. \ref{sec:fabrications}).
While some of the major results were announced in a brief communication \cite{Jaeuk2019a}, there we focused on
applications to sphere packings without detailed derivations. 
In this work, we treat a broader class of sphere packings as well as packings of nonspherical particles and provide detailed mathematical derivations.
We also report associated simulation results that are not contained in Ref. \cite{Jaeuk2019a}.

We present basic mathematical definitions and concepts in Sec. \ref{sec:background}. Then, we precisely describe the tessellation-based procedure in Sec. \ref{sec:Tiling-based procedure}. In Sec. \ref{sec:GeneralTheory} we derive the small-$k$ scalings of the spectral densities for the constructed packings. 
Subsequently, we verify our theoretical results by numerical simulations using Voronoi tessellations and sphere tessellations in Secs. \ref{sec:byVoronoiTessellations} and \ref{sec:Coated-sphere}, respectively.
Then, we discuss the feasibility of fabricating our designs in modern technologies in Sec. \ref{sec:fabrications}. 
Finally, we provide concluding remarks in Sec. \ref{sec:conclusion}.

\section{Background and Definitions}\label{sec:background}

The microstructure of a two-phase system can be described by the \textit{phase indicator function} associated with phase $i=1,2$ \cite{Torquato_RHM}: 
\begin{equation}
\fn{\mathcal{I}^{(i)}}{\vect{r}} =
\begin{cases}
1, & \vect{r} \in \mathrm{phase~}i\\
0, & \mathrm{otherwise}.
\end{cases}
\end{equation}
If the system is statistically homogeneous, then its one-point correlation function is independent of position $\vect{r}$, and identical to the phase-volume fraction $\phi_i$, i.e., $\E{\fn{\mathcal{I}^{(i)}}{\vect{r}}} = \phi_i$, where $\E{\cdot}$ represents an ensemble average.
The \textit{autocovariance} function can be defined in terms of the mean-zero fluctuating indicator function, $\fn{\mathcal{J}^{(i)}}{\vect{r}}\equiv \fn{\mathcal{I}^{(i)}}{\vect{r}} - \phi_i$, as follows:
\begin{equation}
\fn{\chi_{_V}}{\vect{r}} \equiv \E{\fn{\mathcal{J}^{(i)}}{\vect{r}'}\fn{\mathcal{J}^{(i)}}{\vect{r}'+\vect{r}} },
\end{equation}
which is identical for each phase and tends to zero as $r$ increases if the system does not have long-range order.
Its Fourier transform $\spD{\vect{k}}\equiv \int_{\R^d} \d{\vect{y}}e^{-i \vect{k}\cdot\vect{y}} \fn{\chi_{_V}}{\vect{y}}$,  called the \textit{spectral density}, is a nonnegative real-valued function of a wavevector $\vect{k}$.
In experiments, the spectral densities are directly obtainable from elastic scattering intensities \cite{Debye1949} when the wavelength of radiation is larger than atomic distance, but shorter than the length scale of domains.
In numerical simulations, the spectral densities are calculated from  realizations of the media under the periodic boundary conditions as follows:
\begin{equation}\label{eq:spectral density}
\spD{\vect{k}} = \frac{1}{\abs{\mathcal{V}_F}} \E{\abs{\fn{\tilde{\mathcal{J}}^{(i)}}{\vect{k}}}^2},
\end{equation}
where $\abs{\mathcal{V}_F}$ is the volume of the simulation box, a wavevector $\vect{k}$ is a reciprocal lattice vector of the simulation box, and $\E{\cdot}$ represents an ensemble average, where $\fn{\tilde{\mathcal{J}}^{(i)}}{\vect{k}}$ is the Fourier transform of $\fn{\mathcal{J}^{(i)}}{\vect{r}}$ \cite{Torquato1999_3}.

In the context of two-phase media, a packing can be regarded as domains of a ``particle" phase ($N$ generally shaped particles $\tens{P}_1,~\tens{P}_2, \cdots,~\tens{P}_N$) that are dispersed throughout a continuous ``matrix" (void) phase. For such a packing in a periodic fundamental cell $\mathcal{V}_F$, the random variable 
$\fn{\tilde{\mathcal{J}}}{\vect{k}}$ associated with the particle phase (dropping the superscript) can be expressed as follows:
\begin{align}\label{eq:FourierTransform_particle}
\fn{\tilde{\mathcal{J}}}{\vect{k}} & = \sum_{j=	1}^N \fn{\tilde{m}}{\vect{k};\tens{P}_j}e^{-i \vect{k}\cdot\vect{r}_j} - \phi \int_{\mathcal{V}_F} \d{\vect{y}} e^{-i\vect{k}\cdot\vect{y}}  \\
& = \sum_{j=1}^N \fn{\tilde{m}}{\vect{k};\tens{P}_j}e^{-i \vect{k}\cdot\vect{r}_j} - \phi \abs{\mathcal{V}_F} \delta_{\vect{k},\vect{0}}, \label{eq:FourierTransform_Dirac}
\end{align}
where $\vect{r}_j$ is the centroid of $\tens{P}_j$, $\fn{\tilde{m}}{\vect{k};\tens{P}_j}$ is its \textit{form factor} (i.e., the Fourier transform of the ``particle indicator function" $\fn{m}{\vect{r};\tens{P}_j}$ with respect to $\vect{r}_j$), and $\delta_{\vect{k},\vect{0}}$ represents a Kronecker delta symbol.
Since the forward scattering term $\phi \abs{\mathcal{V}_F} \delta_{\vect{k},\vect{0}}$ in Eq. \eqref{eq:FourierTransform_Dirac} always vanishes at nonzero reciprocal lattice vectors $\vect{k}$'s, this term is often ignored in numerical calculations.

For some special particle shapes, closed-form expressions of $\fn{\tilde{m}}{\vect{k};\tens{P}_j}$ are known.   
For a spherical particle of radius $a$, the associated form factor is 
\begin{equation}\label{eq:FourierTransform_sphere}
\fn{\tilde{m}}{\vect{k};a} = \left(2\pi a /k\right)^{d/2}\fn{J_{d/2}}{ka},
\end{equation}
where $k\equiv \abs{\vect{k}}$ and $\fn{J_n}{x}$ is the Bessel function of order $n$.
For a cubic particle of side length $L$, the corresponding function is
\begin{equation}\label{eq:FourierTransform_cubic}
\fn{\tilde{m}}{\vect{k};L} = L^d\prod_{l=1}^d \fn{\mathrm{sinc}}{k_l L/2},
\end{equation}
where $k_l$ represents $l$th component of a wavevector $\vect{k}$ and 
\begin{equation}
\fn{\mathrm{sinc}}{x} \equiv
\begin{cases}
\frac{\sin{x}}{x}, & x\neq 0\\
1, & x=0.
\end{cases}
\end{equation}
We note that Eqs. \eqref{eq:FourierTransform_sphere} and \eqref{eq:FourierTransform_cubic} have their global maxima at the origin, whose values are identical to their particle volumes.

From Eqs. \eqref{eq:spectral density} and \eqref{eq:FourierTransform_particle}, one can straightforwardly derive an expression of the spectral density for sphere packings of the identical particle radius $a$ \cite{Torquato_RHM, Torquato2016_gen}:
\begin{equation}\label{eq:spectraldensity_spherePacking}
\spD{\vect{k}} = \rho \abs{\fn{\tilde{m}}{\vect{k};a}}^2 \fn{S}{\vect{k}},
\end{equation}
where $\rho$ is number density and $\fn{S}{\vect{k}}$ is the structure factor defined as  
\begin{equation}\label{eq:def_StructureFactor}
\fn{S}{\vect{k}}\equiv \frac{1}{N}\E{\abs{\sum_{j=1}^N e^{-i\vect{k}\cdot\vect{r}_j} - N \delta_{\vect{k},\vect{0}}}^2}.
\end{equation}
Equation \eqref{eq:spectraldensity_spherePacking} gives
\begin{equation} \label{eq:spectral_at_0}
\spD{\vect{0}} = \phi^2 \fn{S}{\vect{0}}/\rho,
\end{equation}
and thus, one obtains a hyperuniform packing by decorating a hyperuniform point pattern with spheres of an equal size \cite{Torquato2016_gen}.
In this paper, however, we will not discuss such hyperuniform constructions. 

In the case of a one-component many-particle system in thermal 
equilibrium in which the particles do not overlap, the \textit{fluctuation-compressibility relation} $\fn{S}{\vect{0}} = \rho \kappa_T  k_B T$ \cite{Torquato_RHM, Hansen_statmechReference} and Eq. \eqref{eq:spectral_at_0} yield
\begin{equation}\label{eq:spectral_at_0_equilibrium}
\spD{\vect{0}} = \phi^2 \kappa_T k_B T,
\end{equation}
where $\kappa_T \equiv \rho^{-1} (\partial \rho / \partial p )_T$ is the isothermal compressibility, $p$ is the pressure, $k_B$ is the Boltzmann constant, and $T$ is temperature.
Equation \eqref{eq:spectral_at_0_equilibrium} implies that any compressible ($\kappa_T >0$) one-component system in thermal equilibrium cannot be hyperuniform at a positive temperature \cite{Torquato2015_stealthy, Torquato2018_review}. 

The local volume-fraction variance $\vv{R}$ associated with spherical windows of radius $R$ is defined as \cite{Torquato2018_review, Zachary2009a} 
\begin{equation}\label{eq:localVolFracVariance}
\vv{R} \equiv \E{\fn{\tau ^2}{\vect{x};R}} -{\phi}^2,
\end{equation}
where $\fn{\tau}{\vect{x};R}$ denotes the local volume fraction of the particle phase inside the spherical window of radius $R$ centered at position $\vect{x}$.

From numerical simulations alone, it is difficult to ascertain whether a system is perfectly hyperuniform because the infinite-sample-size limit is never achievable and $\spD{k}$ usually has large relative statistical uncertainties at small wavenumbers.
For these reasons, it is desirable to employ alternative criteria to determine whether a system is \textit{effectively} hyperuniform.
A useful empirical criterion to deem a system to be hyperuniform is that the hyperuniformity metric $H$ is less than $10^{-2}$ or $10^{-3}$ \cite{Atkinson2016, Torquato2018_review, Duyu2018_2}, where $H$ is defined by 
\begin{equation}\label{eq:H-metric}
H \equiv \frac{\spD{\vect{k}\to\vect{0}}}{\spD{\vect{k}_\mathrm{peak}}},
\end{equation}
where $\spD{\vect{k}_\mathrm{peak}}$ is the spectral density at the first dominant (non-Bragg) peak.
Note that this criterion is different from its counterpart for point patterns because of the presence of the form factor in the spectral density \cite{Duyu2018_2}; see Eq. \eqref{eq:spectraldensity_spherePacking}.

\section{Tessellation-Based Procedure}\label{sec:Tiling-based procedure}

\begin{figure}[th]
\includegraphics[width = 0.4 \textwidth]{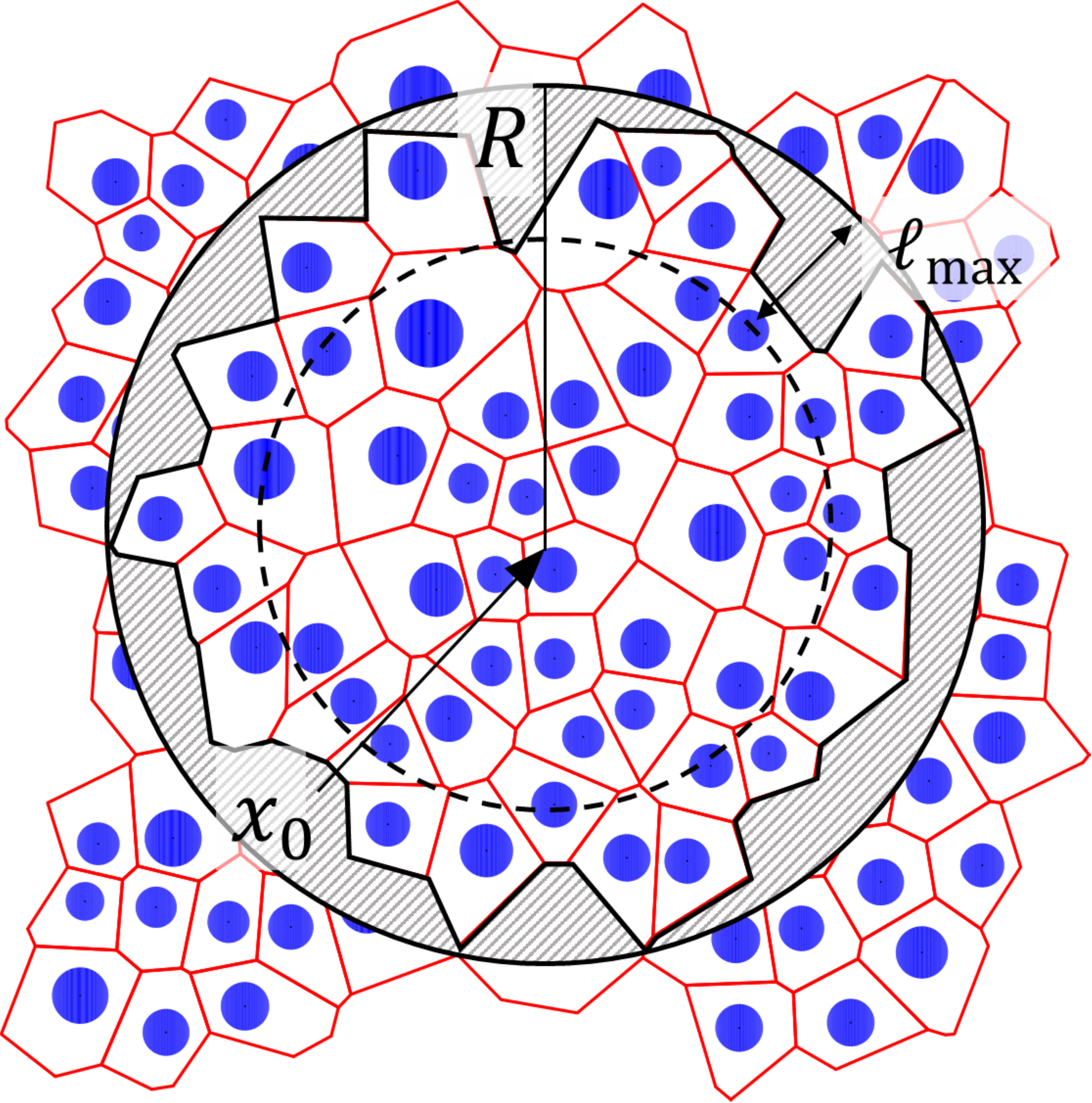}
\caption{(Color online) An illustration of local volume-fraction fluctuations for a disordered packing constructed via the tessellation-based procedure.
For an observation window of radius $R$ and center $\vect{x}_0$, volume-fraction fluctuations arise only from partially covered cells, which are highlighted in gray shade. 
Due to the limited region of fluctuations, this packing becomes hyperuniform in the large sample-size limit; see main text for details. 
\label{fig:illustrations of density fluctuations}}
\end{figure}

Here, we precisely describe the tessellation-based procedure in $d$-dimensional Euclidean space $\R^d$.
For a periodic cubic fundamental cell $\mathcal{V}_F$ of side length $L$ in $\R^d$, the procedure is performed as follows:
\begin{enumerate}
\item Divide the simulation box with $N$ disjoint cells $\tens{C}_1$, $\cdots$, $\tens{C}_N$ [Figs. \ref{fig:schematics}(a) and \ref{fig:schematics}(b)] in which the maximal characteristic linear cell size $\ell_{\max} \equiv \max_{j=1}^N \left \{ \max_{\vect{r}, \vect{r}'\in \tens{C}_j} \{\abs{\vect{r}- \vect{r}'}\} \right\}$ is much smaller than $L$, i.e., 
\begin{equation}\label{def:bounded-cell condition}
\ell_{\max}  \ll L.
\end{equation}
We call this the ``bounded-cell" condition. \label{step1}
\item 
For a specified local-cell packing fraction $0<\phi<1$, place hard particles of arbitrary shapes of total volume $\phi \abs{\tens{C}_j}$ within the $j$th cell $\tens{C}_j$, and then repeat the same process over all cells [see Figs. \ref{fig:schematics}(c) and \ref{fig:schematics}(d) for illustrative examples with disks].
 \label{step2}
\end{enumerate}
Application of this procedure results in a packing in which the local-cell packing fraction $\phi$ is identical to global packing fraction. 
Given an initial tessellation, this construction is realizable only when the local-cell packing fraction $\phi$ in step \ref{step2} is smaller than or equal to the maximal packing fraction $\phi_{\max}$, i.e.,
\begin{equation}\label{eq:max_vol_fraction}
\phi \leq \phi_{\max}\equiv \min_{j=1}^N \left\{ \frac{\abs{\tens{P}_j}_{\max}}{\abs{\tens{C}_j} }	\right\},
\end{equation}
where $\abs{\tens{C}_j}$ and $\abs{\tens{P}_j}_{\max}$ represent volumes of the $j$th cell and the largest particle of a certain shape which is inscribed in this cell, respectively.

Roughly speaking, the maximal packing fraction $\phi_{\max}$ becomes larger when each cell tends to fully enclose a larger particle.
Clearly, the largest particle that a cell can fully circumscribe should be congruent to the cell, and thus the maximal packing fraction will take its largest value ($\phi_{\max} = 1$) only if the shape of each particle is similar to its circumscribing cell.
For this reason, when only considering spherical particles, packing fraction of the multiscale coated-disks model [Fig. \ref{fig:schematics}(d)] can span up to unity, i.e., $\phi_{\max} =1$.

The rationale behind our methodology can be intuitively understood by considering how the volume of the particle phase within a spherical window of radius $R$ fluctuates around its global mean value $\phi\fn{v_1}{R}$.
As shown in Fig. \ref{fig:illustrations of density fluctuations}, our methodology ensures that the cells only in the gray-shaded region can contribute to the fluctuations, which are proportional to volume of this boundary region. 
Since the bounded-cell condition ensures that thickness of this boundary region is smaller than $\ell_{\max}$, the resulting variance in local phase-volume grows on the order of $R^{d-1}$ for sufficiently large windows ($R\gg \ell_{\max} $). 
This implies that $\vv{R}\sim R^{-d-1}$, meaning that this packing is strongly hyperuniform (class I). 
This rationale also explains the hyperuniformity of the constructed packings, regardless of shapes and number of particles inside each cell.  
However, for simplicity, we henceforth focus on the cases in which each cell contains exactly one particle.

\section{General Theoretical Analyses}\label{sec:GeneralTheory}
 
Here, we derive an asymptotic expression for the spectral density for the constructed packings in the small-wavenumber limit.
Consider an initial tessellation $\{\tens{C}_j \}_{j=1}^N$ of a cubic periodic fundamental cell $\mathcal{V}_F$ of side length $L$ in $\R^d$.
Since a fundamental cell is the union of all cells $\tens{C}_1$, $\cdots$, $\tens{C}_N$ of the tessellation, the Fourier transform of $\mathcal{V}_F$ can be decomposed as follows: \begin{align}
\abs{ \mathcal{V}_F} \delta_{\vect{k},\vect{0}}& = \int_{\mathcal{V}_F} \d{\vect{y}} e^{-i \vect{k}\cdot\vect{y}} = \int_{\mathcal{V}_F} \d{\vect{y}} e^{-i\vect{k}\cdot\vect{y}} \sum_{j=1}^N \fn{m}{\vect{y}-\vect{x}_j;\tens{C}_j} \nonumber \\
&  = \sum_{j=1}^N e^{-i\vect{k}\cdot \vect{x}_j}\fn{\tilde{m}}{\vect{k};\tens{C}_j} \label{eq:FT fundamental cell}, 
\end{align}
where $\delta_{\vect{k},\vect{0}}$ represents the Kronecker delta symbol, $\vect{k}$ is a reciprocal lattice vector of $\mathcal{V}_F$, and for the $j$th cell $\tens{C}_j$, $\vect{x}_j$, $\fn{m}{\vect{r};\tens{C}_j}$, and $\fn{\tilde{m}}{\vect{k};\tens{C}_j}$ represent its centroid, the indicator function with respect to $\vect{x}_j$, and the form factor, respectively.

The application of our procedure to this tessellation yields a particle packing that consists of $N$ particles $\tens{P}_1$, $\cdots$, $\tens{P}_N$ whose centroids are $\vect{r}_1$, $\cdots$, $\vect{r}_N$, respectively, with the identical local-cell packing fraction $\phi$.
Using the decomposition of Eq. \eqref{eq:FT fundamental cell} and the spectral density given in Eqs. \eqref{eq:spectral density} and \eqref{eq:FourierTransform_particle}, we obtain the following general expression: 
\begin{align}
&\spD{\vect{k}} = \frac{1}{\abs{\mathcal{V}_F}} \abs{\fn{\tilde{\mathcal{J}}}{\vect{k}}}^2 \nonumber \\
=& \frac{1}{\abs{ \mathcal{V}_F} } \Bigg| \sum_{j=1}^N e^{-i \vect{k}\cdot\vect{x}_j}  \Big[ \fn{\tilde{m}}{\vect{k};\tens{P}_j}e^{-i \vect{k}\cdot\Delta\vect{X}_j}  -\phi \fn{\tilde{m}}{\vect{k};\tens{C}_j}	\Big] \Bigg|^2 , \label{eq:chi_v_step2}
\end{align}
where $\Delta \vect{X}_j \equiv \vect{r}_j - \vect{x}_j$, and $\fn{\tilde{m}}{\vect{k};\tens{P}_j}$ is the form factor of a particle $\tens{P}_j$. 

Due to the bounded-cell condition, $\abs{\vect{r}} < \ell_{\max} \ll L$ for every $\vect{r}$ in each  cell $\tens{C}_j$, or equivalently, one can consider small wavevectors satisfying that $\abs{\vect{k}\cdot \vect{r}} \sim \abs{\vect{r}}/L \ll 1$. 
Thus, the form factor of $\mathcal{V}=(\tens{C}_j\text{ or }\tens{P}_j$) can be well approximated by its Taylor series about $\vect{k} = \vect{0}$: \begin{align} 
\fn{\tilde{m}}{\vect{k};\mathcal{V}}
=& \abs{ \mathcal{V}}\Bigg[1 - \frac{k_\alpha k_\beta}{2} \fn{ \mathcal{M}_{\alpha \beta}}{\mathcal{V}} \nonumber \\
&+ \frac{ik_\alpha k_\beta k_\gamma}{6} \fn{\mathcal{M}_{\alpha\beta\gamma}}{\mathcal{V}}\Bigg]+\order{k^4}, \label{eq:FT fundamental cell2} 
\end{align}
where the Einstein summation convention is employed, 
\begin{equation}
\fn{ \mathcal{M}_{\alpha_1 \alpha_2 \cdots \alpha_n}}{\mathcal{V}} \equiv \frac{1}{\abs{\mathcal{V}}} \int_{ \mathcal{V}} \d{\vect{r}} r_{\alpha_1} r_{\alpha_2} \cdots r_{\alpha_n} \label{eq:moment},
\end{equation}
is the $m$th moment of the mass distribution of $\mathcal{V} (=\tens{C}_j \text{ or } \tens{P}_j)$ that is normalized by its volume $\abs{\mathcal{V}}$, and $r_{\alpha_j}$ represents  the $\alpha_j$th Cartesian component of a vector $\vect{r}$.
We note that since Eq. \eqref{eq:moment} refers to the moments with respect to the centroid of $\mathcal{V}$, its first moment is identically zero.


Using Eq. \eqref{eq:FT fundamental cell2}, the term $\fn{\tilde{\mathcal{J}}}{\vect{k}}$ given in Eq. \eqref{eq:chi_v_step2} can be written as follows (see Appendix \ref{sec:App_FT_phase_indicator} for details):
\begin{align}\label{eq:F.T._phase_indicator}
\fn{\tilde{\mathcal{J}}}{\vect{k}} & = \fn{\tilde{\mathcal{J}}_{(1)}}{\vect{k}} + \fn{\tilde{\mathcal{J}}_{(2)}}{\vect{k}} +\fn{\tilde{\mathcal{J}}_{(3)}}{\vect{k}} + \order{k^4},
\end{align}
where 
\begin{align}
\fn{\tilde{\mathcal{J}}_{(1)}}{\vect{k}} 
=&
\phi \sum_{j=1}^N (e^{-i\vect{k}\cdot\Delta\vect{X}_j}-1) \abs{\tens{C}_j}e^{-i\vect{k}\cdot\vect{x}_j},  \label{def:J1}
\end{align}
and $\fn{\tilde{\mathcal{J}}_{(2)}}{\vect{k}}$ and $\fn{\tilde{\mathcal{J}}_{(3)}}{\vect{k}}$ are defined in Eq. \eqref{eq:weight function}.
Since the leading order terms of $\fn{\tilde{m}}{\vect{k};\tens{P}_j }$ and $\fn{\tilde{m}}{\vect{k};\tens{C}_j}$ exactly cancel each other due to the local constraint $\abs{\tens{P}_j}=\phi \abs{\tens{C}_j}$ for all $j$, these three terms $\fn{\tilde{\mathcal{J}}_{(1)}}{\vect{k}}$, $\fn{\tilde{\mathcal{J}}_{(2)}}{\vect{k}}$, and $\fn{\tilde{\mathcal{J}}_{(3)}}{\vect{k}}$ exhibit the power-law scalings in the small-wavenumber limit.
Assuming typical tessellations that exhibit $\abs{\fn{\tilde{\mathcal{J}}_{(2)}}{\vect{k}}}\sim \abs{\vect{k}}^2$ and $\abs{\fn{\tilde{\mathcal{J}}_{(3)}}{\vect{k}}}\sim \abs{\vect{k}}^3$, the first term can have two scalings [either $\abs{\fn{\tilde{\mathcal{J}}_{(1)}}{\vect{k}}}\sim \abs{\vect{k}}$ or $\fn{\tilde{\mathcal{J}}_{(1)}}{\vect{k}}\sim \order{\abs{\vect{k}}^2}$], depending on the particle displacements $\Delta\vect{X}_j$ with respect to their cell centroids.
Therefore, the scaling of $\fn{\tilde{\mathcal{J}}_{(1)}}{\vect{k}}$ determines that of the spectral density \eqref{eq:chi_v_step2}.

Specifically, whenever $\abs{\fn{\tilde{\mathcal{J}}_{(1)}}{\vect{k}}}\sim \abs{\vect{k}}$, which can be achieved when particle displacements $\Delta\vect{X}_j$ are uncorrelated with one another \footnote{One such set of structures are perturbed crystal packing that result from stochastically displacing particles within a cell of a periodic Voronoi tessellation \cite{Gabrielli2004, Jaeuk2018}.}, the spectral density of the constructed packings tends to zero quadratically in $\abs{\vect{k}}$; specifically, 
\begin{equation}\label{eq:scaling_general}
\spD{\vect{k}} \sim \phi^2 \abs{\vect{k}}^2,
\end{equation}
which corresponds to class I hyperuniformity.
When $\abs{\fn{\tilde{\mathcal{J}}_{(1)}}{\vect{k}}} \sim \order{\abs{\vect{k}}^2}$, the packings are more strongly hyperuniform with a new scaling given by 
\begin{equation}
\label{eq:scaling_stat_isotropy}
\spD{\vect{k}} \sim \phi^2 \abs{\vect{k}}^4 .
\end{equation}
For example, this scaling can be achieved when $\Delta\vect{X}_j = \vect{0}$ for all $j$ [i.e., $\fn{\tilde{\mathcal{J}}_{(1)}}{\vect{k}}=0$] or when $\abs{\fn{\tilde{\mathcal{J}}_{(1)}}{\vect{k}}}\sim \abs{\vect{k}}^2$ due to certain spatial correlations in $\Delta\vect{X}_j$ [see Sec. \ref{sec:byVoronoiTessellations}].
Therefore, the manipulation of particle displacements $\Delta\vect{X}_j$ enables us to engineer either quadratic or quartic scalings of the spectral density.

It is noteworthy that the appearance of $\phi^2$ factor in Eqs. \eqref{eq:scaling_general} and \eqref{eq:scaling_stat_isotropy} is common for small-$\abs{\vect{k}}$ scalings of the spectral densities of all statistically homogeneous sphere packings: see Eq. \eqref{eq:spectraldensity_spherePacking}.   
Here, we note that the theoretical results \eqref{eq:scaling_general} and \eqref{eq:scaling_stat_isotropy} can be straightforwardly generalized to the cases of multiple particles are added in each cell. 
This is achieved by dividing cells such that each subdivided cell should circumscribe a single particle with an identical local-cell packing fraction.

Now, we consider a special case where all particles are similar to the associated cells in the sense that the particles have identical shapes and orientations with their cells, but have different sizes ($\tens{P}_j = \phi^{1/d}\tens{C}_j$ for $j=1,\cdots, N$).
Then, the $n$th moments of particles and cells can be related as $ \fn{\mathcal{M}_{\alpha_1 \cdots \alpha_n}}{\tens{P}_j}=\phi^{n/d}\fn{\mathcal{M}_{\alpha_1 \cdots \alpha_n}}{\tens{C}_j}$.
Substituting this expression into Eq. \eqref{eq:F.T._phase_indicator}, we obtain 
\begin{align}\label{eq:F.T._phase_indicator_similarParticles}
\fn{\tilde{\mathcal{J}}}{\vect{k}}&= \fn{\tilde{\mathcal{J}}_{(1)}}{\vect{k}} \\ 
+ &
\underbrace{\phi(1-\phi^{2/d})\frac{k_\alpha k_\beta}{2}\sum_{j=1}^N \fn{\mathcal{M}_{\alpha\beta}}{\tens{C}_j} \abs{\tens{C}_j} e^{-i \vect{k}\cdot\vect{x}_j}}_{\fn{\tilde{\mathcal{J}}_{(2)}}{\vect{k}}} +\order{k^3}.\nonumber 
\end{align}
Again, whenever $\abs{\fn{\tilde{\mathcal{J}}_{(1)}}{\vect{k}}}\sim \abs{\vect{k}}$, the spectral density of the constructed packings shows a scaling $\spD{\vect{k}} \sim \phi^2 \abs{\vect{k}}^2$. 
However, in the special case of $\Delta\vect{X}_j=\vect{0}$ for all $j$ (i.e., $\tilde{\mathcal{J}}_{(1)}=0$), the resulting scaling is 
\begin{equation}\label{eq:scaling_stat_isotropy_similar}
\spD{k}\sim \phi^2 (1-\phi^{2/d})^2 \abs{\vect{k}}^4.
\end{equation}
 Both hyperuniform cases belong to the class I. 

Here, we should note that all theoretically predicted scalings of $\spD{k}$, given in Eqs. \eqref{eq:scaling_general}, \eqref{eq:scaling_stat_isotropy}, and \eqref{eq:scaling_stat_isotropy_similar}, are analytic at the origin, i.e., the power exponents are even positive integers.
This implies that autocovariance functions $\fn{\chi_{_V}}{\abs{\vect{r}}}$ of the constructed packings must decay to zero exponentially fast (or faster) as $\abs{\vect{r}}\to \infty$ \cite{Torquato2018_review}.


\begin{figure*}[ht]
\subfloat[]{\label{fig:MaxGapDist}
\includegraphics[height = 0.2\textwidth]{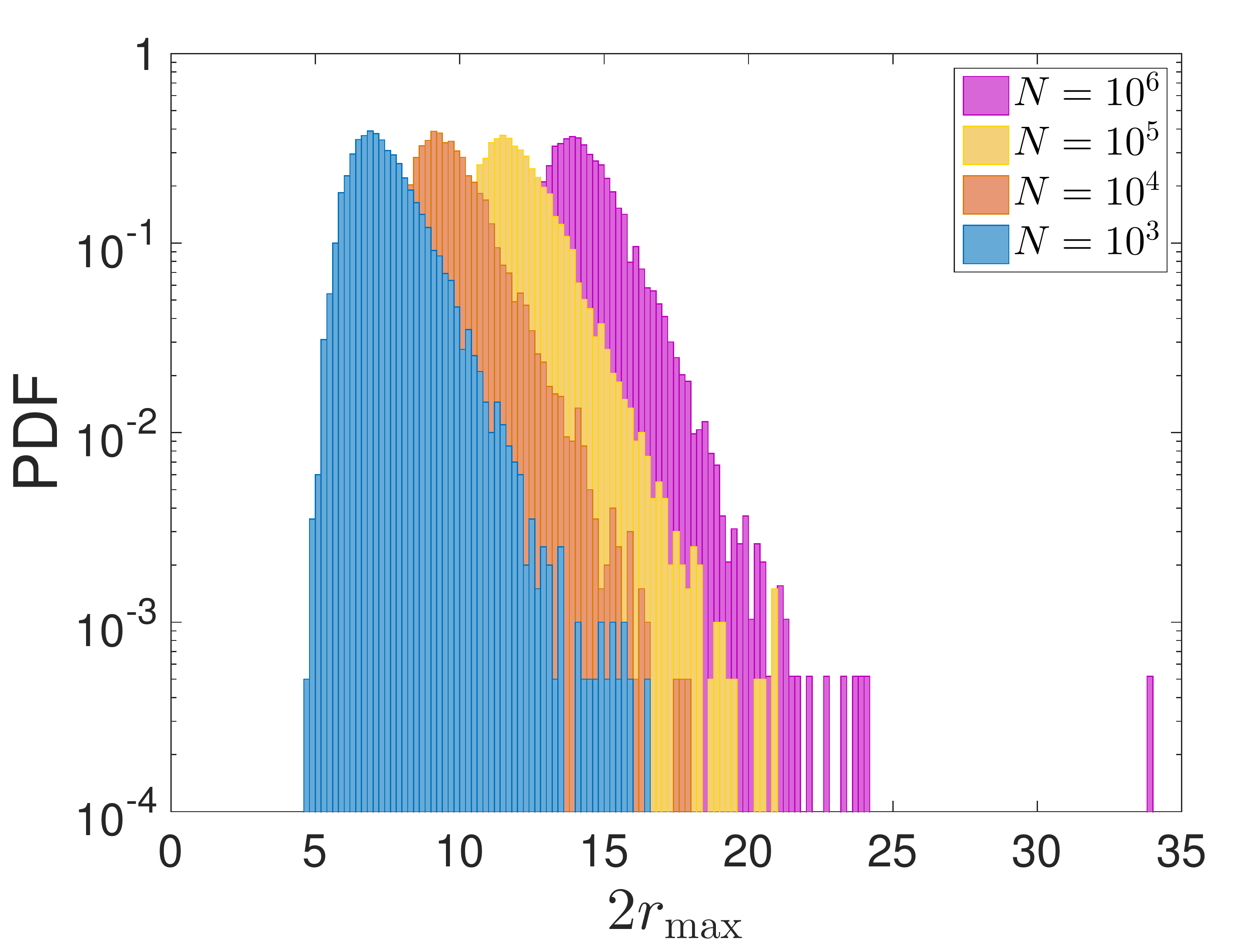}}
\subfloat[]{\label{fig:MaxGapDist_scaled}
\includegraphics[height = 0.2\textwidth]{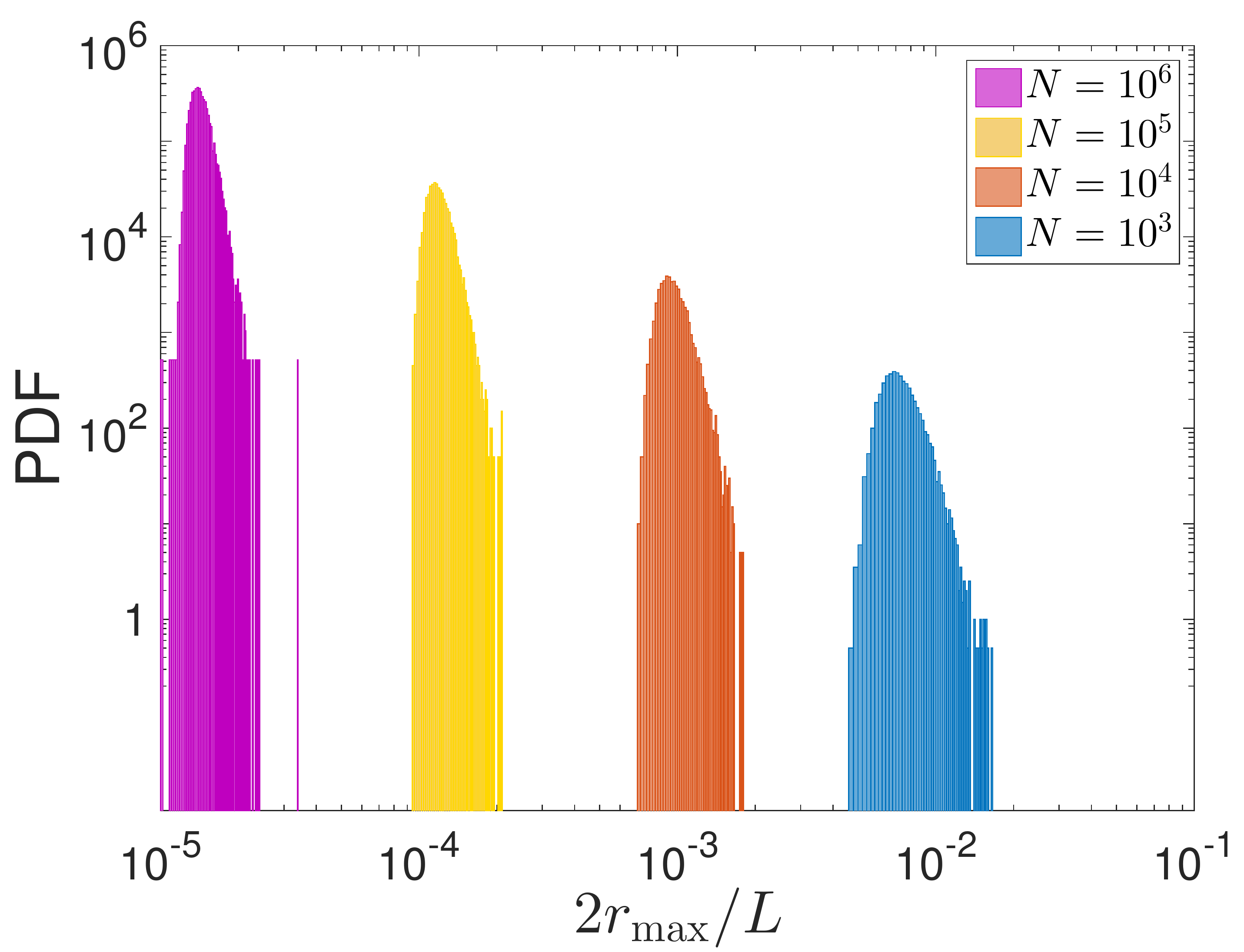}}
\subfloat[]{\label{fig:Ev_Poisson}
\includegraphics[height = 0.2\textwidth]{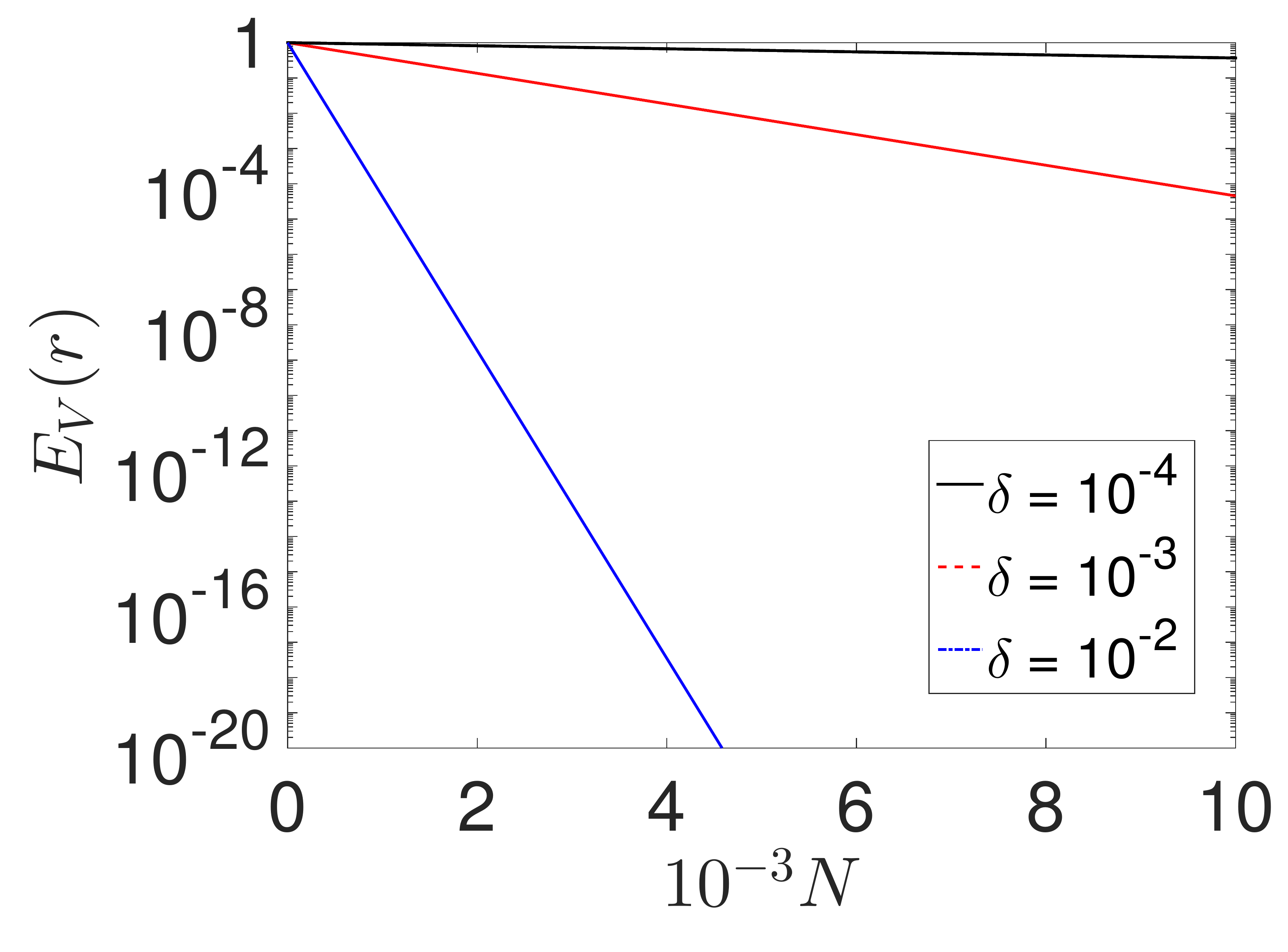}}
\caption{(Color online) Numerical simulations of probability distributions of the largest hole volume ($2r_{\max}$) in  each sample of 1D Poisson point patterns: \subcap{fig:MaxGapDist} on a   semilog scale and \subcap{fig:MaxGapDist_scaled} on a log-log scale.
The $x$-axis in \subcap{fig:MaxGapDist_scaled} represents the relative size $\delta=\fn{v_1}{r_\mathrm{max}}/V$ of the largest holes to the box volume $V(=L)$. For each particle number $N$, the distribution is obtained from $10^4$ independent samples. 
\subcap{fig:Ev_Poisson} Semi-log plot of hole probability $\fn{E_V}{r}$ of finite-size Poisson point patterns as a function of system size $N$. 
Values are computed from Eq. \eqref{eq:Ev_Poisson}.
\label{fig:Illust.boundedCellCond}}
\end{figure*}

\section{Hyperuniform Packings from Voronoi Tessellations}\label{sec:byVoronoiTessellations}
  
In this section, we formulate an efficient numerical algorithm that is based on our tessellation-based methodology to generate very large disordered hyperuniform packings in $\R^2$ and $\R^3$.
This can be accomplished from various types of tessellations \cite{Hiroshi1985, Gellatly1982, Deuschel2012, Florescu2009, Torquato2018_3, Gabbrielli2012}, but we focus on Voronoi tessellations of disordered nonhyperuniform point patterns in this section.
For a given point pattern, the Voronoi cell of a point is defined as the region of space closer to this point than any other points, and the Voronoi tessellation is the collection of all Voronoi cells \cite{Torquato_RHM}.
The computational time for the Voronoi tessellation of a point pattern of $N$ particles in $\R^d$ is at most of the order of $\order{N\log N + N^{\ceil{(d-1)/2}}}$ \cite{Aurenhammer_VoronoiDiagram}.
Due to such small computation cost, the implementation of our methodology in the case of Voronoi tessellations enables a remarkably efficient mapping that converts very large nonhyperuniform point patterns or packings into very large disordered hyperuniform packings.  

We choose the progenitor point patterns for the Voronoi tessellations not only to meet the bounded-cell condition but so that the resulting packing is easy to fabricate. 
From a practical viewpoint, it is useful to employ similar particle shapes and sizes to construct the packings.  
All of these conditions can be readily fulfilled by considering progenitor point patterns derived from dense hard-sphere packings.
In what follows, we elaborate the bounded-cell condition.

\subsection{The bounded-cell condition}\label{sec:bounded-cellcondition}
  
The bounded-cell condition in the initial tessellation is a central requirement to ensure hyperuniformity via the tessellation-based procedure. 
For Voronoi tessellations, the largest cell size $\ell_{\max}$ is on the order of the largest nearest neighbor distance, which in turn is of the order of the largest radius $r_{\max}$ of \textit{holes} (i.e., spherical regions that are empty of particle centers); $\ell_{\max}\sim r_{\max}$. 
Therefore, the bounded-cell condition \eqref{def:bounded-cell condition} can be satisfied if the relative size $r_{\max}/L$ of the largest hole to the sample size [or equivalently, their volume ratio $\delta\equiv \fn{v_1}{r_{\max}}/L^d$] is much smaller than a certain small value.

The information about largest hole size is contained in the void-exclusion or \textit{hole probability} function $\fn{E_V}{r}$, which gives the probability for finding a hole of radius $r$ when it is randomly placed in a point-particle system in the thermodynamic limit \cite{Torquato_RHM}.
Clearly, if the hole probability has \textit{compact support}, i.e., $\fn{E_V}{r}=0$ at any $r>\mathcal{D}$ for a certain length $\mathcal{D}$, then Voronoi tessellations of the associated point patterns always meet the bounded-cell condition for relatively small sample sizes (say $10^{d+1}$ in space dimension $d$). 
Examples of such systems include all crystals, disordered stealthy hyperuniform point patterns \cite{Zhang2017, Ghosh2017_holeConjecture}, and the saturated random sequential addition (RSA) packings \cite{Torquato_RHM, Zhang2013}.
Specifically, RSA is a time-dependent process that irreversibly, randomly, and sequentially adds nonoverlapping spheres into space.
In the infinite-time limit the resulting packing does not have any available space to add further particles, called \textit{saturated}.

However, many other disordered point patterns, including Poisson point patterns, equilibrium hard-sphere liquids, and unsaturated RSA packings, can possess arbitrarily large holes in the thermodynamic limit.
Finite-size samples of these systems tend to have larger holes as the sample size grows [see Fig. \subref*{fig:MaxGapDist} for 1D Poisson point patterns], but the hole size relative to the sample size decreases as the sample size grows [see Fig. \subref*{fig:MaxGapDist_scaled}], implying that samples, in fact, tend to meet the bounded-cell condition \eqref{def:bounded-cell condition}.

We now rigorously show that for sufficiently large statistically homogeneous point patterns, Voronoi tessellations of almost every realization should obey the bounded-cell condition, as long as their hole probabilities are smaller than or equal to that of the Poissonian counterparts for large hole radii. 
For this purpose, we will show that Voronoi tessellations of almost every Poisson point pattern obey the bounded-cell condition for sufficiently large sample sizes. 
We begin by considering{\color{red},}~for finite-size samples, the hole probability $\fn{E_V}{r}$ which can be interpreted as the probability that a finite sample possesses at least a single hole of radius greater than $r$ (i.e., $\ell_{\max}\sim r_{\max} \geq r$). 
Thus, $\fn{E_V}{L[\delta / \fn{v_1}{1}]^{1/d}}$ is, in turn, the probability that the Voronoi tessellation of a single sample does not meet the bounded-cell condition \eqref{def:bounded-cell condition}.
For Poisson point patterns of $N$ particles and volume $V=L^d$ in $\R^d$, its hole probability can be straightforwardly obtained as follows \footnote{
In a Poisson point pattern, the probability that a single particle is placed in the space such that a hole of radius $r$ is empty of a particle is $(V-\fn{v_1}{r})/V$ because all positions are equally probable. Since particle positions are mutually independent, the hole probability $\fn{E_V}{r}$ that $N$ particles are placed in the space such that this hole is empty of particles is simply $[(V-\fn{v_1}{r}/V]^N$.}:  
\begin{equation}\label{eq:Ev_Poisson}
\fn{E_V}{r} = [V-\fn{v_1}{r}]^N / V^N = (1-\delta)^N,
\end{equation}
where $\delta\equiv \fn{v_1}{r}/V$.
This quantity converges to a well-known expression in the thermodynamic limit \cite{Torquato_RHM} $\fn{\exp}{-\rho\fn{v_1}{r}}$, where $\rho$ is the number density.
Equation \eqref{eq:Ev_Poisson} decays exponentially fast for a given $\delta< 1$ as particle number $N$ increases; see Fig. \subref*{fig:Ev_Poisson}.
This implies that for sufficiently large sample sizes, almost every realization of a Poisson point pattern should obey the bounded-cell condition.
Thus, the constructed packings from Voronoi tessellations of Poisson point patterns should be hyperuniform, which is consistent with a recent study of random fields \cite{Kampf2015}.

We will employ correlated and statistically homogeneous point patterns as the progenitor configurations, which are even more likely to obey the bounded-cell condition than Poisson-point counterparts at the same number density. 
In the ensuing discussion, we show that such nonhyperuniform point patterns can be converted into disordered hyperuniform packings.   

\subsection{Spherical particles}\label{sec:Voronoi_spherical}

\begin{figure}[t]
\begin{center}
\includegraphics[width = 0.35\textwidth]{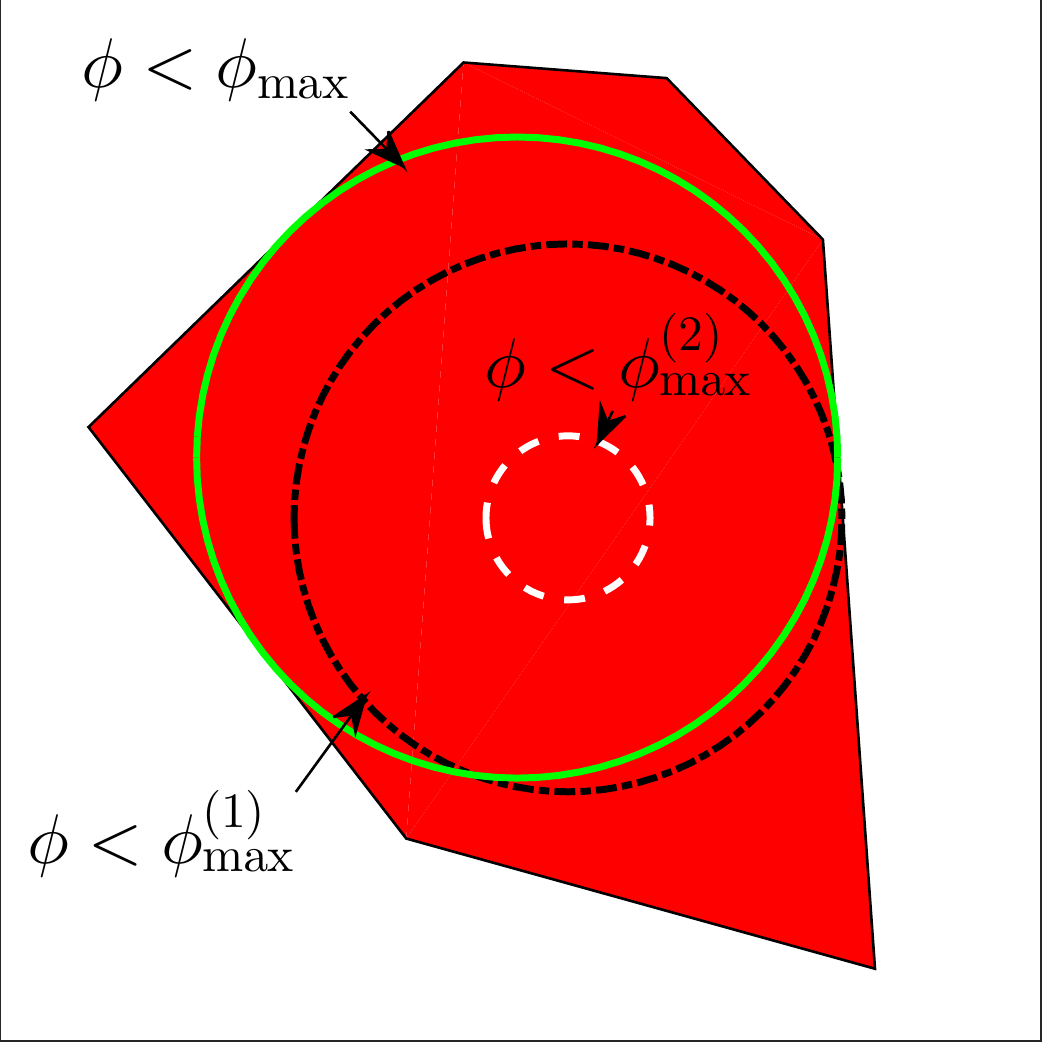}
\end{center}
\caption{(Color online) Schematic for three types of the maximal packing fractions $\phi_{\max}^{(2)} \leq \phi_{\max}^{(1)}\leq \phi_{\max}$.
For a Voronoi cell of an initial particle (white circle), the black circle and the green one represent the largest inscribed circles when the circle center is fixed or is free to move, respectively.
These three circles (white, black, and green ones) illustrate the largest particles $\abs{\tens{P}}_j$ of this cell for three distinct maximal packing fractions $\phi_{\max}^{(2)}$, $\phi_{\max}^{(1)}$, and $\phi_{\max}$, respectively. 
We note that the initial particle size is exaggerated for a clear visualization.
 \label{fig:schem_max_packingfraction}}
\end{figure}

Here, we numerically implement our procedure by solely rescaling particle volumes without changing particle centers and shapes.
It results in disordered sphere packings whose Voronoi tessellations are identical to those of their progenitor packings and we show that they are exactly hyperuniform of class I.

Due to the fixed particle centers, we define two alternative maximal packing fractions $\phi_{\max}^{(1)}$ and $\phi_{\max}^{(2)}$; see Fig. \ref{fig:schem_max_packingfraction}. For $\phi_{\max}^{(1)}$, the volume $\abs{\tens{P}_j}_{\max}$ of the largest particle of a Voronoi cell in definition \eqref{eq:max_vol_fraction} is that of the smallest spherical particle inscribed in the cell.
For $\phi_{\max}^{(2)}$, the largest particle is the particle in the progenitor packing.
By definition, we have the following inequalities: $\phi_{\max}\geq \phi^{(1)}_{\max} \geq \phi^{(2)}_{\max}$, where $\phi^{(1)}_{\max} = \phi^{(2)}_{\max}$ occurs only when every particle in the progenitor has a neighbor in contact, and $\phi_{\max} = \phi^{(1)}_{\max}$ occurs only when every particle in the progenitor is inscribed in its Voronoi cell. 
Obviously, if a constructed packing has the packing fraction $\phi \leq \phi_{\max}^{(2)}$, none of its particles can be larger than those in the progenitor packing.
For 2D and 3D saturated RSA packings, values of $\phi^{(1)}_{\max}$ are around $0.35$ or $0.25$, respectively.
These values tend to decrease as the packing fraction of the progenitor packings is smaller, and they become arbitrarily small for Poisson point patterns (i.e., RSA with the zero packing fraction).
Additional values of $\phi^{(1)}_{\max}$ and $\phi^{(2)}_{\max}$ are summarized in Sec. IB in the Supplemental Material \cite{Supplementary}. 

\begin{figure}[t]
\begin{center}
\subfloat[]{\label{fig:ex_2D_RSA_unsat}
 \includegraphics[width = 0.23\textwidth]{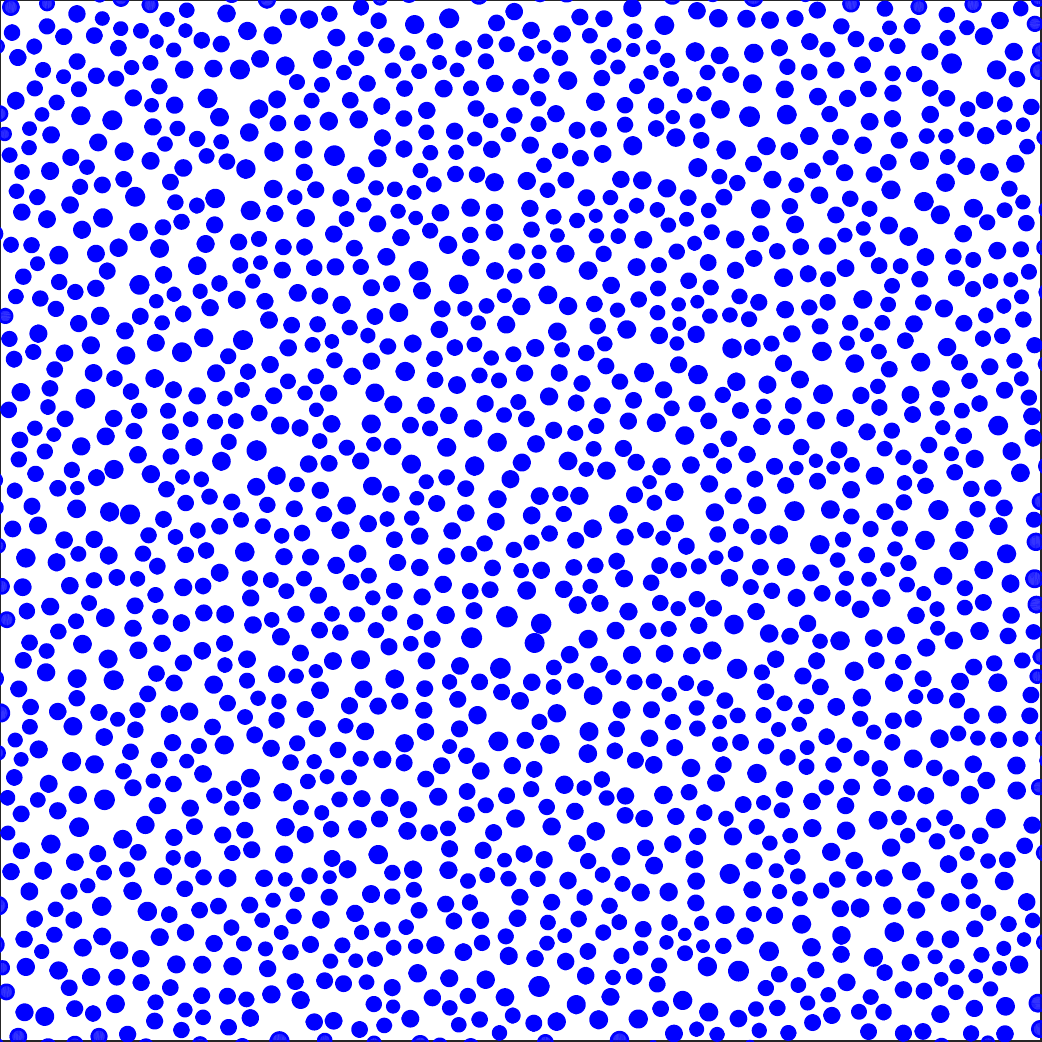}}
 \subfloat[]{\label{fig:ex_3D_RSA_sat}
 \includegraphics[width = 0.23\textwidth]{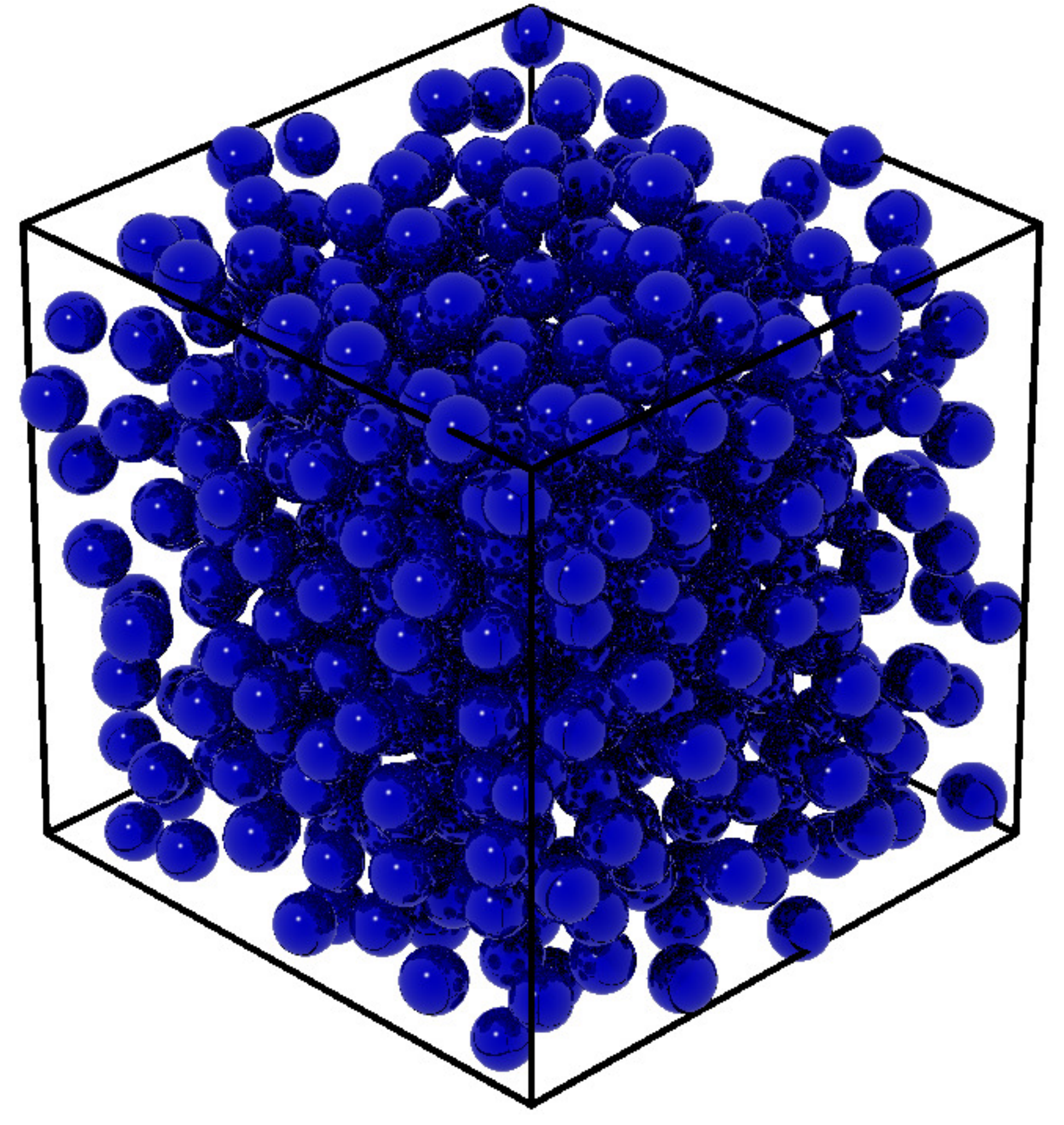}}
\end{center}
\caption{(Color online) \subcap{fig:ex_2D_RSA_unsat} A portion of a hyperuniform disk packing that was converted from a 2D RSA packing with the packing fraction $\phi_{\mathrm{init}} = 0.41025$.
\subcap{fig:ex_3D_RSA_sat} A portion of a hyperuniform sphere packing that was converted from a 3D saturated RSA packing. 
\label{fig:ex_constructed}}
\end{figure}
 
\begin{figure}
\includegraphics[width=0.4\textwidth]{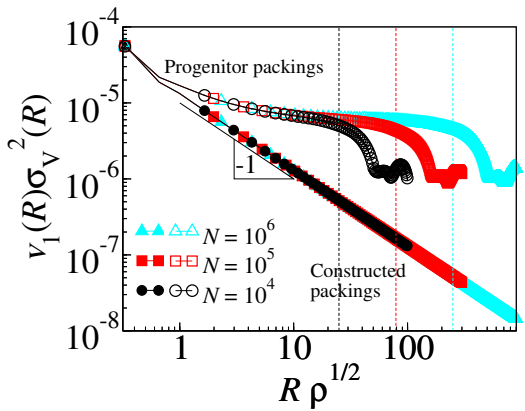}
\caption{(Color online) Log-log plot of the scaled local volume-fraction variance $\fn{v_1}{R}\vv{R}$ of the progenitor and constructed packings. 
The scaling of the latter clearly shows that it is hyperuniform of class I.
The progenitor packings are 2D saturated RSA packings and the Monte Carlo technique is employed.
Three vertical grids represent a quarter of side length of the simulation boxes (i.e., $R=L/4$) for $N=10^4$, $10^5$, and $10^6$, respectively.
\label{fig:2D_RSA_local_volumefraction_variance}}
\end{figure}

\begin{figure*}[ht]
\subfloat{\label{fig:2DRSA_Voronoicellvolume}
\includegraphics[width = 0.3\textwidth]{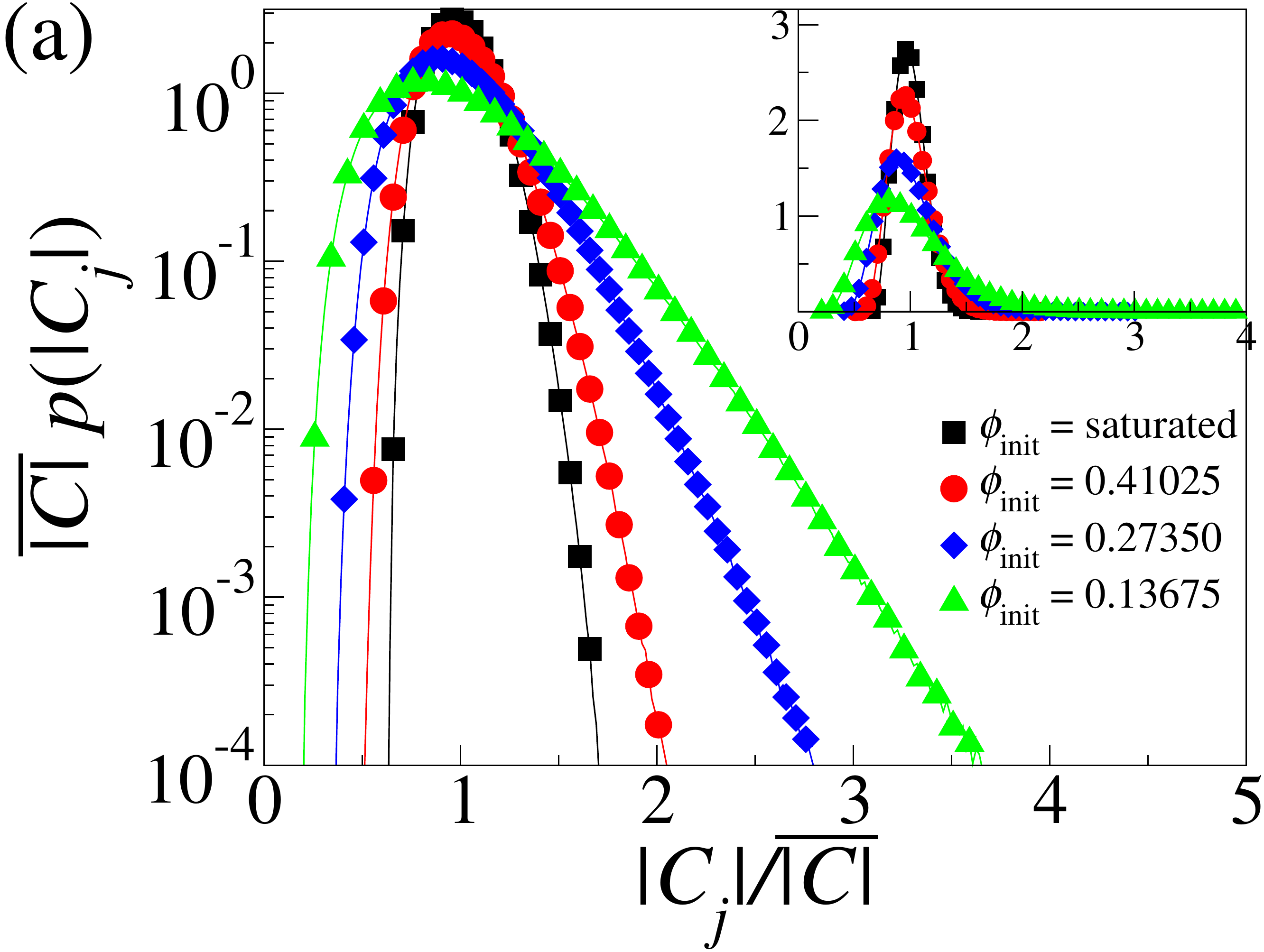}
}
\subfloat{\label{fig:chivk_2DRSA}
\includegraphics[width = 0.3\textwidth]{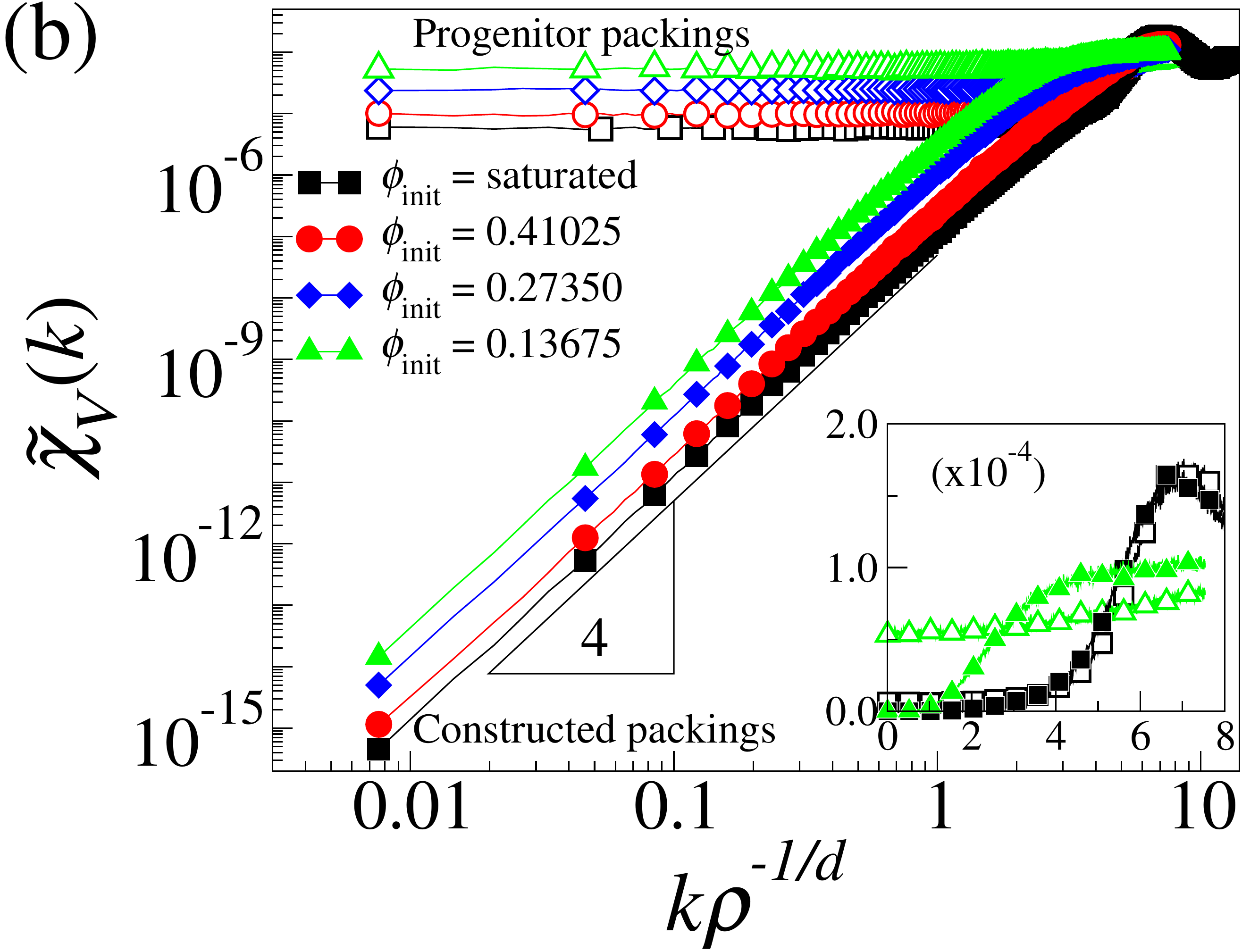}}
\subfloat{\label{fig:2DRSA_chivk_size}
\includegraphics[width = 0.3\textwidth]{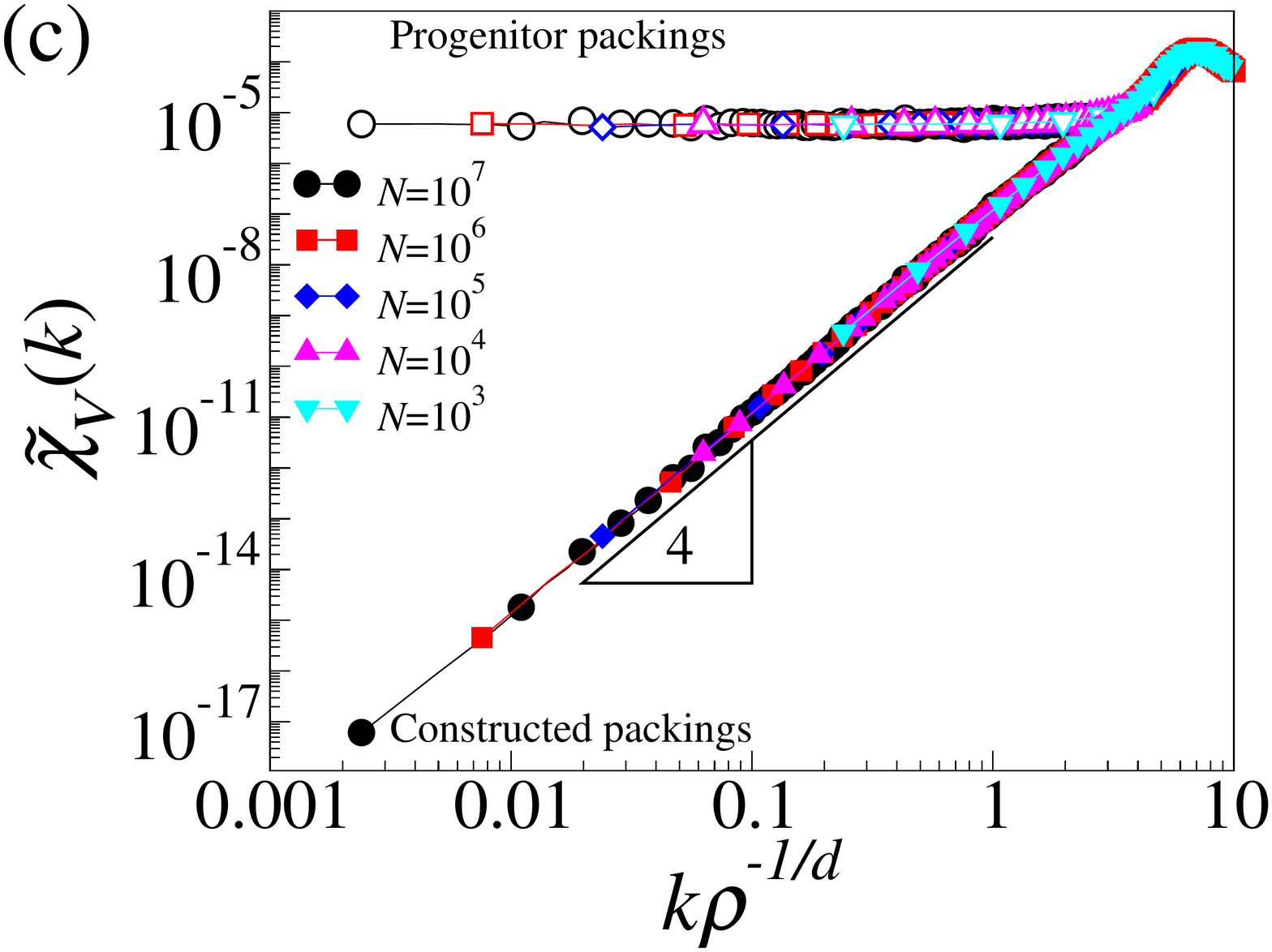}}

\subfloat{\label{fig:3DRSA_Voronoicellvolume}
\includegraphics[width = 0.3\textwidth]{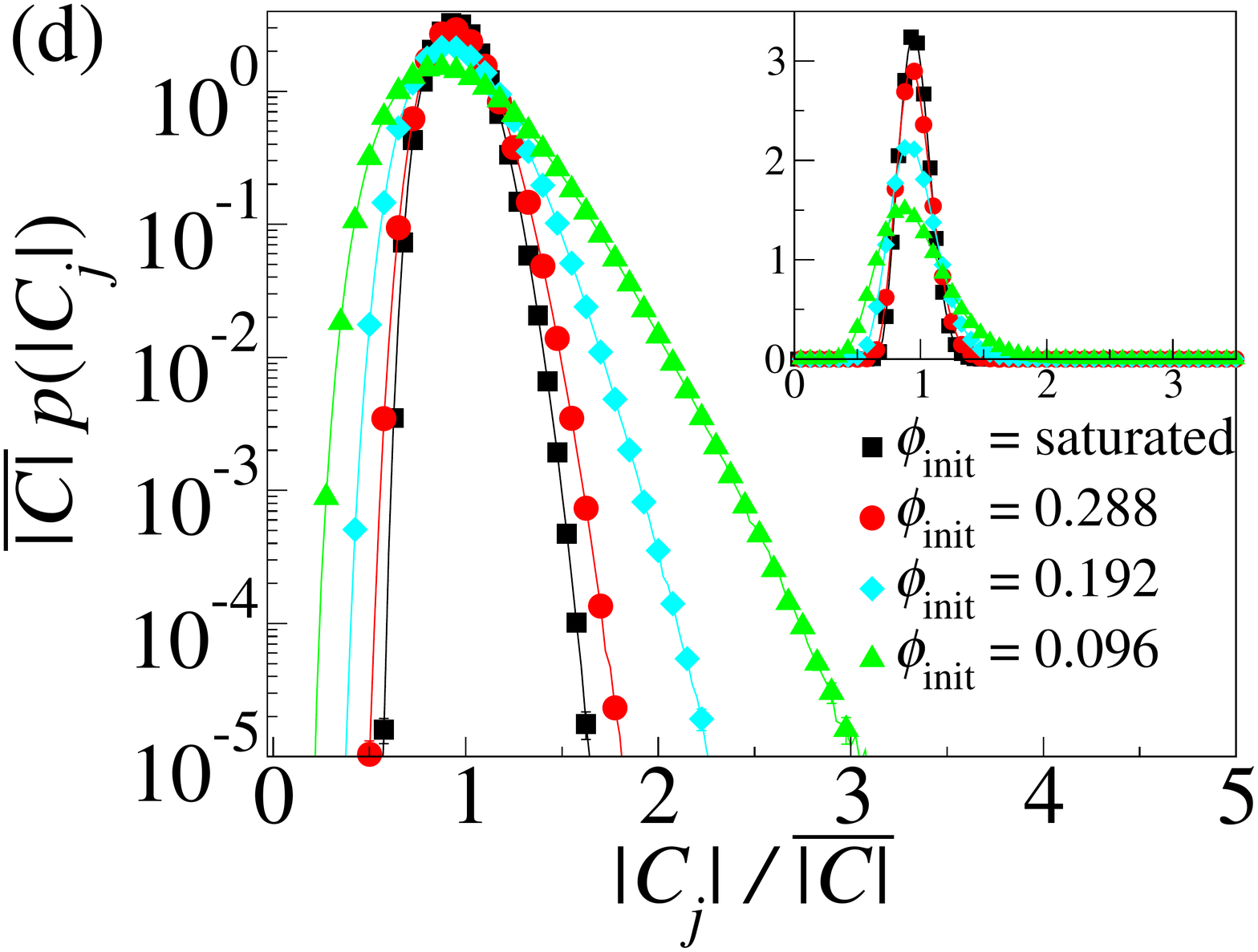}}
\subfloat{\label{fig:3DRSA_chivk}
\includegraphics[width = 0.3\textwidth]{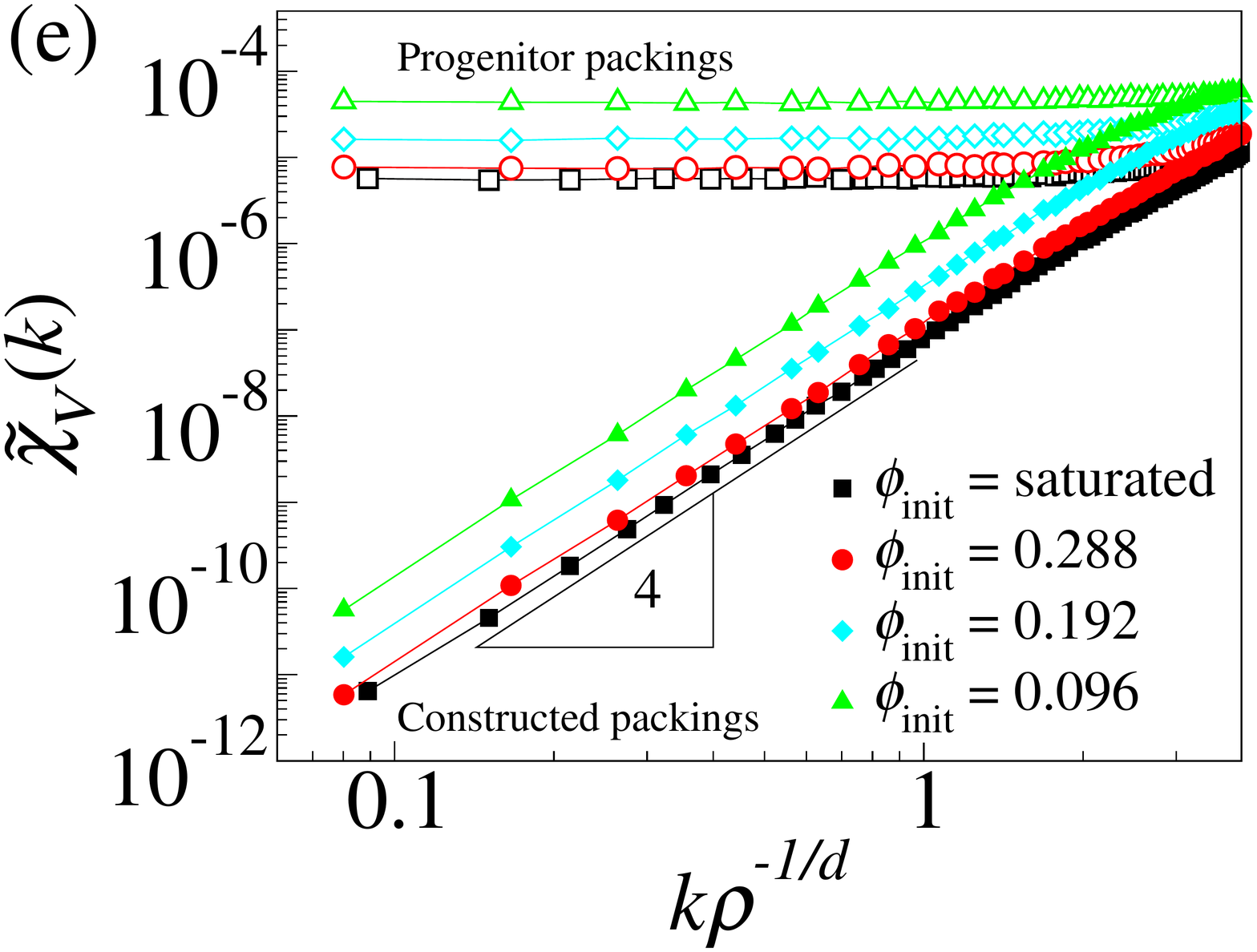}}
\subfloat{\label{fig:3DRSA_chivk_size}
\includegraphics[width = 0.3\textwidth]{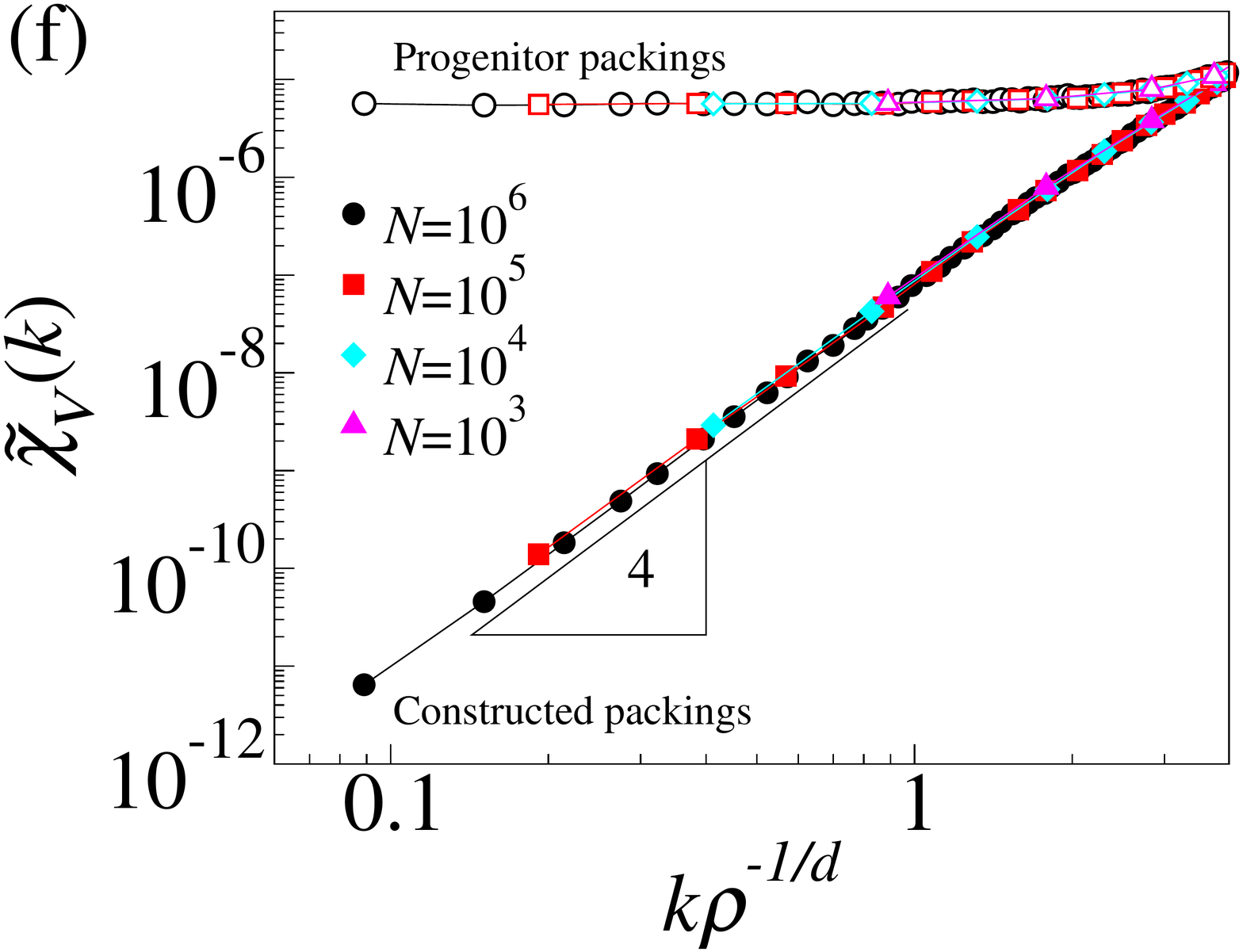}}

\caption{(Color online) Simulation results for sphere packings from Voronoi tessellations of RSA packings in (a-c) $\R^2$ and (d-f) $\R^3$. 
(a, d) Probability density $\fn{p}{\abs{\tens{C}_j}}$ of Voronoi cell volumes versus the scaled cell volume $\abs{\tens{C}_j}/\overline{\abs{\tens{C}}}$ plotted on a semi-log scale (larger panel) and a linear scale (inset). Here, $\overline{\abs{\tens{C}}}$ represents the expected cell volume. 
(b, e) The spectral densities versus wavenumber $k$ for small wavenumbers plotted on a log-log scale. 
Inset in \subcap{fig:chivk_2DRSA} is on a linear scale.  
Here, we note that all packings are rescaled to the common packing fraction $\phi=0.01$.
(c, f) Log-log plots of the spectral densities for progenitor packings (saturated RSA) and associated constructed packings according to sample size $N$. 
\label{fig:Results_RSA}}
\end{figure*}

We employ three different types of nonhyperuniform progenitor point patterns in both $\R^2$ and $\R^3$: RSA packing \cite{Torquato_RHM, Zhang2013}, equilibrium hard-sphere liquids, and lattice-packings with point vacancies.
First, we employ RSA packings for various values of initial packing fractions $\phi_\mathrm{init}$.
These RSA packings are efficiently generated by the voxel-list algorithm \cite{Zhang2013}.
Second, as progenitor packings, we use equilibrium hard-sphere liquid configurations with a range of values of the initial packing fraction $\phi_\mathrm{init}$; see the Supplemental Material \cite{Supplementary} for employed parameters.
In 2D, its $\spD{\vect{0}}$ can be well approximated from Eq. \eqref{eq:spectral_at_0_equilibrium} and a formula for the pressure \footnote{The isothermal compressibility $\kappa_T \equiv -V^{-1}\pd{V}{p}{T}$ of 2D equilibrium hard-sphere liquids with the packing fraction $\phi$ is computed from the equation of states $p/(\rho k_B T) = 1+2\phi \fn{g_2}{D}$, where $\fn{g_2}{D}$ is the contact value of the radial pair-correlation \cite{Torquato_RHM}. The contact value is $\fn{g_2}{D}=\fn{G}{D}=(1-0.436\phi)/(1-\phi)^2$ for $0\leq \phi \leq \phi_f$; see Ref. \cite{Torquato1995_2}} as follows: 
\begin{equation}\label{eq:S0_2D_HSF}
\spD{\vect{0}} \approx \frac{\phi^2(1-\phi_\mathrm{init})^3}{ \rho (1 +\phi_\mathrm{init} + 0.38406{\phi_\mathrm{init}}^2 -0.12802{\phi_\mathrm{init}}^3)},
\end{equation} 
where $\phi$ is the packing fraction of the decorated spheres.

\begin{table}[b]
\caption{Extrapolated values of $\spD{0}$ and hyperuniform metric $H$ of the constructed packings.
While values in the second column are initial packing fractions $\phi_\mathrm{init}$ for RSA and equilibrium hard-sphere liquids (HSL), those represent the vacancy concentration $c$ for imperfect $\mathbb{Z}^d$.
For the progenitor packings, $H\sim 0.1 - 0.01$.
We do not compute $H$ for the imperfect lattices because of their Bragg peaks. 
See the Supplemental Material \cite{Supplementary} for additional data.  \label{tab:changes_spectralDensity}}
\begin{ruledtabular}
\begin{tabular}{c l| l r | l l}
\multicolumn{2}{c|}{Progenitors/Parameters} & \multicolumn{2}{c|}{$\spD{0}$} & \multicolumn{2}{c}{$H$} \\
\hline
\multirow{4}{*}{\parbox{0.07\textwidth}{2D RSA\\$N=10^7$}} &
$\phi_{\mathrm{sat}}\approx 0.5471 $ 
           & 2.29(54)&$\times10^{-18}$	&1.36(32)&$\times10^{-14}$\\
&~~0.41025 & 4.7(15)&$\times10^{-18}$	&3.4(11)&$\times10^{-14}$\\
&~~0.27350 & 1.15(60)&$\times10^{-17}$	&1.01(52)&$\times10^{-13}$\\
&~~0.13675 & 3.1(18)&$\times10^{-17}$	&3.0(17)&$\times10^{-13}$\\ 
\hline
\multirow{4}{*}{\parbox{0.07\textwidth}{3D RSA\\$N=10^6$}} &
$\phi_\mathrm{sat}\approx 0.3841$
		 & 1.97(31)&$\times10^{-12}$	&1.51(24)&$\times10^{-8}$\\
&~~0.288 & 1.39(31)&$\times10^{-12}$	&1.27(28)&$\times10^{-8}$\\
&~~0.192 & 3.77(85)&$\times10^{-12}$	&3.90(88)&$\times10^{-8}$\\
&~~0.096 & 1.44(32)&$\times10^{-11}$	&1.66(36)&$\times10^{-7}$\\
\hline
\multirow{3}{*}{\parbox{0.07\textwidth}{2D HSL\\$N=10^5$}} &
~~0.65  & 2.03(32)&$\times10^{-14}$	&6.03(98)&$\times10^{-11}$\\
&~~0.40 & 6.7(19)&$\times10^{-14}$	&4.5(13)&$\times10^{-10}$\\
&~~0.20 & 2.48(76)&$\times10^{-13}$	&2.20(67)&$\times10^{-9}$\\
\hline
\multirow{3}{*}{\parbox{0.07\textwidth}{3D HSL\\$N=10^6$}} &
~~0.45  & -1.4(32)&$\times10^{-13}$	&7.0(162)&$\times10^{-10}$\\
&~~0.30 & 9.2(47)&$\times10^{-13}$	&7.6(39)&$\times10^{-9}$\\
&~~0.20 & 7.7(32)&$\times10^{-13}$	&8.0(33)&$\times10^{-9}$\\
\hline
\multirow{3}{*}{\parbox{0.12\textwidth}{Imperfect $\Z^2$\\$N_s=10^8$}} &
~~0.10  & 6.9(50)&$\times10^{-20}$ 	&\multicolumn{2}{c}{-}\\
&~~0.20 & 2.1(15)&$\times10^{-19}$ 	&\multicolumn{2}{c}{-}\\
&~~0.40 & 2.49(95)&$\times10^{-18}$	&\multicolumn{2}{c}{-}\\
\hline
\multirow{3}{*}{\parbox{0.12\textwidth}{Imperfect $\Z^3$\\$N_s=27\times 10^6$}} &
~~0.10  & 4.1(71)&$\times10^{-15}$	&\multicolumn{2}{c}{-}\\
&~~0.20 & 5.0(20)&$\times10^{-14}$	&\multicolumn{2}{c}{-}\\
&~~0.40 & 2.42(76)&$\times10^{-13}$ &\multicolumn{2}{c}{-}\\
\end{tabular}
\end{ruledtabular}
\end{table}

\begin{figure*}
\subfloat{\label{fig:2D_HSF_volDist}
\includegraphics[width = 0.3\textwidth]{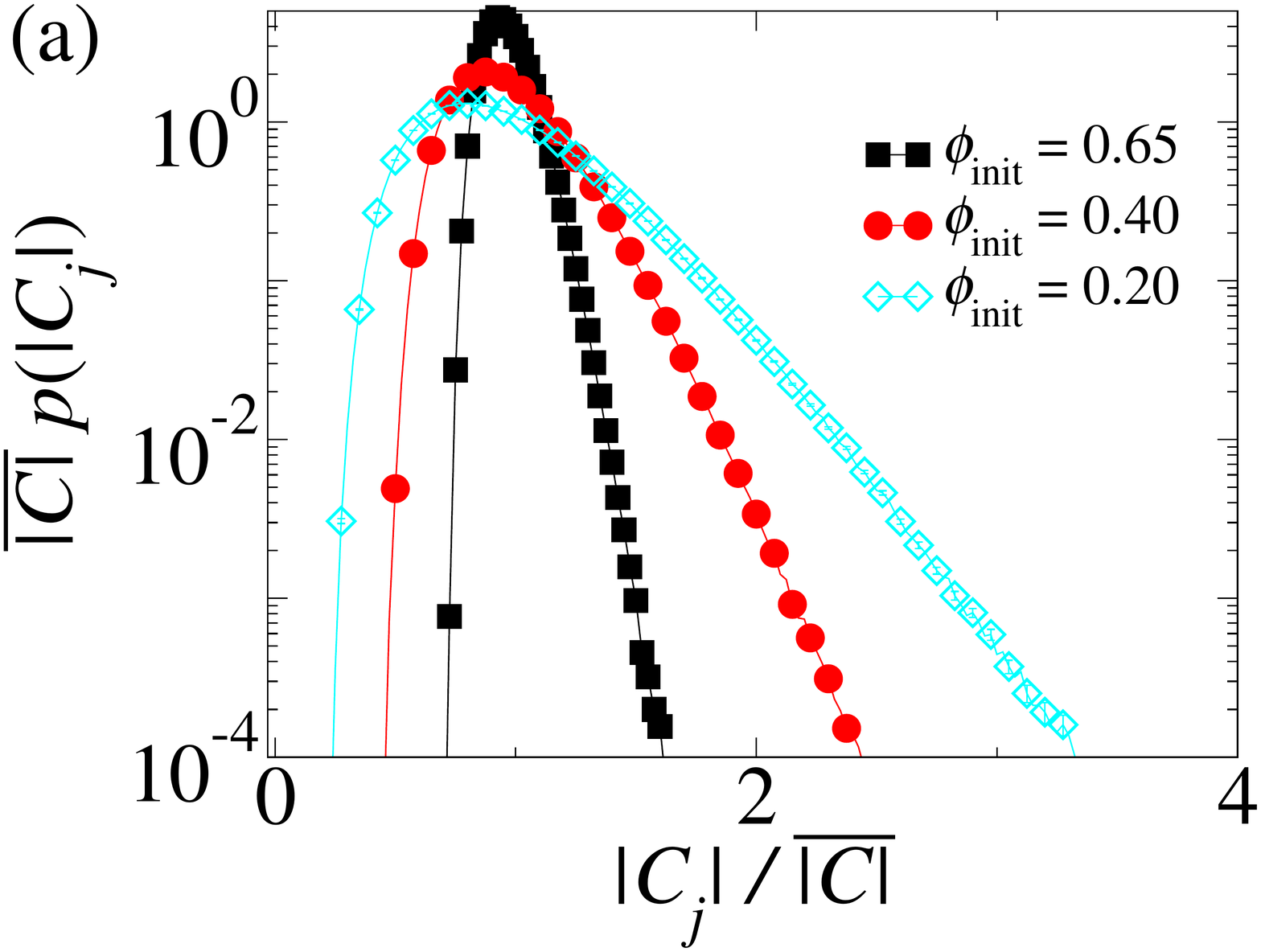}}
\subfloat{\label{fig:2D_HSF_chiv}
\includegraphics[width = 0.3\textwidth]{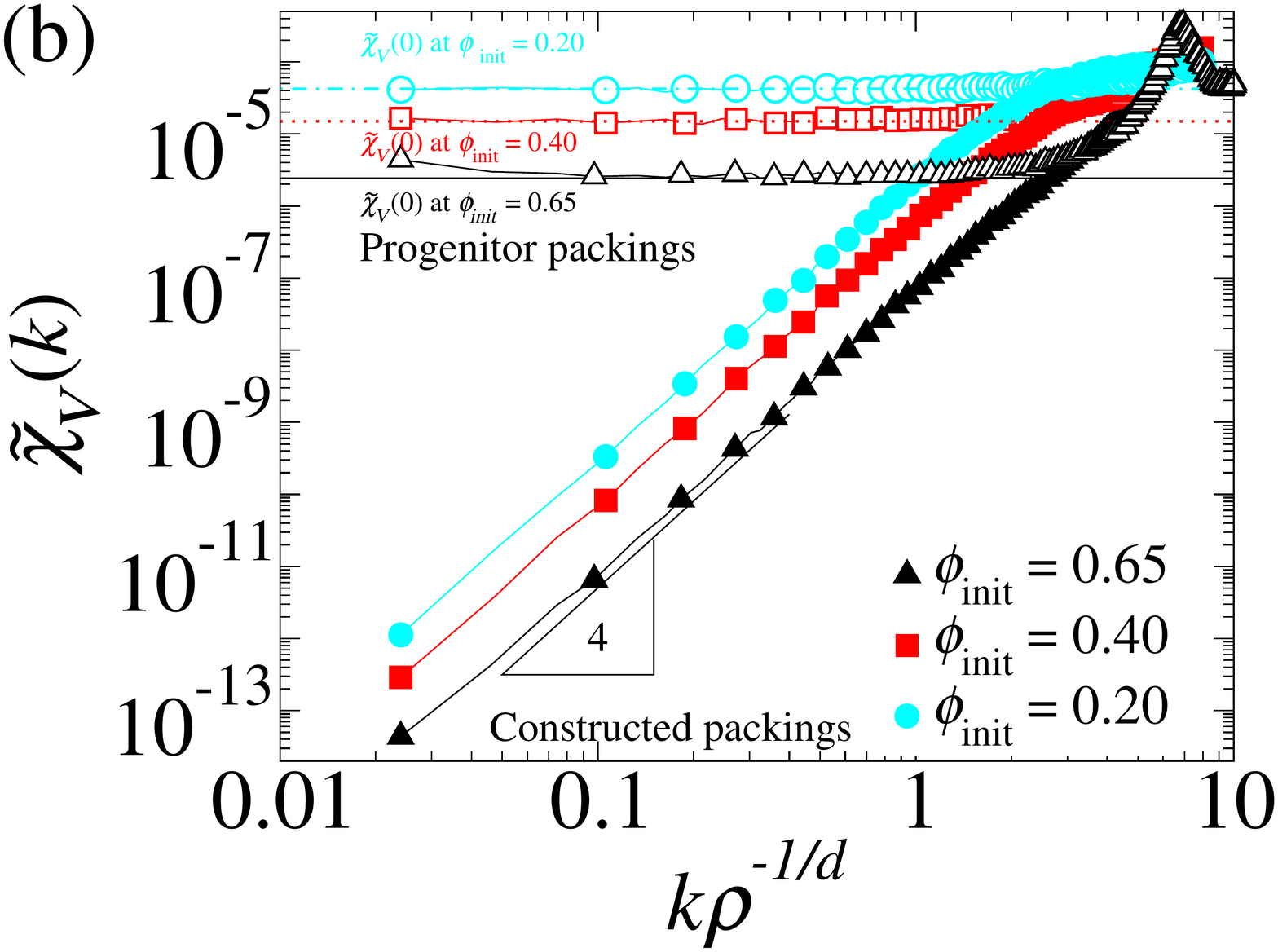}
}
\subfloat{\label{fig:2D_HSF_chiv2}
\includegraphics[width = 0.3\textwidth]{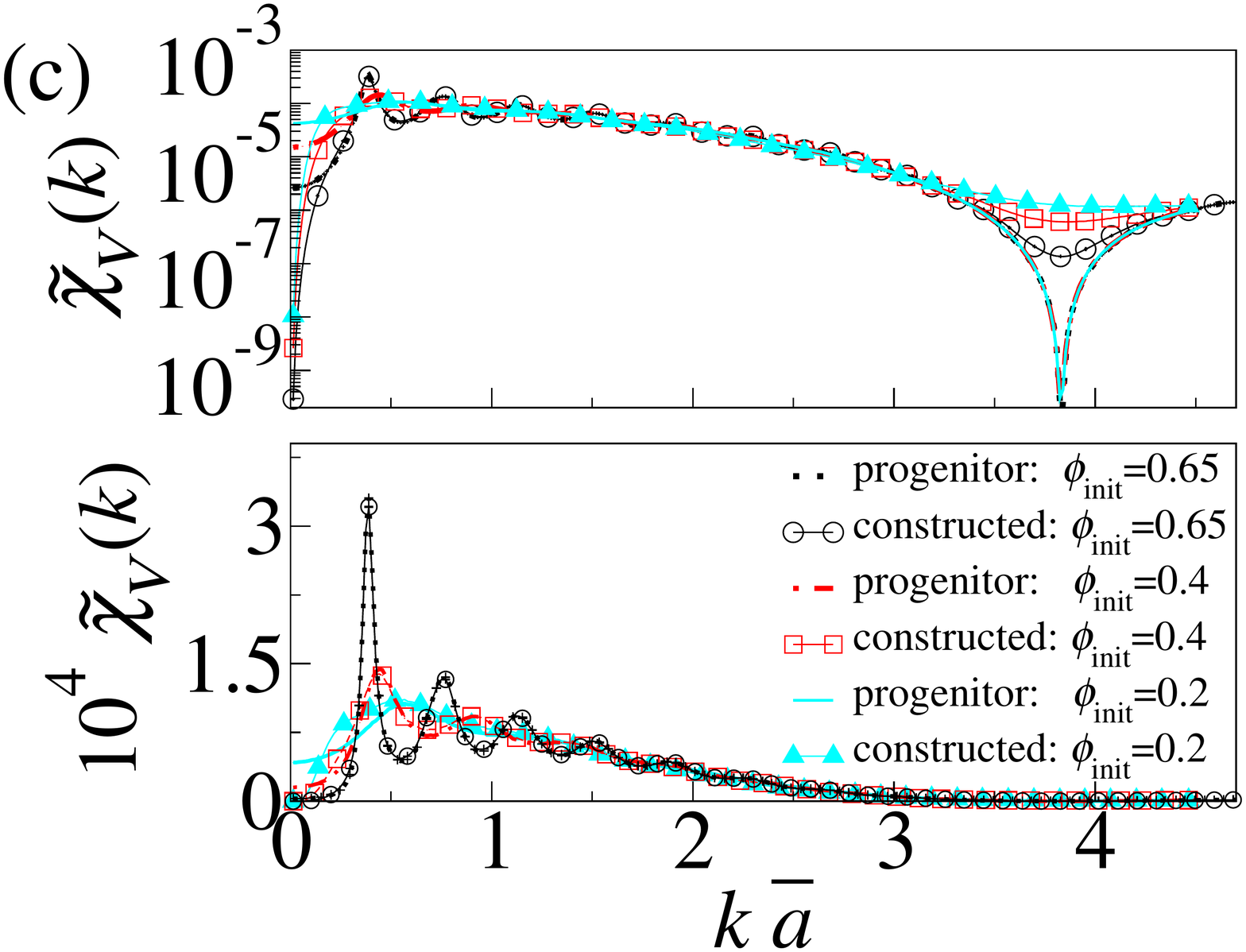}}
\caption{(Color online) Simulation results for disk packings converted from 2D equilibrium hard-disk liquids of $N=10^5$. 
\subcap{fig:2D_HSF_volDist} Probability density $\fn{p}{\abs{\tens{C}_j}}$ of Voronoi cell volumes versus the scaled cell volumes $\abs{\tens{C}_j}/\overline{\abs{\tens{C}}}$ plotted on a semi-log scale. 
\subcap{fig:2D_HSF_chiv} The spectral densities versus wavenumber $k$ for small wavenumbers plotted on a log-log scale. Theoretical values of $\spD{0}$ are obtained from Eq. \eqref{eq:S0_2D_HSF}.  
\subcap{fig:2D_HSF_chiv2} The spectral densities versus wavenumber $k$ for intermediate and large wavenumbers plotted on a semi-log (upper) and a linear (lower) scales. Here, sample size is $N=10^3$ and $\overline{a}$ is the mean particle radius of the constructed packings. 
\label{fig:Results_HSF}}
\end{figure*} 

The last type of progenitor packings are vacancy-riddled square and simple cubic lattices in two and three dimensions, respectively. 
These progenitor systems are characterized by the number $N_s$ of initial lattice sites and the fraction $c$ of point vacancies to $N_s$.
While these imperfect lattices still have Bragg peaks, they are not hyperuniform \cite{Durian2017, Jaeuk2018} and it is easy to generate extremely large samples ($N\sim 10^8$). 
Furthermore, for high values of $c$, their cell-volume distributions are similar to those for Poisson point patterns.
Therefore, this investigation will immediately demonstrate our discussion in Sec. \ref{sec:bounded-cellcondition}.

We compute the Voronoi tessellations of the three aforementioned types of progenitor packings via VORO++ library \cite{voro++} (see Appendix \ref{app:voronoi} for the implementation), and then perform our methodology without changing particle centers.
To get a visual sense of the resulting hyperuniform packings, we show representative but small hyperuniform sphere packings in 2D and 3D derived from RSA initial conditions in Fig. \ref{fig:ex_constructed}.  

We compute the local volume-fraction variances $\vv{R}$ of certain progenitor and constructed packings by Monte Carlo sampling of windows \cite{Torquato_RHM, Quintanilla1997} for many values of $R$ up to $L$ (i.e., side length of the simulation box).
For illustrative purposes, Fig. \ref{fig:2D_RSA_local_volumefraction_variance} plots such variances for progenitor packings that are 2D saturated RSA packings of various system sizes ($N=10^4$, $10^5$, and $10^6$).
We note that these results for $\vv{R}$ are reliable only up to a window radius $R<L/4$ \cite{Dreyfus2015,Chieco2017} due to the finite-size effects.
The constructed packings exhibit a common scaling $\vv{R}\sim R^{-(d+1)}$, as predicted from the heuristic rationale associated with Fig. \ref{fig:illustrations of density fluctuations}. 
This rationale also predicts that such a scaling starts from window sizes $R\gtrsim 5\ell_{\max},$ where $\ell_{\max}$ is the maximal cell length defined in step \ref{step1} [cf. Eq. \eqref{def:bounded-cell condition}], from which the region where volume-fraction fluctuations arise (the gray-shaded region in Fig. \ref{fig:illustrations of density fluctuations}) can be effectively regarded as window boundary.
Since the Voronoi tessellations of 2D saturated RSA that have a small $\ell_{\max}$ ($\lesssim 1.7 \rho^{-1/2}$), the class I hyperuniform scaling is well-established in length scales over several orders of magnitude (even beyond the reliable regime) for even relatively small samples ($N=10^4$) and the largest ones ($N=10^6$); see Fig. \ref{fig:2D_RSA_local_volumefraction_variance}. 

\begin{figure*}
\begin{center}
\subfloat{\label{fig:volDist_2DImLat}
\includegraphics[width = 0.3\textwidth]{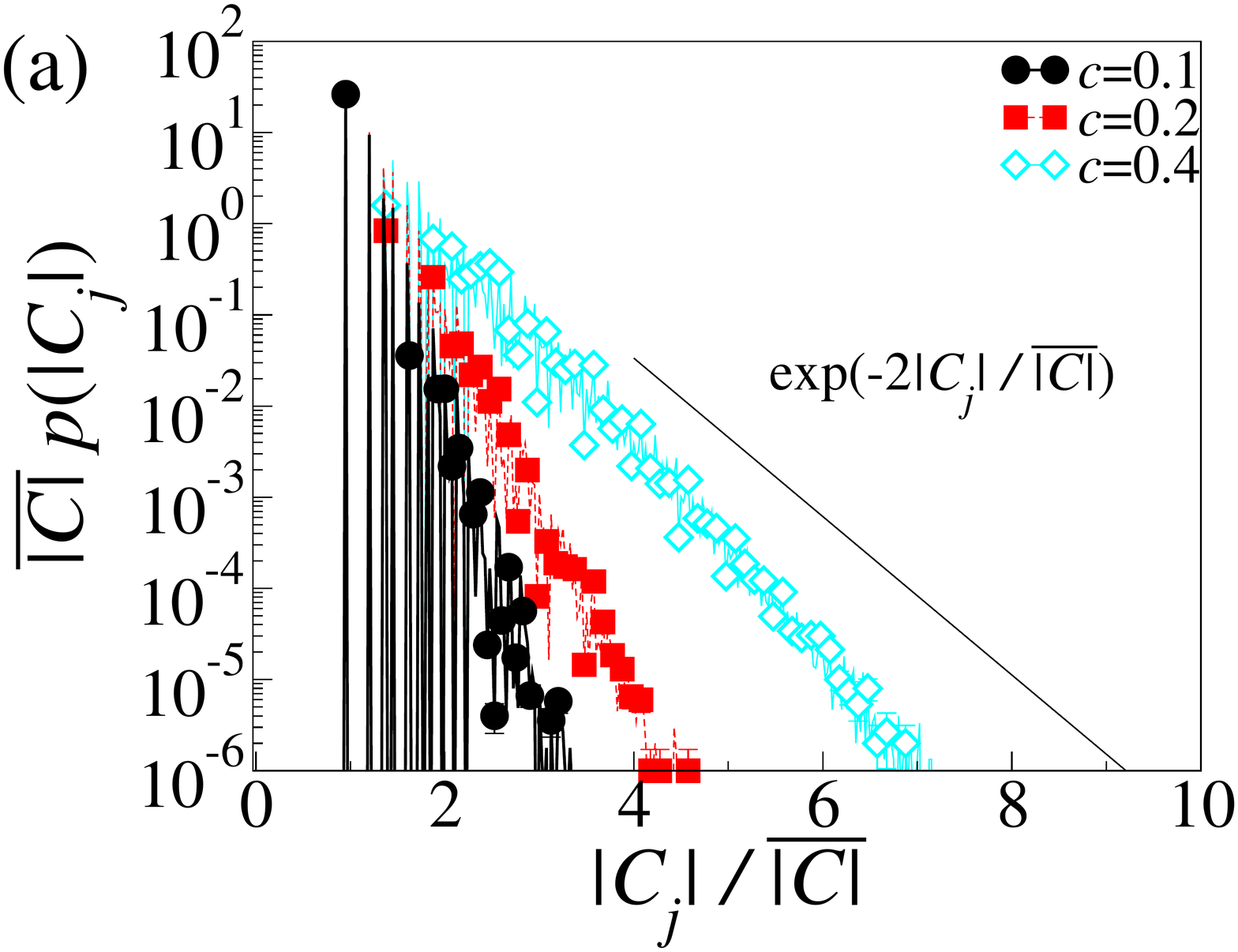}
}
\subfloat{\label{fig:chiv_2DImLat}
\includegraphics[width = 0.3\textwidth]{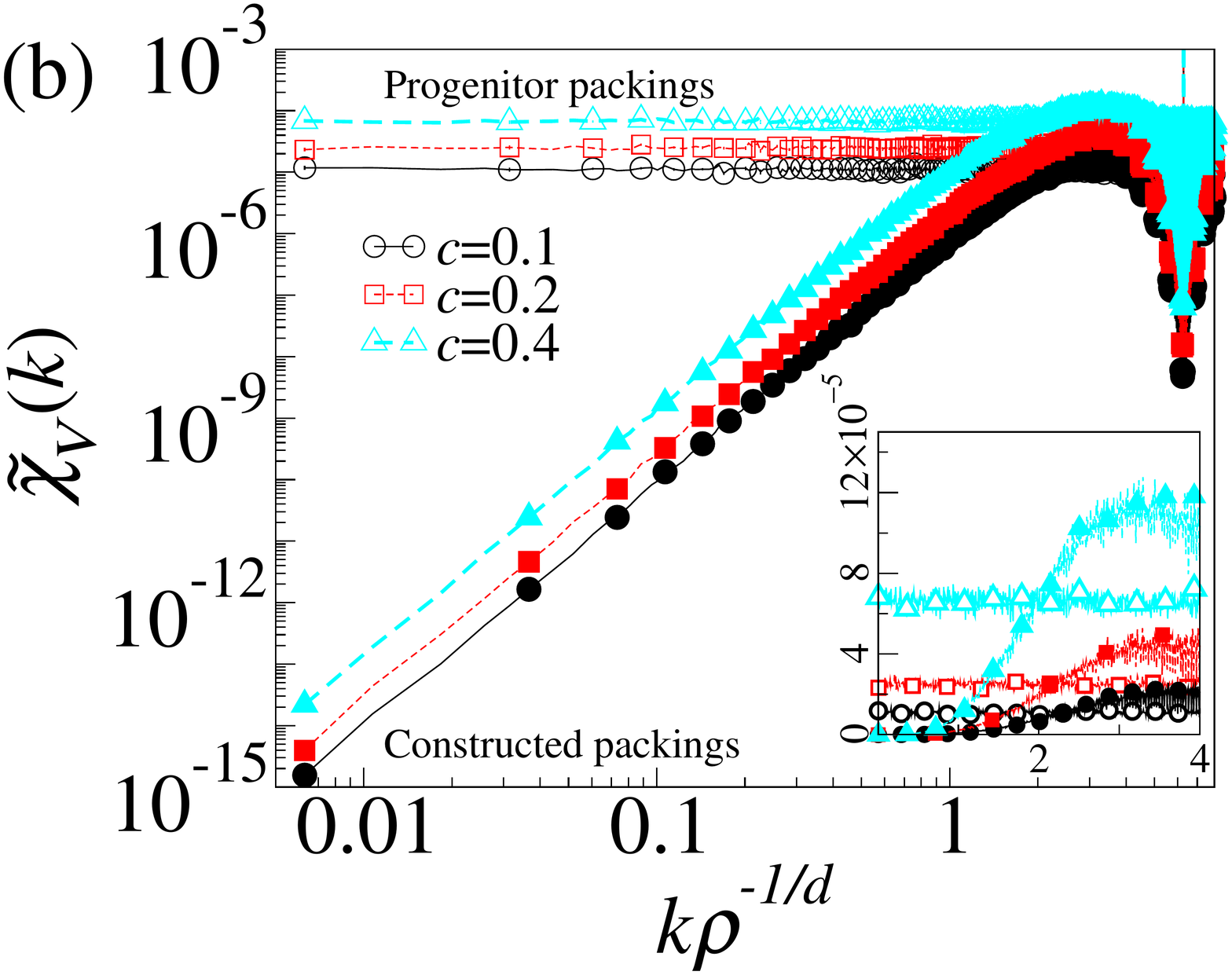}
}
\subfloat{\label{fig:chiv_2DImLat_size}
\includegraphics[width = 0.3\textwidth]{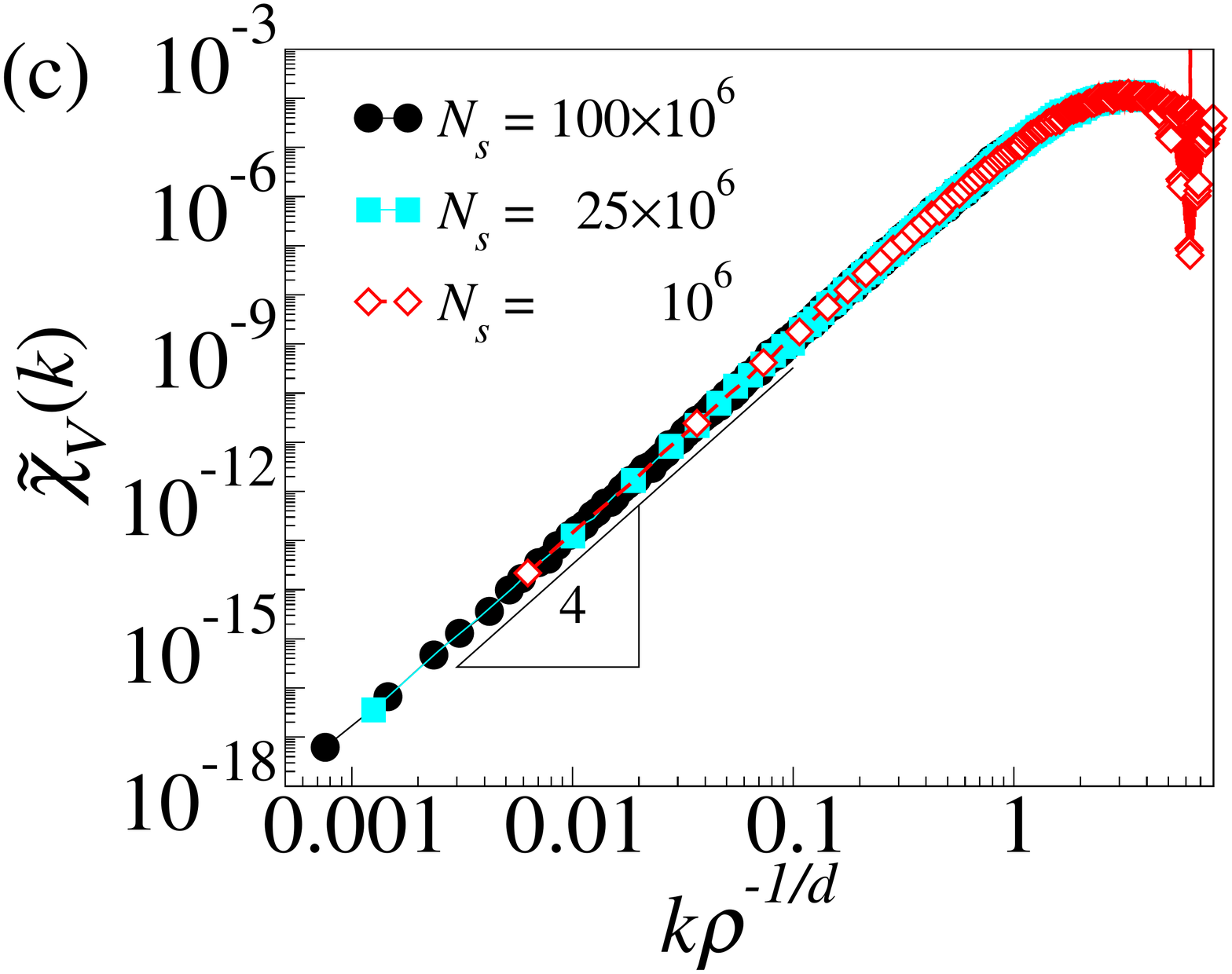}
}
\end{center}
\caption{(Color online) Simulation results for constructed disk packings from 2D lattice packing for various values of vacancy concentration $c$. 
(a) Semi-log plots of probability density functions of Voronoi cell volumes ($N_s=10^6$). (b) Log-log plots of the spectral densities for progenitor packings of various vacancy concentrations and the constructed packings ($N_	s=10^6$). (c) Log-log plots of the spectral densities of the constructed packings of various sample sizes ($c=0.4$).  
\label{fig:Results_ImLat}} 
\end{figure*}

Figures \ref{fig:Results_RSA}, \ref{fig:Results_HSF}, and \ref{fig:Results_ImLat} summarize the simulation results for packings that were converted from RSA, equilibrium hard-sphere liquids, and vacancy-riddled lattices, respectively.
By construction, these constructed packings have particle-volume distributions that are identical to those of Voronoi cell volumes in their progenitors; see Figs. \subref*{fig:2DRSA_Voronoicellvolume}, \subref*{fig:3DRSA_Voronoicellvolume}, \subref*{fig:2D_HSF_volDist}, and \subref*{fig:volDist_2DImLat}.
Furthermore, since particle centers are fixed in the procedure, the progenitor packings and their associated constructed packings have identical Voronoi tessellations and thus obviously have identical local statistics associated with Voronoi cells. 
This implies that local statistics alone generally may or may not determine hyperuniformity, which we elaborate in Sec. \ref{sec:conclusion}.

Isotropic spectral densities of progenitor and constructed packings are computed by using Eqs. \eqref{eq:spectral density} and \eqref{eq:FourierTransform_sphere}.
To compare them, we rescale all packings to a common packing fraction $\phi=0.01$; see Figs. \subref*{fig:chivk_2DRSA}, \subref*{fig:3DRSA_chivk}, \subref*{fig:2D_HSF_chiv}, and \subref*{fig:chiv_2DImLat}.
For small wavenumbers (i.e., $0<k\overline{a} \ll 1$, where $\overline{a} \sim 0.1 \rho^{-1/d}$ is the mean of particle radii), as shown in these figures, the spectral densities of the progenitors are significantly different from those of the constructed packings.
While progenitor packings are clearly nonhyperuniform (i.e., $H\sim 0.1-0.01$ for RSA packings and equilibrium hard-sphere liquids), the constructed packings are class I hyperuniform with a common power-law scaling $\spD{\abs{\vect{k}}}\sim \abs{\vect{k}}^4$.
Similar to the case of $\vv{R}$ shown in Fig. \ref{fig:2D_RSA_local_volumefraction_variance}, this power-law scaling is particularly well-established over several orders of magnitude, even for relatively small samples [$N=10^3$ as shown in Figs. \subref*{fig:2DRSA_chivk_size} and \subref*{fig:3DRSA_chivk_size}].
This $k^4$-scaling results from $\abs{\fn{\tilde{\mathcal{J}}_{(1)}}{\vect{k}}}\sim \abs{\vect{k}}^2$, which comes from special correlations in $\Delta\vect{X}_j$; see Sec. \ref{sec:GeneralTheory}.
Some values of $\spD{0}$ and $H$ are summarized in Table \ref{tab:changes_spectralDensity}; see, for additional data, Sec. IC in the Supplemental Material \cite{Supplementary}.

However, for intermediate wavenumbers (i.e., $1<k\overline{a}<3$), the spectral densities of the constructed packings largely resemble those of their progenitors; see, for example, Fig. \subref*{fig:2D_HSF_chiv2}.
This result reflects that local statistics of both packings are identical at corresponding length scales.  
At large wavenumbers (i.e., $k\overline{a}>3$), however, the spectral densities of progenitor and constructed packings again become different from each other because the spectral density at this regime mainly depends on the particle-size distributions; see Fig. \subref*{fig:2D_HSF_chiv2}.

\subsection{Nonspherical particles}\label{sec:Voronoi_cubic}

\begin{figure}[h]
\centering
\subfloat{
\includegraphics[width =0.22 \textwidth]{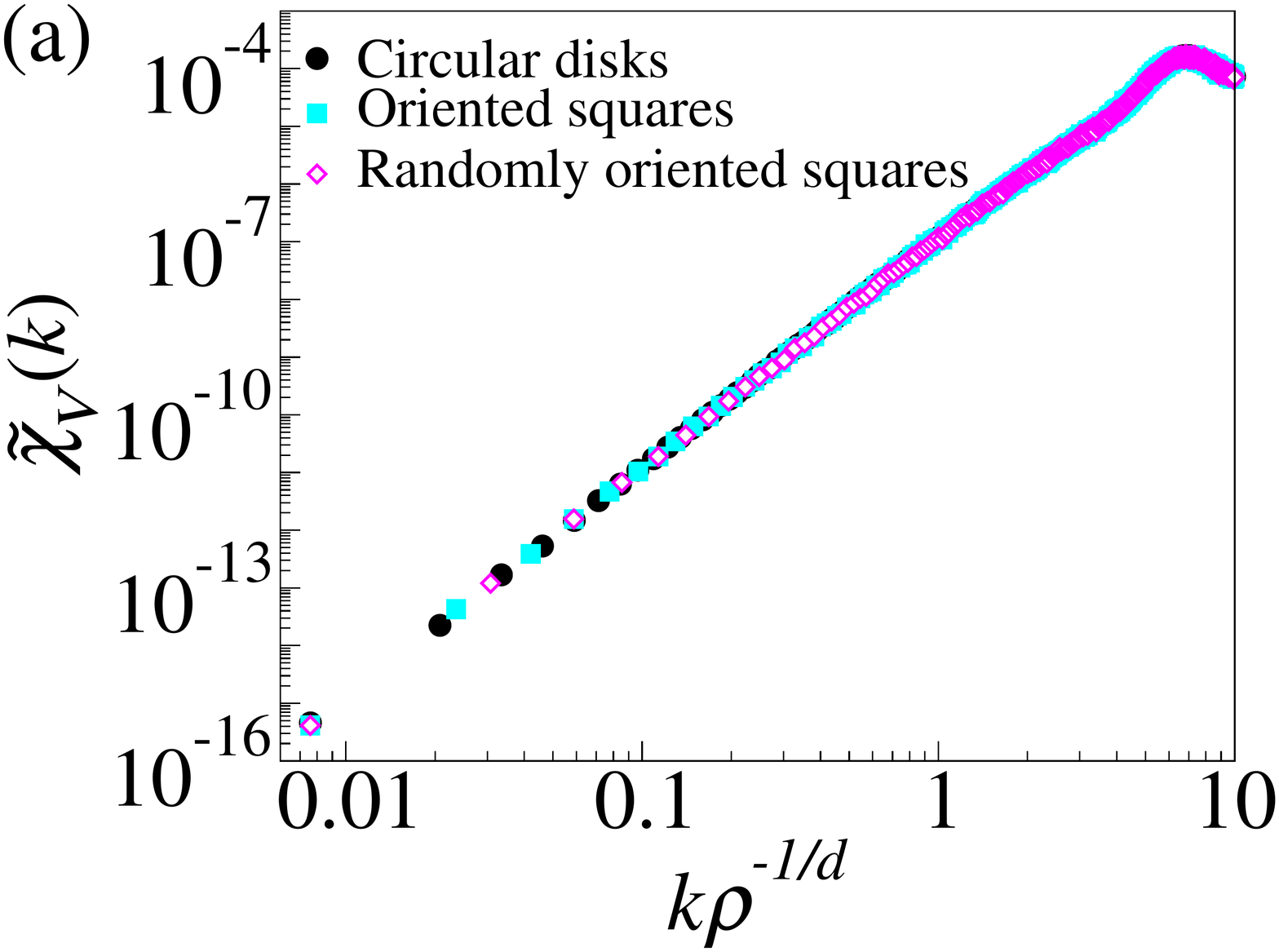} \label{fig:2D_RSA_chiv_squares}}
\subfloat{
\includegraphics[width =0.22 \textwidth]{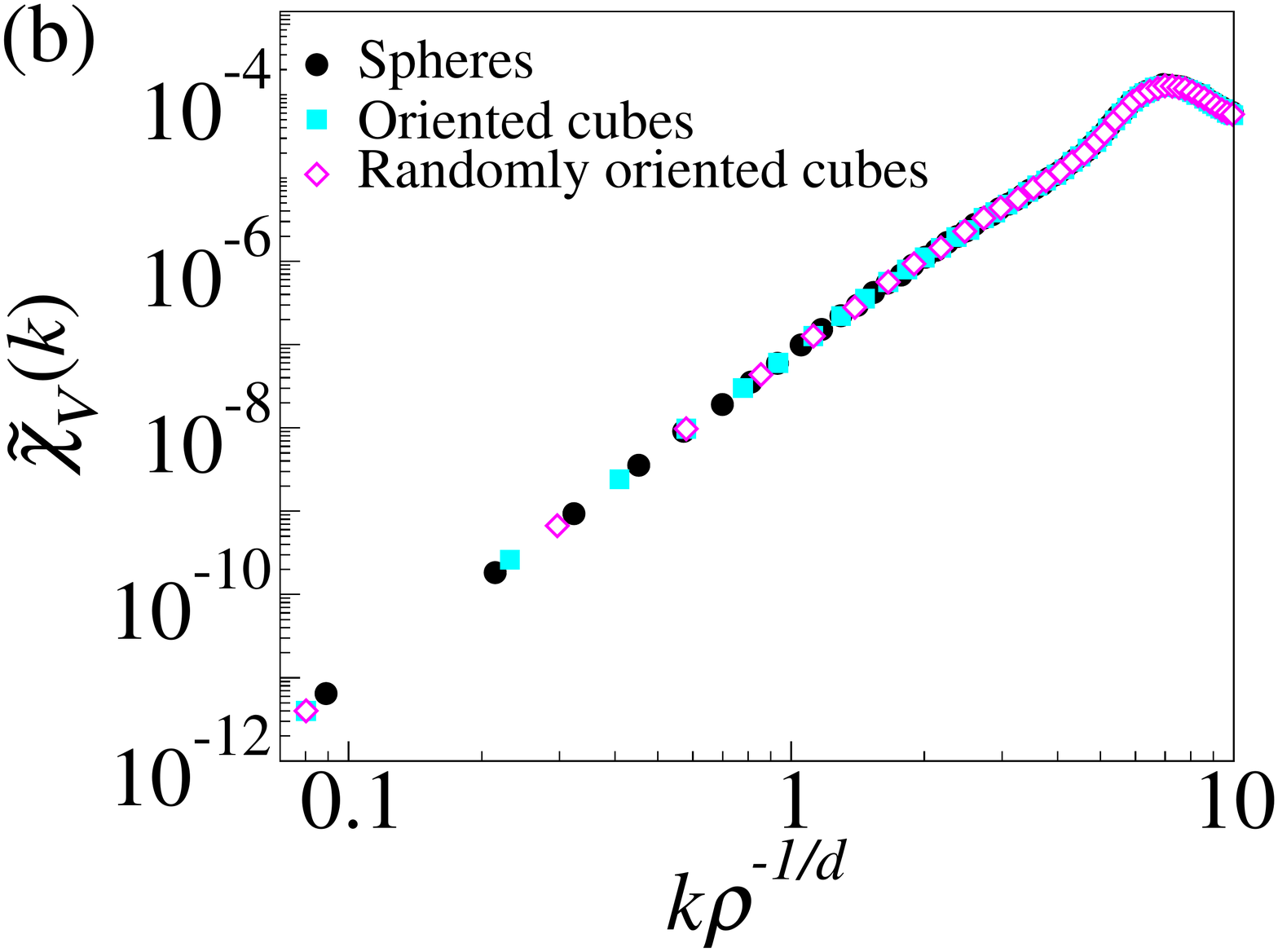} \label{fig:3D_RSA_chiv_cubes}}
\caption{(Color online) Simulation results for $\spD{k}$ of the constructed packings depending on particle shapes.
Here, the progenitor point patterns are the centers of saturated RSA packings with $N=10^6$ in (a) two and (b) three dimensions.  
\label{fig:chiv_cubic particles} }
\end{figure}

As shown in the derivation in Sec. \ref{sec:GeneralTheory}, the  tessellation-based procedure  can result in class I hyperuniformity even for nonspherical particle shapes, including regular and random polyhedrons.
However, since it is generally nontrivial to compute the form factors of those particle shapes, we consider a simple geometric shape, i.e., a $d$-dimensional hypercubic particles; see Eq. \eqref{eq:FourierTransform_cubic}. 

For simplicity, we take advantage of the sphere packings constructed in Sec. \ref{sec:Voronoi_spherical} by replacing the spherical particles with the inscribed cubic particles.
In this case, the corresponding maximal packing fraction for cubes will be related to that of spheres as follows: $\phi_{\max}^\mathrm{cube} = \frac{\fn{\Gamma}{1+d/2} 2^{d-1}}{(d \pi)^d} \phi_{\max}^{\mathrm{sphere}}$. 
We consider two cases: (identically) oriented particles and randomly oriented particles.
Figure \ref{fig:chiv_cubic particles} summarizes the simulation results for 2D and 3D packings converted from saturated RSA packings of $N\approx 10^6$.
We note that the results for squares or cubes are largely indistinguishable from those for spheres because both particle shapes have comparable length scales and the Fourier transforms of both a cube \eqref{eq:FourierTransform_cubic} and a sphere \eqref{eq:FourierTransform_sphere} are isotropic near the origin and have similar profiles. 
However, these spectral densities will exhibit different behaviors at intermediate and large wavenumbers, depending on particles shapes and orientations.

\section{Multiscale Coated-Spheres Model}\label{sec:Coated-sphere}

The tessellation-based methodology can be applied to any type of tessellations of space besides the Voronoi tessellations, as long as they obey the bounded-cell condition \eqref{def:bounded-cell condition}.
One interesting example is a tessellation of space that consists of nonoverlapping spherical cells.
Since a monodisperse sphere packings in $d>1$ cannot occupy all space, spherical cells in such tessellations should have a polydispersity in size down to the infinitesimally small; see Fig. \ref{fig:schematics}(b).
For these (multiscale) sphere tessellations, the bounded-cell condition can be guaranteed by making the ratio of the largest cell-volume to the sample volume, $\delta = v_{\max}/\abs{\mathcal{V}_F}$, sufficiently small. 
Implementing our methodology from these tessellations should result in hyperuniform packings regardless of particle shapes, positions, and numbers per cell.

A simple but important case of such disordered hyperuniform packings are ones in which each cell includes only a single particle to which the cell is concentric; see Fig. \ref{fig:schematics}(d).
Since particle shapes are similar to those for cells, the local-cell packing fraction $\phi$ of those packings can span up to unity, as mentioned in Sec. \ref{sec:Tiling-based procedure}.
In addition, we can use the expression of the second moment of a sphere of radius $R$, ${\mathcal{M}_{\alpha\beta}} = \frac{1}{(2+d)} R^2 \delta_{\alpha, \beta}$, $\Delta\vect{X}_j=\vect{0}$, and Eq. \eqref{eq:F.T._phase_indicator_similarParticles} to obtain the following power-law scaling for the spectral density:
\begin{align}
&\spD{\vect{k}}\nonumber \\
& = 
 \left[\frac{\phi(1-\phi^{2/d})}{2(2+d)}	\right]^2 \frac{1}{\abs{ \mathcal{V}_F}}  \abs{\sum_{j=1}^\infty \fn{v_1}{\mathcal{R}_j} (k\mathcal{R}_j)^2 e^{-i\vect{k}\cdot\vect{x}_j}}^2 + \order{k^5} \nonumber\\
& \sim  \phi^2 (1-\phi^{2/d})^2 k^4, \label{eq:chi_v coated spheres infty}
\end{align}
where radius of cell $j$ is denoted by $\mathcal{R}_j$.

It is crucial to observe that cells and the associated particles form composite spheres that fill all space.
Each composite sphere is comprised of a spherical core (particle) of one phase that is surrounded by a concentric spherical shell of the other phase such that the fraction of space occupied by the core phase is equal to its global phase volume fraction, which is guaranteed by our procedure. 
Thus, this structure is a packing (dispersion) in which spherical particles (blue regions) are ``well-separated" from one another in a fully connected (continuous) void (matrix) phase (red region).
This is traditionally understood as a key characteristic of optimal two-phase structures \cite{Torquato2018_3, Torquato2018_5}. 

Surprisingly, these hyperuniform multiscale structures are identical to the \textit{Hashin-Shtrikman multiscale coated-spheres models} \cite{Hashin1962, Torquato_RHM, Milton1986a, Torquato2004}, which are perhaps the most famous results from the theory of heterogeneous materials.
These special particle composites are optimal for the effective electric (thermal) conductivity and bulk elastic modulus for a given phase fractions and phase properties.
This observation will shed light on the origin of nearly optimal transport properties of disordered stealthy hyperuniform packings \cite{Zhang2016, Chen2017} and cellular networks derived from hyperuniform systems \cite{Torquato2018_3}.

The coated-spheres structures are infinitely degenerate with varying degrees of order/disorder.
The most ordered structures would be ones, derived from certain deterministic sphere tessellations, such as \textit{Apollonian gaskets}, or initial lattice packings of identical spheres in which particles are added in a sequential multiscale manner.

\begin{figure}
\includegraphics[width = 0.4\textwidth]{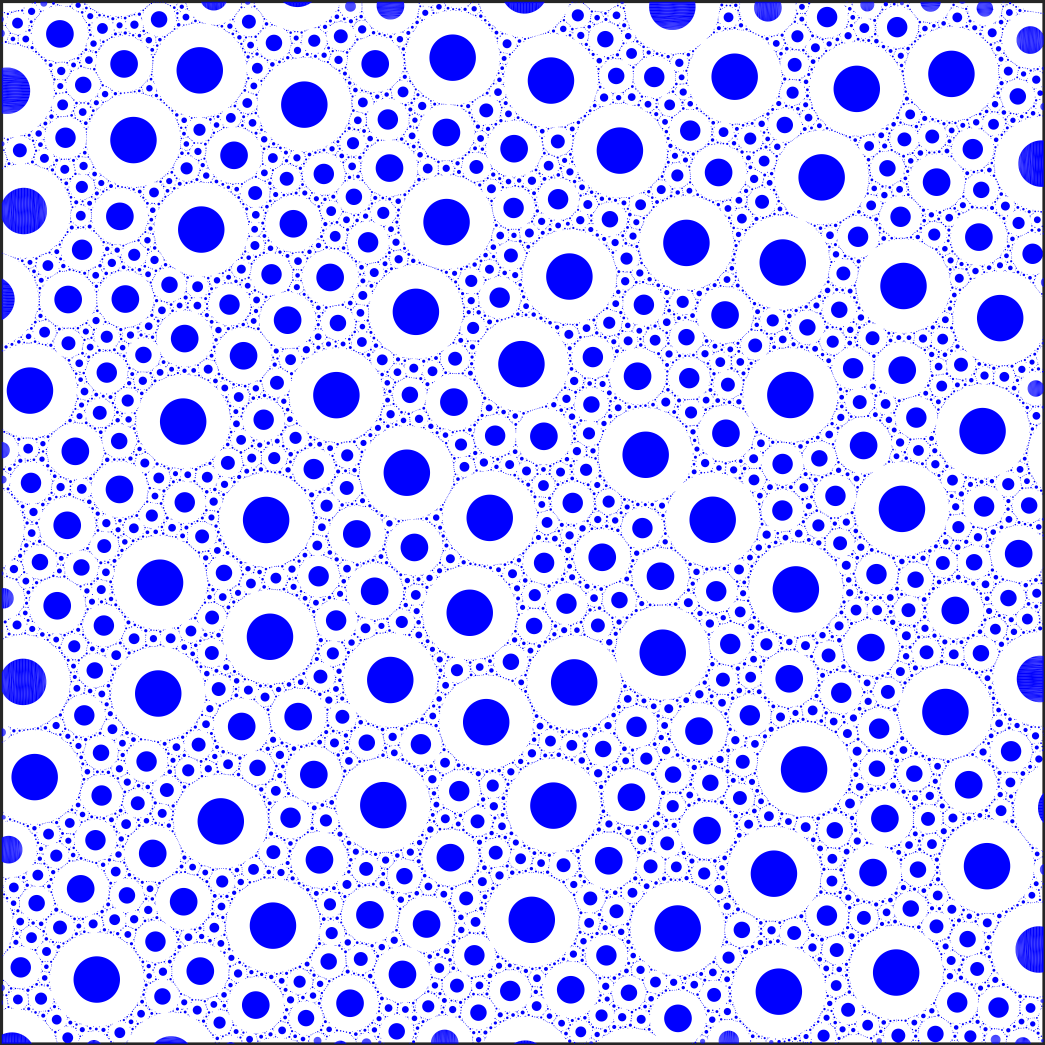}
\caption{(Color online) A representative of disordered coated-disks model with the local-cell packing fraction $\phi=0.25$ derived from a sphere tessellation for a power-law scaling with $p=1.5$ and $m=400$.
Here, inclusions are only displayed; see Sec. II in the Supplemental Material \cite{Supplementary} for an enlarged image.  \label{fig:Ex_CoatedDisks}}
\end{figure}

\subsection{Simulation model}\label{sec:Coated-sphere_model}

In what follows, we describe our model to simulate the coated-spheres model and ascertain their hyperuniformity. 
We first construct very dense disordered sphere packings via a multistage version of the RSA procedure, in which volume of the inserted spheres reduced in every stage.
These dense ``precursor" packings are later used to simulate ideal sphere tessellations.
The coated-spheres model is then simulated by scaling spheres in the precursor packings at a certain ratio.

Specifically, these precursor packings are constructed from an empty simulation box $\mathcal{V}_F$ in $\R^d$ under the periodic boundary conditions.
The construction procedure has two control parameters: the upper bound $v_{\max}$ on cell volumes and a positive decreasing function $\fn{g}{i}$ of positive integers $i$, which satisfies $\fn{g}{1}=1$ and $\sum_{j=1}^\infty \fn{g}{j}< \infty$.
The infinite-sample-size limit can be achieved as the upper bound $v_{\max}$ tends to be infinitesimally small with the simulation box fixed. 
Subsequently, one determines the prescribed number $N$ of spheres that will be inserted in every stage and the maximum cell volume $v^{(1)}$ to fill all space:
\begin{align}
N & \equiv \ceil{\abs{\mathcal{V}_F}/ \left(v_{\max} G	\right)}, \\
v^{(1)} &\equiv \abs{\mathcal{V}_F}/ \left(N G	\right),
\end{align}
where $G\equiv \sum_{j=1}^\infty \fn{g}{j}$ and $\ceil{x}$ is the ceiling function.
Every sphere in the $m$th stage has volume $v^{(m)} = v^{(1)} \fn{g}{m}$ and the associated diameter $D_m$.
Using these parameters, the precursor packings are constructed by the following steps: 
\begin{enumerate}
\item In the first stage ($m=1$), irreversibly, randomly, and sequentially add (i.e., via RSA procedure) nonoverlapping spheres of a diameter $D_1$.
The insertion stops when $N$ spheres are added, unless the packing becomes saturated.
\item In the $m$th stage ($m>1$), nonoverlapping spheres of a diameter $D_m$ ($<D_{m-1}$) are added via the RSA procedure. Make sure that newly inserted spheres do not overlap with the spheres in the previous stages $1,2,\cdots,m-1$.
The insertion stops when the packing reaches to a prescribed covering fraction $N/{\abs{\mathcal{V}_F}} \sum_{i=1}^m v^{(i)}$, unless the packing becomes saturated.
\item The procedure stops when it reaches to a prescribed number of stages; see Fig. \ref{fig:schematics}(b).
\end{enumerate}

Note that the aforementioned voxel-list algorithm \cite{Zhang2013} is implemented in step 1 and 2.
Using the $m$th stage precursor packings, we simulate the coated-spheres model of the inclusion volume fraction $\phi$ by reducing these spheres at a volume ratio $\phi$.
Figure \ref{fig:Ex_CoatedDisks} shows a representative but small disordered multiscale coated-disks construction. 

We note that due to saturation at each stage, the number $N_m$ of spheres inserted in the $m$th stage can be different from the prescribed number $N$.
Let $\mathcal{N}_m\equiv \sum_{i=1}^m N_i$ and $\eta_m  = \sum_{i=1}^m N_i v^{(i)} /\abs{\mathcal{V}_F}$ denote the total number of spheres at the end of the $m$th stage and the fraction of space covered by these spheres, called the \textit{covering fraction}, respectively.
By construction, $\eta_m$ cannot exceed the prescribed covering fraction in the $m$th stage: 
\begin{equation}\label{eq:covering fraction}
\eta_m \leq \frac{N}{\abs{\mathcal{V}_F}}\sum_{i=1}^m v^{(i)}.
\end{equation}
Thus, in a finite $m$th stage, the precursor packings will not cover all space but have gaps that can be only covered by smaller spheres in the next stages.
As the number of stages tends to be infinite, those gaps are eventually covered by spheres of size down to the infinitesimally small, i.e., $\eta_m\to1$ as $m\to\infty$.

In this work, we consider two different types of volume scalings: a power-law scaling and an exponential scalings. 
For the power-law scaling, the cell volume in the $m$ stage is determined by
\begin{equation}\label{eq:power-law_cell}
 v^{(m)}/v^{(1)} = 1/m^p,
\end{equation}
for a given scaling exponent $p>1$.
From this relation, it is straightforward to derive the relation between maximum cell volume $v^{(1)}$ and the prescribed insertion number $N$, and the prescribed covering fraction:
\begin{align}
\abs{\mathcal{V}_F} &= N v^{(1)}\fn{\zeta}{p}, \label{eq:power-law_TotalVolume}\\
1-\eta_m &= \frac{1}{\fn{\zeta}{p}} \sum_{j=m+1}^\infty 1/j^p \label{eq:power-law_CoveringFraction}, 
\end{align}
where $\fn{\zeta}{p}$ denotes the Riemann zeta function. 

For the exponential scaling, cell volume in the $m$th stage is determined by
\begin{equation}\label{eq:exponential-law_cellvolume}
v^{(m)}/v^{(1)} = 1/q^{m-1},
\end{equation}
for a given scaling base $q>1$.
The analogues of Eqs. \eqref{eq:power-law_TotalVolume} and \eqref{eq:power-law_CoveringFraction} are 
\begin{align}
\abs{\mathcal{V}_F} &= Nv^{(1)}q/(q-1),\label{eq:exponentialScaling_TotalVolume}\\
1-\eta_m & = 1/{q^m}. \label{eq:exponentialScaling_CoveringFraction}
\end{align}

\subsection{Theoretical analyses}
While in the limit of $m\to \infty$ our coated-spheres model is predicted to be strongly hyperuniform [see Eq. \eqref{eq:chi_v coated spheres infty}], our model in a finite $m$th stage will be nearly hyperuniform, rather than perfectly hyperuniform, due to the uncovered gaps ($\eta_m <1$). 
To estimate the degree of hyperuniformity of our model in the $m$th stage, we consider the associated spectral density $\fn{\tilde{\chi}^{(m)}_{_V}}{\vect{k}}$, given by 
\begin{widetext}
\begin{align}
\fn{\tilde{\chi}_{_V} ^{(m)}}{\vect{k}} & = \frac{1}{\abs{\mathcal{V}_F}} \E{ \abs{\sum_{j=1}^{\mathcal{N}_m} \fn{\tilde{m}}{\vect{k};\phi^{1/d }\mathcal{R}_j}  e^{-i \vect{k}\cdot\vect{x}_j}  - \phi \int_{\mathcal{V}_F} \d{\vect{y}} e^{-i\vect{k}\cdot\vect{y}}   }^2} \label{eq:chiv_multi_1-1},
\end{align}
where $\phi$ denotes the local-cell packing fraction.
Without any prior knowledge of cell-volume distribution, its rigorous bound can be obtained by using some simple inequalities (see Appendix \ref{sec:upperbound}):
\begin{align}\label{eq:chiv_multi_UB}
\fn{\tilde{\chi}_{_V} ^{(m)}}{\vect{k}} \leq  2 \fn{F}{\vect{k};\phi}  + 2 \phi^2 \abs{\mathcal{V}_F} (1-\eta_m)^2,
\end{align}
where 
\begin{equation}\label{eq:chiv_multi_scaling}
\fn{F}{\vect{k};\phi} \equiv \frac{1}{\abs{\mathcal{V}_F}} \left[\frac{\phi (1-\phi^{2/d})}{2(2+d)}	\right]^2 \E{\abs{ \sum_{j=1}^{\mathcal{N}_m}  \fn{v_1}{\mathcal{R}_j}(k\mathcal{R}_j)^2	e^{-i\vect{k}\cdot \vect{x}_j} +\order{k^4}}^2}  \propto [\phi^2(1-\phi^{2/d})^2] \abs{\vect{k}}^4, ~~\abs{\vect{k}}\to 0.
\end{equation}
\end{widetext}
Here, the constant term $\phi^2 \abs{\mathcal{V}_F} (1-\eta_m)^2$ is the largest volume-fraction fluctuations that can arise from uncovered gaps in the $m$th stage coated-spheres model.
Importantly, the constant term depends on $\phi(1-\eta_m)$, i.e., the deviation between the local-cell packing fraction $\phi$ and the global packing fraction $\phi\eta_m$, which will vanish as the uncovered gaps become completely filled in the limit of $m\to\infty$.
Thus, in this limit, the upper bound implies that our model becomes perfectly hyperuniform of class I, consistent with Eq. \eqref{eq:chi_v coated spheres infty}.

\begin{figure}[ht]
\subfloat{\label{fig:2DZeta_coveringfraction}\includegraphics[width =0.24 \textwidth]{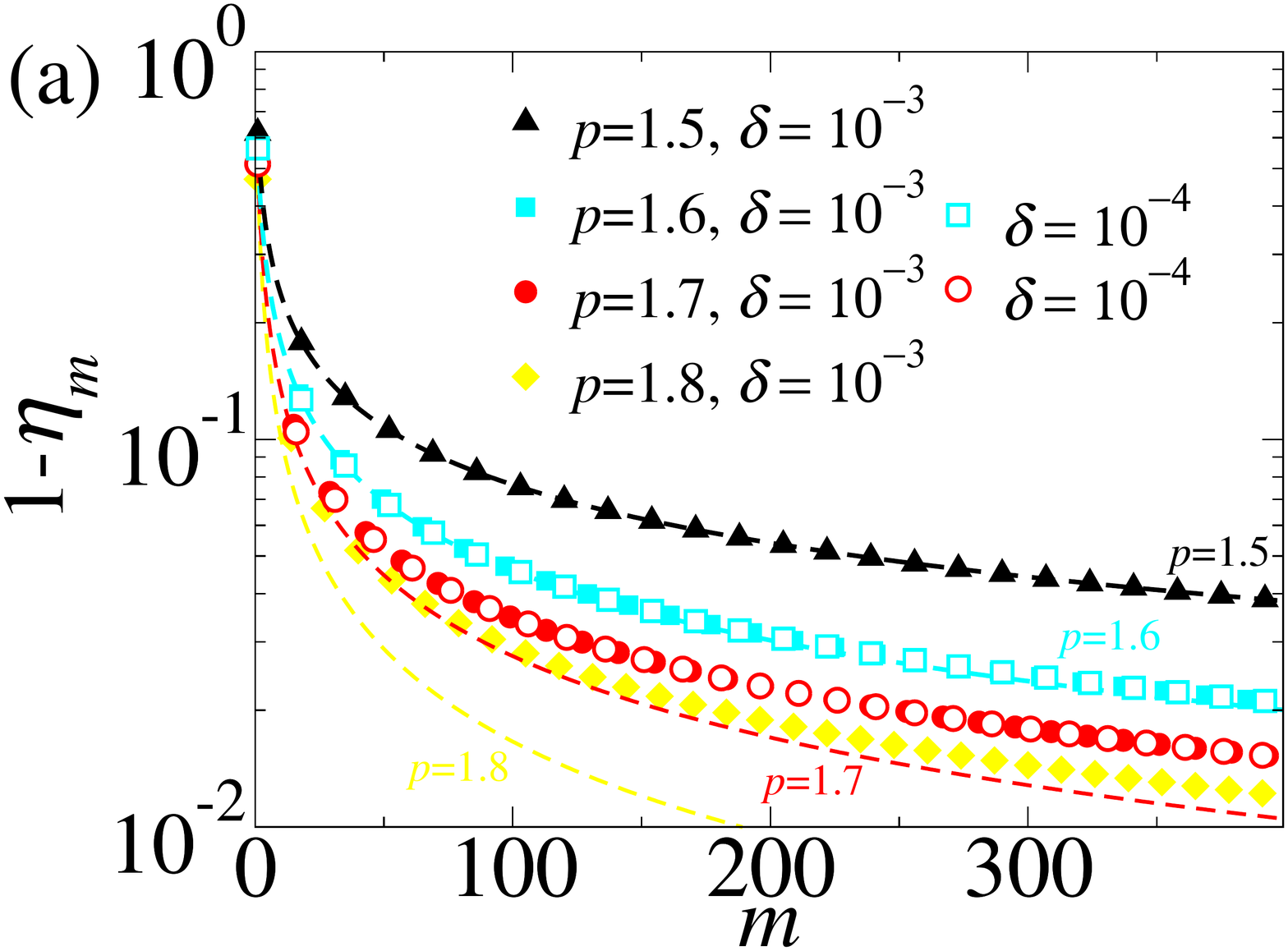}}
\subfloat{\label{fig:2DZeta_totalNumber}
\includegraphics[width = 0.24 \textwidth]{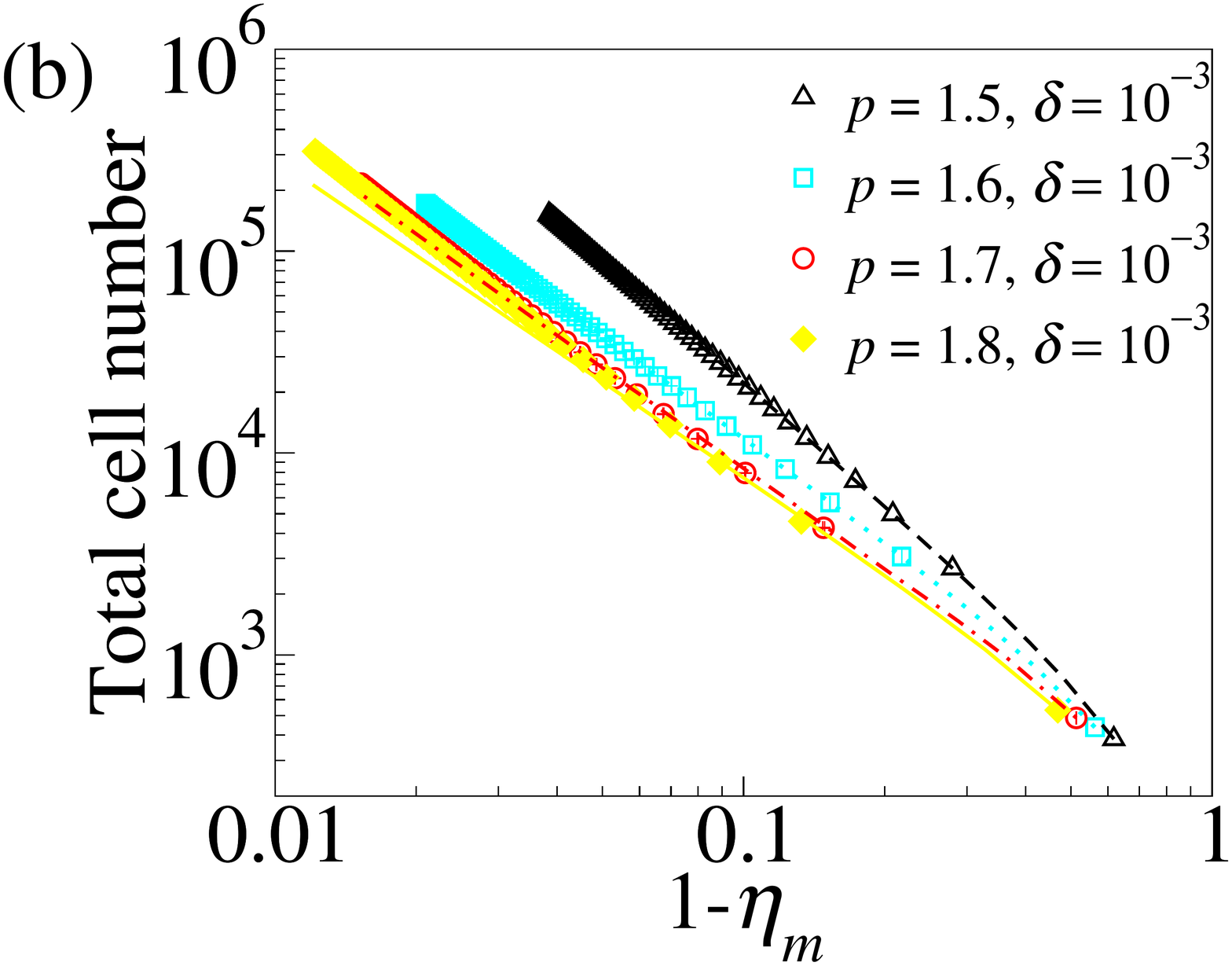}
}

\subfloat{\label{fig:2DPower_coveringfraction}\includegraphics[width=0.24 \textwidth]{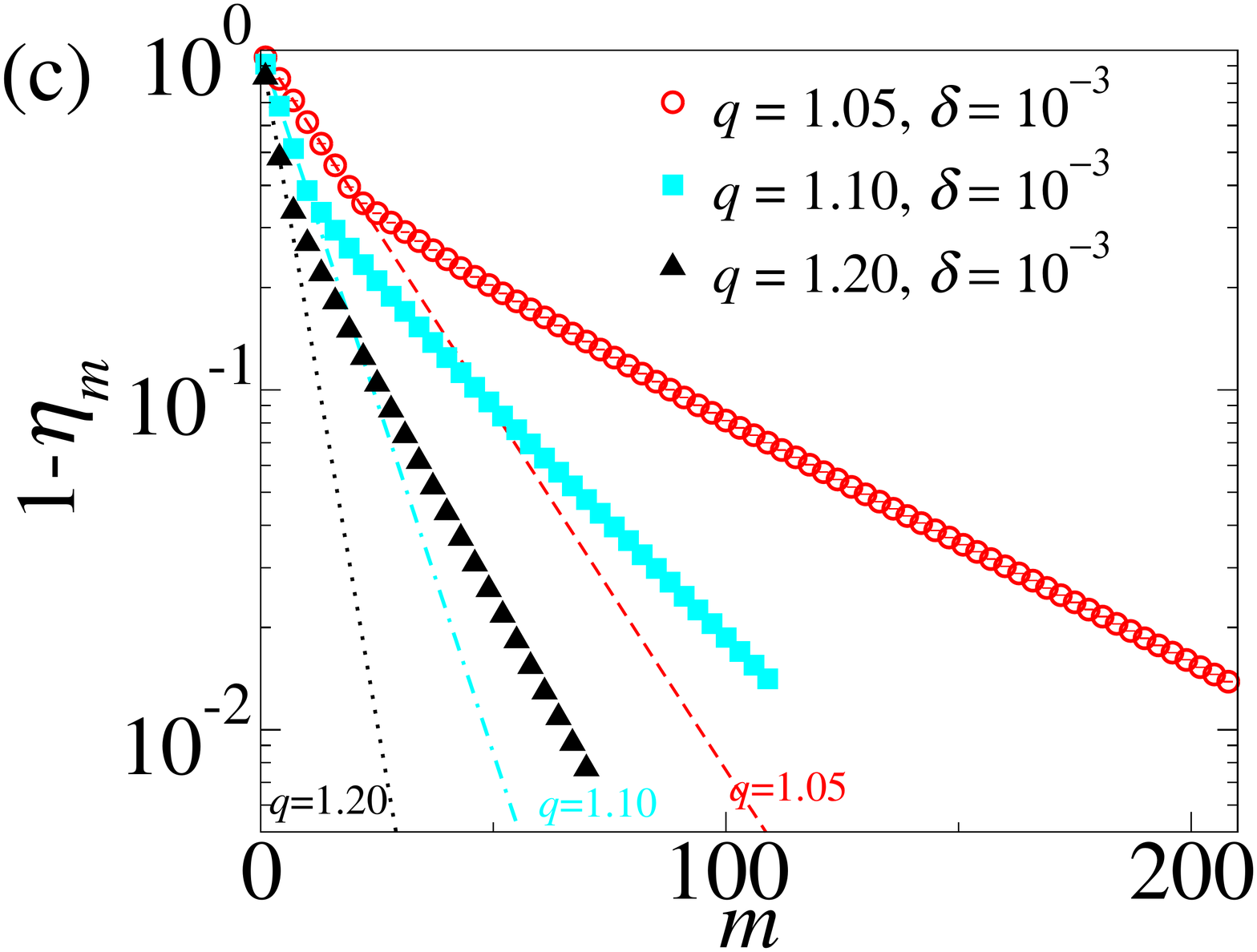}}
\subfloat{\label{fig:2DPower_totalnumber}\includegraphics[width=0.24 \textwidth]{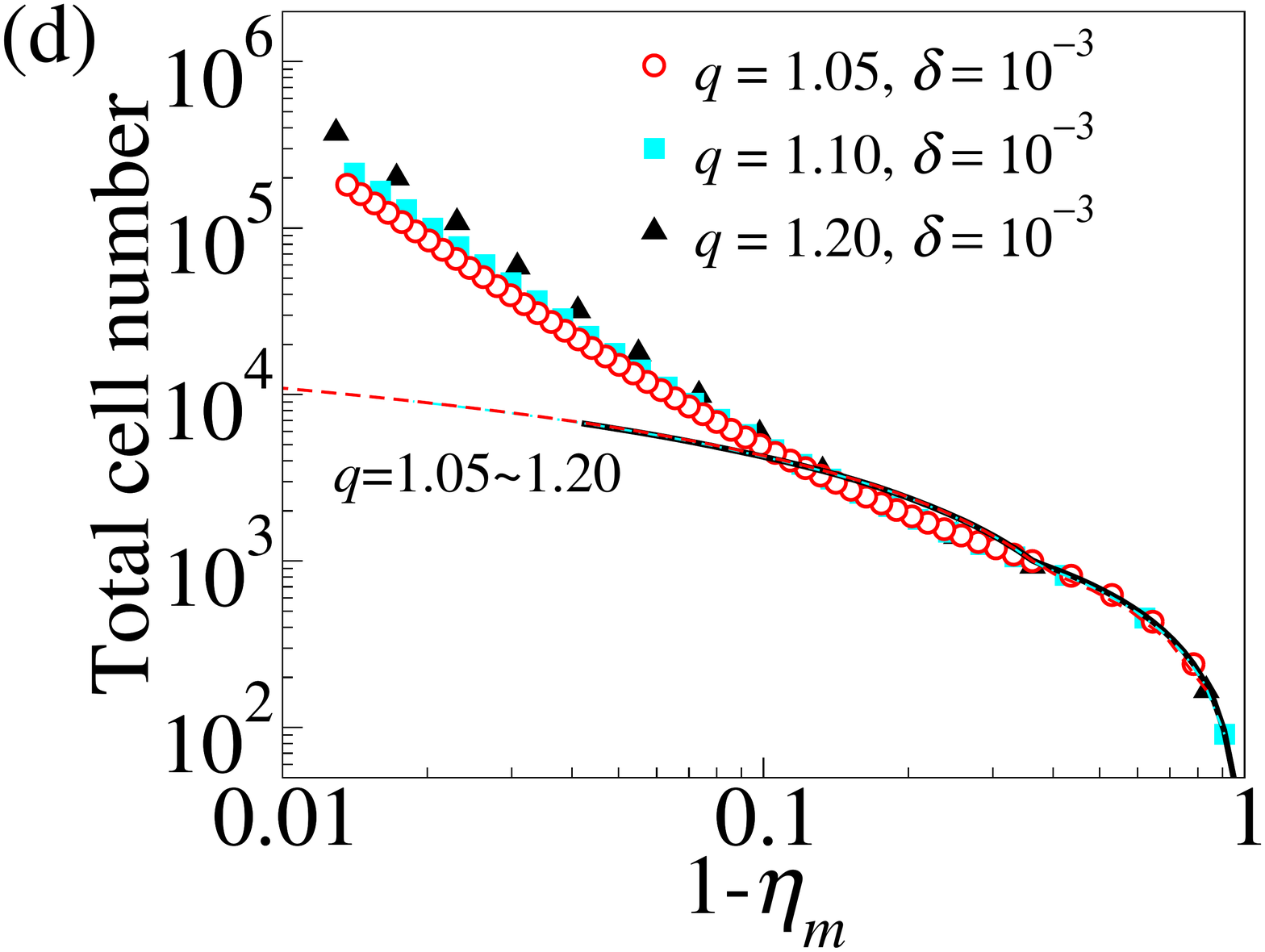}}

\caption{(Color online) Simulation results for multiscale-disk tessellations.
For the power-law scaling \eqref{eq:power-law_cell}, \subcap{fig:2DZeta_coveringfraction} the semi-log plot of fraction of uncovered space $(1-\eta_m)$ versus the number of stages $m$, and \subcap{fig:2DZeta_totalNumber} the log-log plot of total cell number $\mathcal{N}_m$ versus $(1-\eta_m)$.
For the exponential scaling \eqref{eq:exponential-law_cellvolume}, \subcap{fig:2DPower_coveringfraction} the semi-log plot of $(1-\eta_m)$ in the $m$th stage, and \subcap{fig:2DPower_totalnumber} the log-log plot of total cell number $\mathcal{N}_m$ as functions of $(1-\eta_m)$.
Here, we note that dashed lines represent prescribed covering fractions, given by Eqs. \eqref{eq:power-law_CoveringFraction} and \eqref{eq:exponentialScaling_CoveringFraction}, and $\delta \equiv v_{\max}/\abs{\mathcal{V}_F}$.
\label{fig:results_multiscale-disk_tilings}
}
\end{figure}

Although the upper bound \eqref{eq:chiv_multi_UB} is rigorous, it is a gross overestimation compared to corresponding simulation results.
Thus, we obtain a better estimate of Eq. \eqref{eq:chiv_multi_1-1} by assuming that the uncovered gaps are spatially  uncorrelated, which effectively removes the ensemble average of their cross terms, yielding 
\begin{equation}\label{eq:chiv_multi_approx}
\fn{\tilde{\chi}_{_V} ^{(m)}}{\vect{k}} \approx \fn{F}{\vect{k};\phi} +\frac{\phi^2}{\abs{\mathcal{V}_F}} \E{\sum_{j={\mathcal{N}_m}+1}^\infty  \fn{v_1}{\mathcal{R}_j}^2 },
\end{equation}
where the summation term becomes dominant as $\abs{k}\to0$ because $\fn{F}{\vect{k};\phi}$ shows a power-law scaling; see Appendix \ref{sec:derivation_RandomPhaseApproximation} for details. 

Further estimation of the second term in Eq. \eqref{eq:chiv_multi_approx} requires the cell-size distribution.
Consider two distinct cell-size scalings discussed in Sec. \ref{sec:Coated-sphere_model}: a power-law form \eqref{eq:power-law_cell} and an exponential functional form \eqref{eq:exponential-law_cellvolume}. 
For the power-law scaling, using Eqs. \eqref{eq:power-law_TotalVolume} and \eqref{eq:power-law_CoveringFraction}, the summation in Eq. \eqref{eq:chiv_multi_approx} can be written as
\begin{align}
&\E{\sum_{j={\mathcal{N}_m}+1}^\infty \fn{v_1}{\mathcal{R}_j}^2 } = N \left(v^{(1)}\right)^2 \sum_{j=m+1}^\infty \frac{1}{j^{2p}} \nonumber \\
=&  v^{(1)}\abs{\mathcal{V}_F}\fn{\zeta}{p} (1-\eta_m)^2 \fn{f}{m}, \label{eq:chi0_multi_powerlaw}
\end{align}
where $\fn{f}{m}\equiv \left(\sum_{j=m+1}^\infty j^{-2p}	\right) / \left( \sum_{j=m+1}^\infty j^{-p}	\right)^2$.
Substituting Eq. \eqref{eq:chi0_multi_powerlaw} into Eq. \eqref{eq:chiv_multi_approx} yields an approximation of the residual spectral density in the small-wavenumber limit as follows:
\begin{equation}\label{eq:residual_spectral_powerscaling}
\fn{\tilde{\chi}_{_V} ^{(m)}}{k\to0} \approx  v^{(1)}	\fn{f}{m}\fn{\zeta}{p} \left[\phi(1-\eta_m)\right]^2.
\end{equation}

\begin{figure*}[ht]

\subfloat{\label{fig:2DZeta_chiv}
\includegraphics[width =0.4 \textwidth]{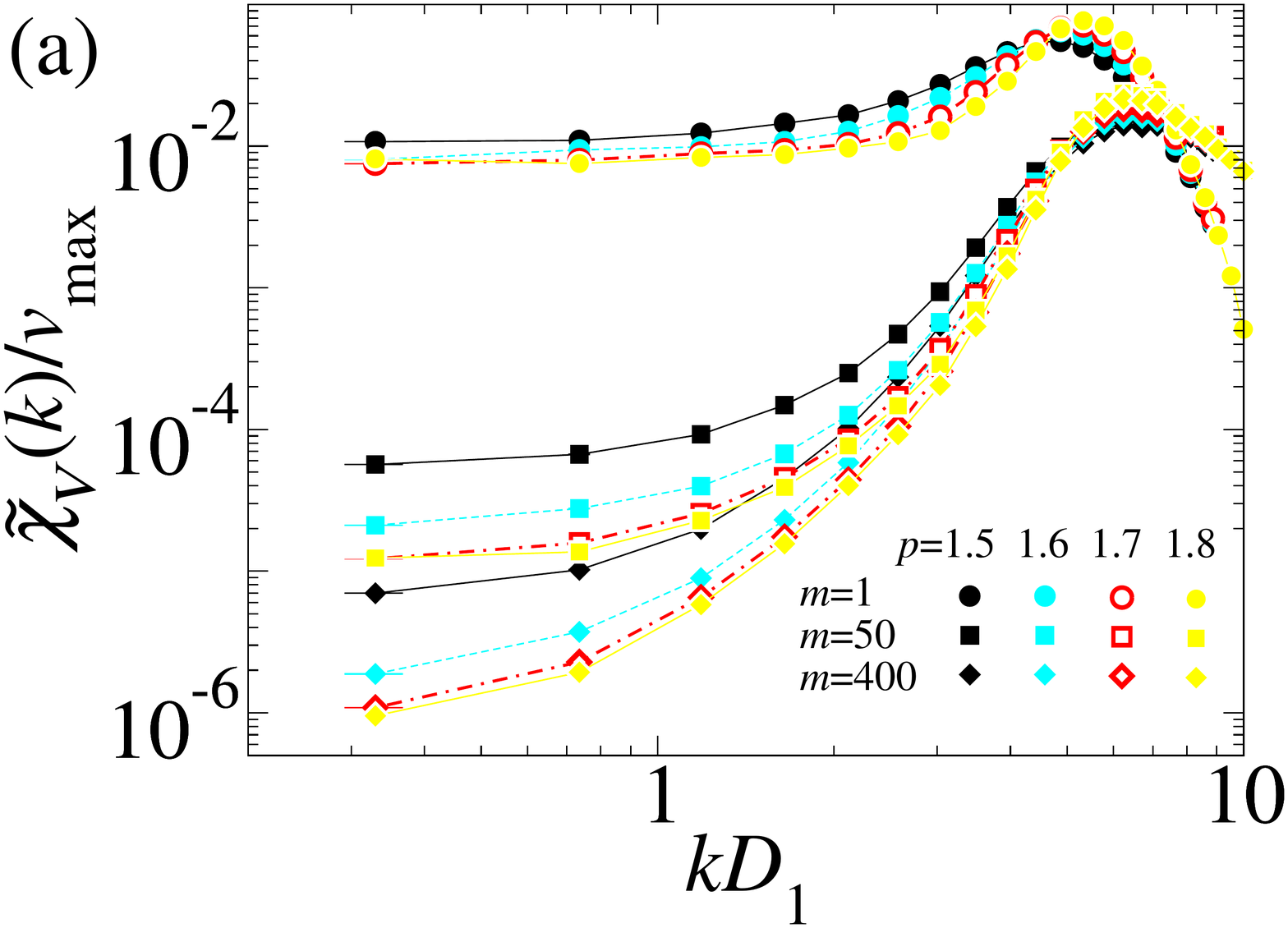}}
\subfloat{\label{fig:2DZeta_chiv0}
\includegraphics[width =0.4 \textwidth]{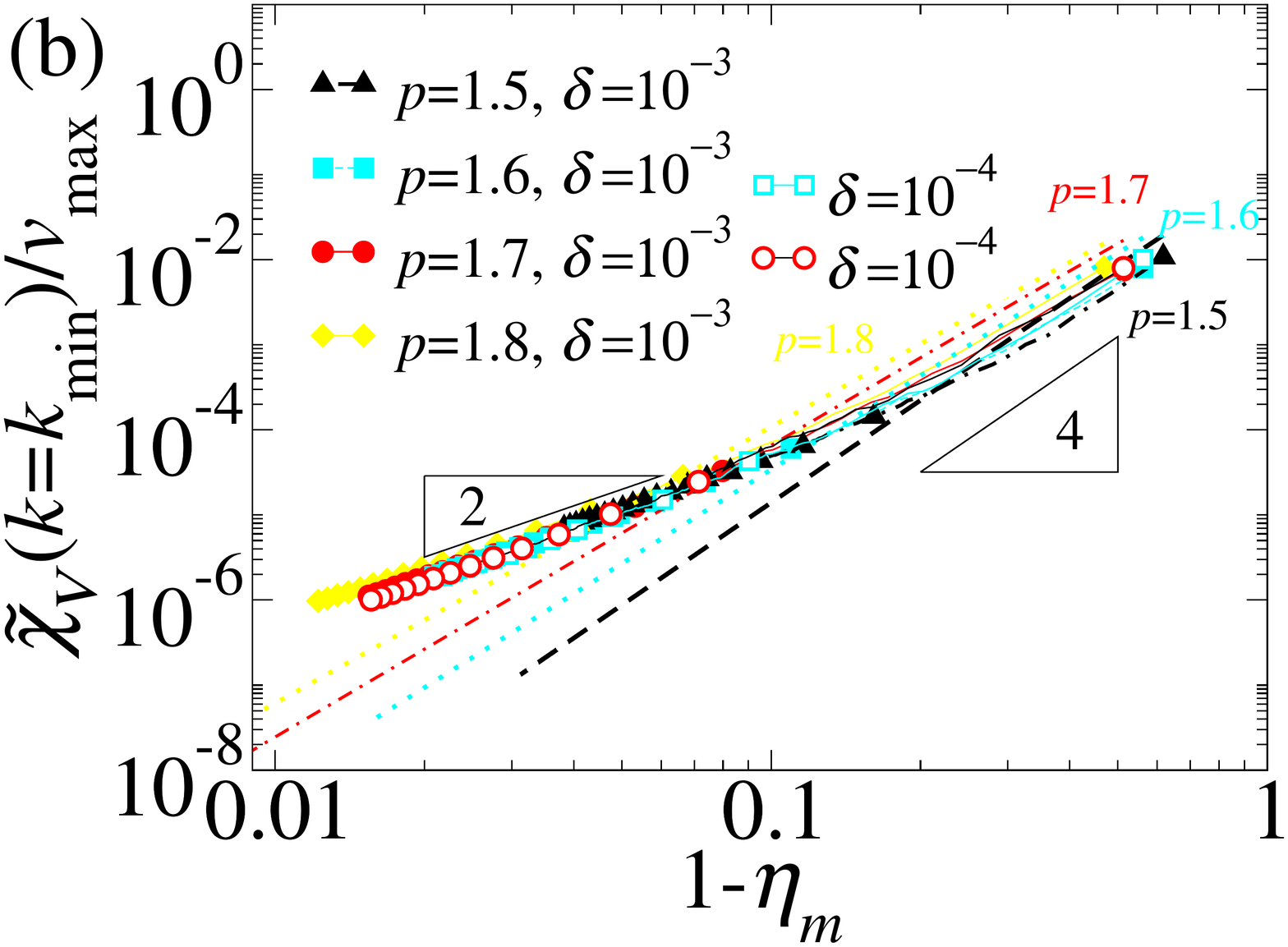}}

\subfloat{\label{fig:2DPower_chiv}\includegraphics[width=0.4\textwidth]{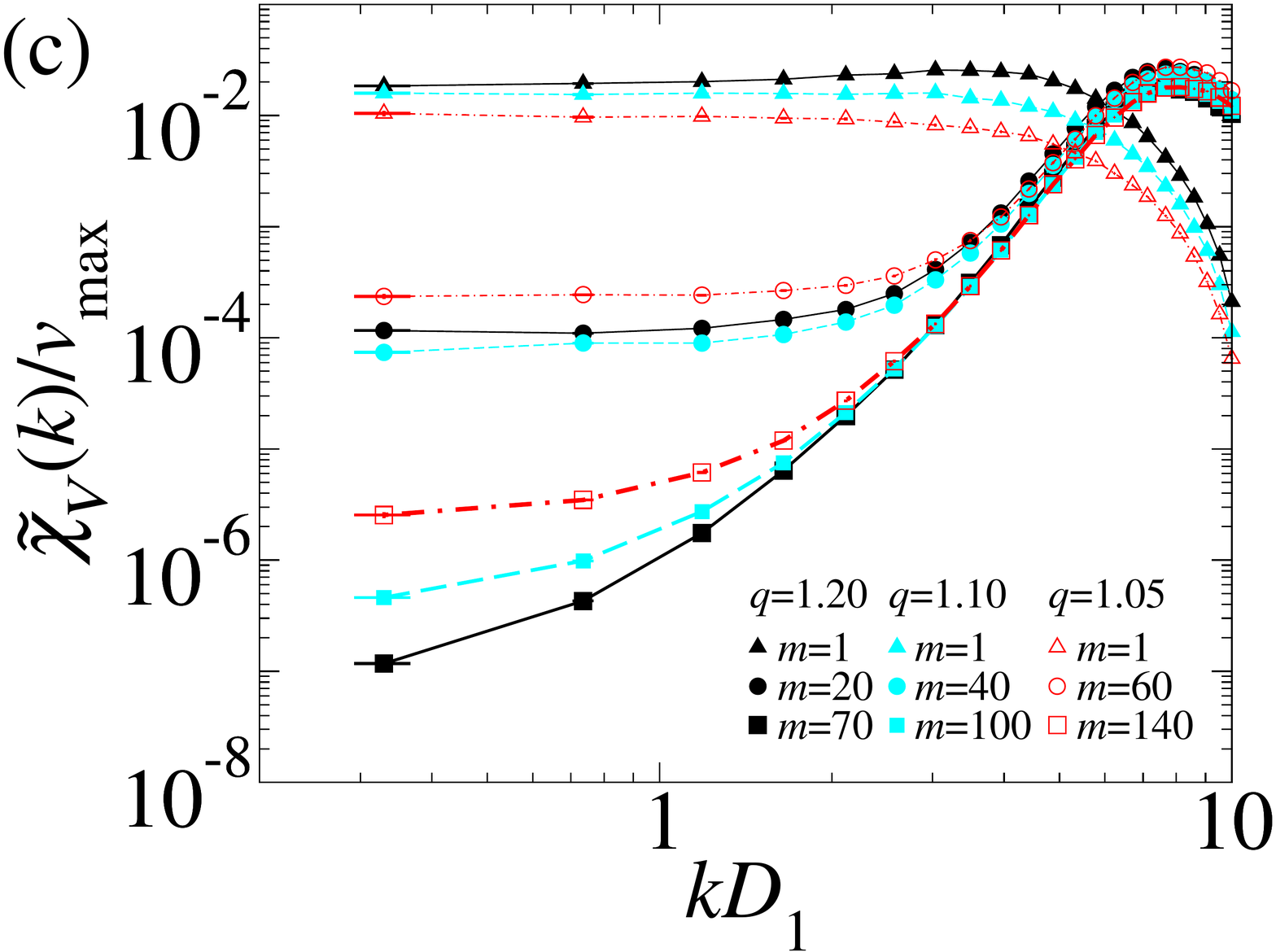}}
\subfloat{\label{fig:2DPower_chiv0}\includegraphics[width=0.4\textwidth]{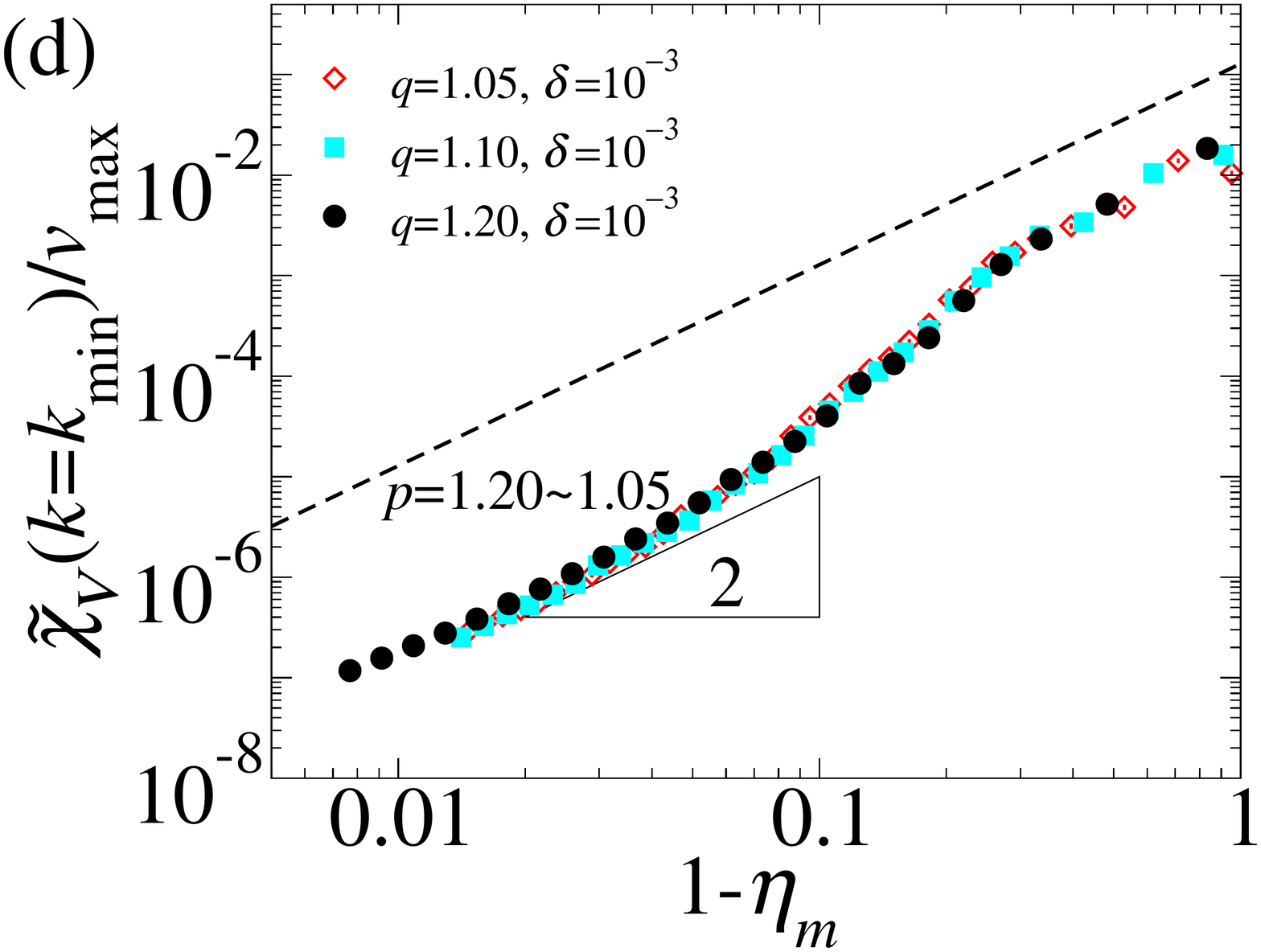}}
\caption{(Color online) Simulation results for the coated-disks models for (a, b) power-law and (c, d) exponential scalings of cell volumes.
(a, c) Log-log plots of the spectral densities $\fn{\tilde{\chi}_{_V}^{(m)}}{\vect{k}}$ versus wavenumber $k$ for our coated-spheres model at various values of stages. 
(b, d) Log-log plots of the associated spectral densities $\fn{\tilde{\chi}_{_V}^{(m)}}{k= k_{\min}}$ at the minimal wavenumber as functions of the fraction of uncovered space $(1-\eta_m)$.
The dashed lines represent the values calculated from Eqs. \eqref{eq:residual_spectral_powerscaling} and \eqref{eq:residual_spectral_exponentialscaling}, respectively.
\label{fig:multiscale-tessellations}}
\end{figure*}

For the exponential scaling, Eqs. \eqref{eq:exponentialScaling_TotalVolume} and \eqref{eq:exponentialScaling_CoveringFraction} are used to simplify Eq. \eqref{eq:chiv_multi_approx} as follows:
\begin{align}
&\E{\sum_{j={\mathcal{N}_m}+1}^\infty \fn{v_1}{\mathcal{R}_j}^2 }  = N \left[v^{(1)} \frac{q}{q^m (q-1)}	\right]^2 \frac{(q-1)^2}{q^2 -1} \nonumber \\
=& \abs{\mathcal{V}_F} v^{(1)}\frac{q}{q+1} (1-\eta_m)^2 \label{eq:chi0_multi_exponential},
\end{align}
which results in the following leading-order term of the spectral density in the $m$th stage:
\begin{equation}\label{eq:residual_spectral_exponentialscaling}
\fn{\tilde{\chi}_{_V} ^{(m)}}{k\to0} \approx v^{(1)} \frac{q}{q+1} \left(\phi(1-\eta_m)\right)^2.
\end{equation}

\subsection{Simulation results}\label{sec:Coated-sphere_results}

For the purpose of illustration and simplicity, we specialize to two dimensions to generate multiscale-disk tessellations for the aforementioned cell-size scalings: power-law and exponential functional forms, as defined by Eqs. \eqref{eq:power-law_cell} and \eqref{eq:exponential-law_cellvolume}, respectively. 
Using the power-law functional forms, the simulations proceed up to the $400$th stage, yielding a scaling exponent $p$ ranging from 1.5 to 1.8, and the ratios of the maximal cell volume to the sample volume $\delta=10^{-3}$ and $10^{-4}$.
For the exponential functional forms, the constructions proceed to achieve around covering fraction $\eta_m\approx 0.99$ for values of a scaling base $q$=$1.05$, $1.10$, and $1.20$, and the ratio $\delta=10^{-3}$. 

For both types of scaling functions, as the scaling parameters increases, the smaller is the number of stages needed to achieve a prescribed covering fraction.
Instead, for even larger scaling parameters, the precursor packings are more likely to be saturated \cite{Shier2013}, which often results in significant computational costs.
This is because as the packing approaches to a saturation state, the voxel-list algorithm that we employ subdivides current voxels with an increasingly finer resolution, which requires increasingly larger computer memory and computational times. 
For this reason, we choose scaling parameters smaller than 2.

To obtain multiscale-disk tessellations that nearly fill space ($\eta_m\approx 0.95$), these procedures need to continue up to around a few hundred stages [Fig. \subref*{fig:2DZeta_coveringfraction} and \subref*{fig:2DPower_coveringfraction}].
For $m\lesssim 10$, cell volumes in the exponential scalings do not change much, compared to those in the power-law scalings.
Thus, the tessellations in the exponential scalings can cover a larger fraction of the simulation box with a smaller number of stages than those for the power-law scaling.
Instead, for the exponential scaling, the precursor packings more easily achieve saturation from relatively low covering fractions ($\eta_m\approx 0.65$).
Then, the simulated covering fraction never keeps up with the prescribed covering fraction \eqref{eq:exponentialScaling_CoveringFraction}; see Fig. \subref*{fig:2DPower_coveringfraction}.
However, the number of total cells $\mathcal{N}_m$ required to reach to a certain covering fraction will be similar for both cases; see Fig. \subref*{fig:2DZeta_totalNumber} and \subref*{fig:2DPower_totalnumber}.

From these precursor polydisperse packings, we simulate the coated-disks model with the inclusion volume fraction $\phi=0.5$. 
To ascertain their hyperuniformity, we compute the associated spectral densities $\fn{\tilde{\chi}_{_V}^{(m)}}{\vect{k}}$. 
As shown in Fig. \subref*{fig:2DZeta_chiv} and \subref*{fig:2DPower_chiv}, the packings in the finite stages (around $\eta_m \approx 0.95$) are effectively hyperuniform.
Importantly, this comes from the fact that the precursor packings have uncovered gaps, and thus the local-cell packing fraction $\phi$ and their global packing fraction $\phi\eta_m$ have a discrepancy.  
The consequent volume-fraction fluctuations are estimated by Eqs. \eqref{eq:residual_spectral_powerscaling} and \eqref{eq:residual_spectral_exponentialscaling}.
Figures \subref*{fig:2DZeta_chiv0} and \subref*{fig:2DPower_chiv0} show that while the numerical results in $\fn{\tilde{\chi}_{_V}^{(m)}}{k = k_{\min}}$ and these theoretical estimations conform to each other for low covering fractions ($\eta_m \lesssim 0.8$), they significantly deviate for relatively high covering fractions. 
This is because the uncovered gaps are no longer spatially uncorrelated in high covering fractions.

However, for high covering fractions, the residual spectral densities $\fn{\tilde{\chi}_{_V}^{(m)}}{k =k_{\min}}$ vanish with an error on the order of $(1-\eta_m)^2$. 
Thus, our upper bound \eqref{eq:chiv_multi_UB}, which although is a gross overestimation, correctly predicts the scaling behavior of $\fn{\tilde{\chi}_{_V}^{(m)}}{k =k_{\min}}$.
Both the theoretical and numerical results show that the coated-spheres model should be strongly hyperuniform in the limit of $m\to \infty$. 

It is noteworthy that as the number $m$ of stages increases, the spectral densities in Fig. \subref*{fig:2DZeta_chiv} and \subref*{fig:2DPower_chiv} tend to resemble one another for intermediate wavenumbers.
In addition, the terms $\fn{\tilde{\chi}_{_V}^{(m)}}{k = k_{\min}}$, as shown in Fig. \subref*{fig:2DZeta_chiv0} and \subref*{fig:2DPower_chiv0}, collapse to a single curve that seemingly only depends on the fraction of uncovered space. 
From these observations, we surmise that the spectral densities of the multiscale coated-spheres structures ($\eta_m \to \infty$) are identical to one another, regardless of the cell-volume distributions of the precursor sphere packings. 
This universal behavior of spectral densities is apparently related to the fact that the coated-spheres structures have identical effective properties.

\section{ Fabrication of Our Designs }\label{sec:fabrications}
 
Disordered hyperuniform structures, constructed by numerical simulations have been produced via modern fabrication technologies \cite{Man2013, Ma2016}.
Here, we explicitly discuss how our designed hyperuniform packings (dispersions) can be fabricated via state-of-the-art technologies, such as 2D photolithographic \cite{Zhao2018} and 3D printing techniques \cite{Wong2012, Shirazi2015, Tumbleston2015}.

Photolithography is a microfabrication technique that uses light to transfer a designed 2D pattern in the photomask on the photo-sensitive chemicals coated on a flat substrate.
Then, after a series of chemical treatments, the desired pattern is engraved on the material, or material is deposed on the pattern.
These methods are widely used in industry and research because of their high efficiency and exact control over the shapes and size of the patterns. 
Instead, diffraction tends to round all sharp corners in the designed pattern, the radius of which is associated with the minimum feature size.
State-of-the-art photolithography techniques are capable of creating patterns on a 30-cm-diameter wafer with the minimum feature size down to 25 nm \cite{Zhao2018}.
With equipment of 1.5 $\mu$m minimum feature size, one can readily fabricate our 2D hyperuniform disk packings (probably derived from Voronoi tessellations) that include more than one million particles.

Three-dimensional hyperuniform packings can be fabricated using 3D additive manufacturing techniques \cite{Wong2012, Shirazi2015, Tumbleston2015}. 
3D printing refers to various processes  that solidify materials layer by layer to create a 3D object (e.g., fused filament fabrication, stereolithography, and selective laser melting). 
A printed structure must be topologically connected such that it can be mechanically self-supporting after the procedure.
Thus, the void (matrix) phase of our 3D hyperuniform packings(dispersions) can be readily printed \cite{Zerhouni2018}. 
Due to recent developments in 3D printing methods, some commercial desktop 3D printers can print a sample with dimensions $125\times125\times125~\mathrm{cm}^3$ in 50 h with around 100 $\mu \mathrm{m}$ $XY$ resolution and 20 $\mu \mathrm{m}$ in $Z$ resolution.
Setting the minimum pore size is 300 $\mu$m, such devices can readily fabricate our 3D hyperuniform sphere packings that include up to 50 million pores.
If pore sizes are on the order of the resolutions, then spherical pores will be suitable to reduce the effect of thermal deformations during the printing process. 

\section{Conclusions and Discussion}\label{sec:conclusion}

In summary, we have introduced the tessellation-based methodology to construct disordered hyperuniform packings (dispersions) in $d$-dimensional Euclidean space $\R^d$.
This procedure is simple to implement because it only involves computing cell volumes, each of which can be performed exactly and in parallel.
Furthermore, it guarantees that the constructed packings are perfectly hyperuniform of class I in the infinite-sample-size limit, as we analytically showed in Eqs. \eqref{eq:scaling_general}, \eqref{eq:scaling_stat_isotropy}, and \eqref{eq:scaling_stat_isotropy_similar}.
We have implemented numerically our methodology from two distinct types of disordered tessellations: Voronoi tessellations of nonhyperuniform progenitor packings and sphere tessellations.

In the case of Voronoi tessellations, we demonstrated that our methodology provides a remarkable mapping that converts virtually all samples of any statistically homogeneous point pattern, whether hyperuniform or nonhyperuniform, into perfectly hyperuniform packings. 
Furthermore, the fact that it is easy to create a Voronoi tessellation of a very large point pattern with the combination of our efficient procedure enables us to construct very large samples (of the order of $10^8$ particles) of perfectly hyperuniform disordered packings. 
Such large system sizes have not been possible for previous numerical methods.

In the case of sphere tessellations, we established that the optimal Hashin-Shtrikman multiscale dispersions are hyperuniform of class I.
In addition to the fact that the spherical inclusions in such dispersions are ``well-separated'' from one another \cite{Torquato2018_3}, hyperuniformity  is apparently another important structural attribute to attain optimal effective transport and elastic properties \cite{Milton_TheoComposites, Torquato_RHM, Hashin1962, Milton1986a, Torquato2004}. 
In this regard, it is noteworthy that some disordered hyperuniform packings \cite{Zhang2016, Chen2017} and cellular networks derived from hyperuniform systems \cite{Torquato2018_3} have been reported to possess (nearly) optimal effective transport and elastic properties. 
Thus, it will be interesting to investigate the physical properties of hyperuniform packings constructed by our procedure.

Furthermore, it is important to observe that our procedure allows many distinct types of tessellations as long as they meet the bounded-cell condition \eqref{def:bounded-cell condition}. 
Besides Voronoi tessellations and sphere tessellations studied in this work, examples include disordered isoradial graphs \cite{Deuschel2012}, Delaunay triangulations, ``Delaunay-centroidal'' tessellations \cite{Florescu2009, Torquato2018_3}, dissected tessellations \cite{Gabbrielli2012}, and various generalizations of Voronoi tessellations, such as Laguerre tessellations \cite{Gellatly1982, Hiroshi1985} and tessellations in Manhattan distances \cite{Krause_TaxicabGeometry}.
Since constructing progenitor packings is the most time-consuming step in our numerical implementations, one would increase sample sizes by employing other tessellations that can be efficiently generated.

We note that similar to our procedure, a 1D model \cite{Gabrielli2004b} and the ``equal-volume tessellation" \cite{Gabrielli2008} enable the generation of hyperuniform point mass patterns and hyperuniform point patterns from certain initial tessellations, respectively.
Specifically, these two models place a point mass \cite{Gabrielli2004b} and point particles \cite{Gabrielli2008} in each cell such that mass densities and number densities in the cells become identical, respectively.
Thus, the hyperuniform systems in both models can be regarded to be a zero-$\phi$ limit of the hyperuniform packings in our procedure.
Importantly, however, both previous models \cite{Gabrielli2004b, Gabrielli2008} do not  consider volume-fraction fluctuations, which is our central concern and accounted for by our procedure.
In addition, the mass density of point masses in this 1D model does not change the structural characteristics of the resulting systems, which again is different from our methodology.
We also note that the equal-volume tessellation is less versatile in its initial tessellations than our procedure  because all cell volumes in an initial tessellation should be integer multiples of a common unit volume.

We have shown that our procedure enables a mapping that converts a nonhyperuniform packing into a hyperuniform one without changing the initial tessellation, i.e., the tessellation is an invariant under the transformation.
In other words, packings that have identical tessellations can either be nonhyperuniform or hyperuniform by simply tuning local characteristics. 
It immediately follows that any tessellation statistic, including the distributions of nearest-neighbor distances and Voronoi-cell volumes, are identical for both the progenitor nonhyperuniform point patterns and their corresponding constructed hyperuniform packings.
These results reinforce previous observations that local structural characteristics may or may not determine the hyperuniformity of systems, e.g., substantial local clustering of particles may not be inconsistent with hyperuniformity \cite{Zachary2011_2} and disordered systems with appreciable short-range order are often not hyperuniform \cite{Zhang2013}; see Ref. \cite{Torquato2018_review}. 
Moreover, our results also show that transitions from a nonhyperuniform state to a hyperuniform state can be achieved without correlated movements of particles/mass across all length scales.
Such transitions stand in contrast to those in some previous hyperuniform systems in thermal equilibrium \cite{Jancovici1981, Uche2004, Batten2008,Zhang2015, Lomba2017, Chen2017} as well as nonequilibrium \cite{Hexner2015, Tjhung2015, Weijs2015, Kurita2011, Ricouvier2017, TJ_algorithm, LS_algorithm}, in which the transitions always involve collective rearrangements of the particles.

It should not go unnoticed that our methodology also allows ones to tune particle shapes, positions, and numbers within each cell with preserving hyperuniformity of the constructed packings. 
For instance, one can engineer these two-phase systems at large length scales (i.e., $k \ll 1$) by choosing nonspherical particles with various aspect ratios and placing them away from the centroids of the associated cells. 
However, at intermediate length scales the two-phase media are structurally similar to their progenitor patterns (see Sec. \ref{sec:Voronoi_spherical}).  
One can tune local structures of the hyperuniform packings to achieve the ``well-separated" condition, which is a necessary requirement to attain the optimal transport and mechanical properties (see Sec. \ref{sec:Coated-sphere}).
Moreover, the small-$k$ scaling of the spectral density can either be $\spD{k}\sim k^2$ or $\spD{k}\sim k^4$ by engineering particle displacements $\Delta \vect{X}_j$ with respect to the associated cell centroids (see Sec. \ref{sec:GeneralTheory}).
Due to this tunability, our methodology allows ones to design an enormous class of hyperuniform packings, including (nearly) optimal structures.
Combining our computational designs with the aforementioned 2D and 3D fabrication techniques \cite{Jaeuk2019a} is expected to accelerate the discovery of novel disordered hyperuniform two-phase materials. 

There are several other open questions for future exploration.
For example, could other types of initial tessellations lead to scaling behaviors of the spectral density besides quadratic or quartic shown here?
To what extent can our procedure be generalized by relaxing the identical local-cell packing faction constraint (e.g., giving certain spatial correlations in local-cell packing fractions)? Can such generalized versions of our procedure construct ``disordered" stealthy (i.e., $\spD{k} = 0$ for $k<K$) \cite{Batten2008, Zhang2015, Zhang2015_2, Torquato2015_stealthy} or class II hyperuniform (i.e., $\spD{k}\sim k$, such as MRJ packings \cite{Jiao2011, Atkinson2016})?
If possible, then this would certainly allow one to obtain more general scaling behaviors.

~

\section*{Acknowledgements}
We thank G. Zhang, T. G. Mason, A. B. Hopkins, and M. A. Klatt for very helpful discussions.  
This work was supported by the National Science Foundation under Grant No. DMR-1714722.

{
\appendix

\section{Voronoi tessellations}\label{app:voronoi}
The Voronoi tessellation of a given progenitor packing is computed via VORO++ library \cite{voro++}.
To enhance the performance, we divide the particle centroids of the progenitor packing into several domains and compute the Voronoi tessellations for each domain in parallel.
To avoid any deformation in Voronoi cells due to domain boundaries, we choose domains in the following steps:
\begin{enumerate}
\item Divide the simulation box into several disjoint subdomains in cubic shape.

\item Add a marginal region surrounding each subdomain, and these two regions form a domain.
Here, the thickness $W$ of the marginal region is chosen as $6\rho^{-1/d}$, where $\rho$ is the number density.
\end{enumerate}
Then, for each domain, we compute the Voronoi cells of points inside ``subdomains."
For point patterns with larger holes, the thickness of the marginal region should be increased accordingly.

Since VORO++ is designed for three-dimensional geometries, its 2D implementation is performed in a quasi 3D simulation box whose height (in $z$ component) is unity.
To avoid any possible ``memory leakage" in this 2D implementation, the number of particles within a domain should be smaller than $10^5$.

\begin{widetext}
\section{Derivation of Eq. \eqref{eq:F.T._phase_indicator}}\label{sec:App_FT_phase_indicator}
First, we rewrite the Fourier transform of the particle indicator function $\fn{\tilde{\mathcal{J}}}{\vect{k}}$, given in Eq. \eqref{eq:chi_v_step2}:
\begin{align*}
\fn{\tilde{\mathcal{J}}}{\vect{k}} =& \sum_{j=1}^N e^{-i\vect{k}\cdot\vect{x}_j } 
\left[ \fn{\tilde{m}}{\vect{k};\tens{P}_j }e^{-i \vect{k}\cdot \Delta\vect{X}_j} -\phi \fn{\tilde{m}}{\vect{k};\tens{C}_j} \right] \\
=&
\sum_{j=1}^N e^{-i\vect{k}\cdot\vect{x}_j }
 \left[	\big( \fn{\tilde{m}}{\vect{k};\tens{P}_j} -\phi \fn{\tilde{m}}{\vect{k};\tens{C}_j}\big) +\fn{\tilde{m}}{\vect{k};\tens{P}_j}\left(	e^{-i \vect{k}\cdot\Delta\vect{X}_j} - 1\right) \right] .
\end{align*}
Here, we substitute the form factors $\fn{\tilde{m}}{\vect{k};\tens{P}_j}$ and $\fn{\tilde{m}}{\vect{k};\tens{C}_j}$ with their Taylor expansions \eqref{eq:FT fundamental cell2}, which yields
\begin{align}
\fn{\tilde{\mathcal{J}}}{\vect{k}} =&  \sum_{j=1}^N \phi\abs{\tens{C}_j} e^{-i\vect{k}\cdot\vect{x}_j} 
 \Bigg\{ -\frac{k_\alpha k_\beta}{2} \left[	\fn{\mathcal{M}_{\alpha\beta}}{\tens{P}_j} - \fn{\mathcal{M}_{\alpha\beta}}{\tens{C}_j}\right] +  \frac{ik_\alpha k_\beta k_\gamma}{6} \left[\fn{\mathcal{M}_{\alpha\beta\gamma}}{\tens{P}_j}-\fn{\mathcal{M}_{\alpha\beta\gamma}}{\tens{C}_j}	\right]\nonumber\\
&+ 
\left[1-\frac{k_\alpha k_\beta}{2} \fn{\mathcal{M}_{\alpha\beta}}{\tens{P}_j }\right] (e^{-i\vect{k}\cdot \Delta\vect{X}_j } -1)	\Bigg\} +\order{k^4} \label{eq:weight function_line 1}  \\
=&
\phi \sum_{j=1}^N \abs{\tens{C}_j}e^{-i\vect{k}\cdot\vect{x}_j} \Bigg\{ \left(e^{-i\vect{k}\cdot\Delta\vect{X}_j}	- 1\right)  -\frac{k_\alpha k_\beta}{2} \left[	\fn{\mathcal{M}_{\alpha\beta}}{\tens{P}_j} - \fn{\mathcal{M}_{\alpha\beta}}{\tens{C}_j}\right] \nonumber \\
&+ \frac{i k_\alpha k_\beta k_\gamma}{6} \left[\fn{\mathcal{M}_{	\alpha\beta\gamma}}{\tens{P}_j}-\fn{\mathcal{M}_{\alpha\beta\gamma}}{\tens{C}_j}\right] -\frac{k_\alpha k_\beta}{2} \fn{\mathcal{M}_{\alpha\beta}}{\tens{P}_j } (e^{-i\vect{k}\cdot\Delta\vect{X}_j} -1)\Bigg\}  +\order{k^4} \nonumber  \\
=&\underbrace{\phi \sum_{j=1}^N   \left(e^{-i\vect{k}\cdot\Delta\vect{X}_j}	- 1\right) \abs{\tens{C}_j}e^{-i\vect{k}\cdot\vect{x}_j}}_{\fn{\tilde{\mathcal{J}}_{(1)}}{\vect{k}}} 
+
\underbrace{(-1)\phi \frac{k_\alpha k_\beta}{2} \sum_{j=1}^N   \left[	\fn{\mathcal{M}_{\alpha\beta}}{\tens{P}_j} - \fn{\mathcal{M}_{\alpha\beta}}{\tens{C}_j}\right] \abs{\tens{C}_j} e^{-i \vect{k}\cdot\vect{x}_j} }_{\fn{\tilde{\mathcal{J}}_{(2)}}{\vect{k}}} \nonumber \\
&+ \underbrace{ \phi \frac{i k_\alpha k_\beta k_\gamma}{6}\sum_{j=1}^N \left[\fn{\mathcal{M}_{	\alpha\beta\gamma}}{\tens{P}_j}-\fn{\mathcal{M}_{\alpha\beta\gamma}}{\tens{C}_j} + 3 (\Delta \vect{X}_j)_\gamma   \fn{\mathcal{M}_{\alpha\beta}}{\tens{P}_j }\right] \abs{\tens{C}_j}e^{-i \vect{k}\cdot\vect{x}_j}}_{\fn{\tilde{\mathcal{J}}_{(3)}}{\vect{k}}}  +\order{k^4} .\label{eq:weight function}
\end{align}
In Eq. \eqref{eq:weight function_line 1}, we used the identical local-cell packing fraction condition, i.e., $\abs{\tens{P}_j}=\phi\abs{\tens{C}_j}$ for $j=1,\cdots, N$.
In Eq. \eqref{eq:weight function}, $(\Delta\vect{X}_j)_\gamma$ is the $\gamma$th Cartesian component of a vector $\Delta\vect{X}_j$, and we apply the Taylor expansion $\left(e^{-i\vect{k}\cdot\Delta\vect{X}_j}	- 1\right) = -ik_\gamma (\Delta\vect{X}_j)_\gamma +\order{k^2}$, which is a good approximation due to the bounded-cell condition \eqref{def:bounded-cell condition}.

\section{Derivations of the spectral density of the $m$th stage coated-spheres model}

Here, we compute upper bounds on the spectral density $\fn{\tilde{\chi}_{_V} ^{(m)}}{\vect{k}}$ defined in Eq. \eqref{eq:chiv_multi_1-1} in three cases of cell-volume distributions: unknown, a power-law scaling, and an exponential scaling.

\subsection{Upper bounds of the spectral density}\label{sec:upperbound}
From Eq. \eqref{eq:chiv_multi_1-1}, 
\begin{align}
\fn{\tilde{\chi}_{_V} ^{(m)}}{\vect{k}} & = \frac{1}{\abs{\mathcal{V}_F}} \E{\abs{\sum_{j=1}^{\mathcal{N}_m} \left[\fn{\tilde{m}}{\vect{k}; \phi^{1/d}\mathcal{R}_j} - \phi \fn{\tilde{m}}{\vect{k};\mathcal{R}_j}	\right]e^{-i\vect{k}\cdot\vect{x}_j}  - \phi \sum_{j={\mathcal{N}_m}+1}^\infty \fn{\tilde{m}}{\vect{k};\mathcal{R}_j}e^{-i\vect{k}\cdot\vect{x}_j} }^2} \label{eq:chiv_multi_1-2}\\
& = \frac{1}{\abs{\mathcal{V}_F}} \E{\abs{ \frac{\phi (1-\phi^{2/d})}{2(2+d)}\sum_{j=1}^{\mathcal{N}_m}  \fn{v_1}{\mathcal{R}_j}(k\mathcal{R}_j)^2	e^{-i\vect{k}\cdot\vect{x}_j} +\order{k^4} - \phi \sum_{j={\mathcal{N}_m}+1}^\infty \fn{\tilde{m}}{\vect{k};\mathcal{R}_j}e^{-i\vect{k}\cdot\vect{x}_j} }^2} \label{eq:chiv_multi_1-3}.
\end{align}
Now, note the following inequality for two complex numbers $A$ and $B$:  
\begin{align}
\abs{A + B}^2 \leq & \left(\abs{A} + \abs{B}	\right)^2 = \abs{A}^2+\abs{B}^2 + 2\abs{A}\abs{B} \label{ineq:step1}\\
\leq&2( \abs{A}^2 +\abs{B}^2),\label{ineq:step2}
\end{align}
which is obtained by successively applying the triangle inequality and then the inequality of arithmetic and geometric means, and thus equality occurs if and only if $A=B$. 
Using this inequality, we obtain a rigorous upper bound on the spectral density described in Eq. \eqref{eq:chiv_multi_1-3}:
\begin{align}
\fn{\tilde{\chi}_{_V} ^{(m)}}{\vect{k}} & \leq \frac{2}{\abs{\mathcal{V}_F}} \left(\frac{\phi (1-\phi^{2/d})}{2(2+d)}	\right)^2 \E{\abs{ \sum_{j=1}^{\mathcal{N}_m}  \fn{v_1}{\mathcal{R}_j}(k\mathcal{R}_j)^2	e^{-i\vect{k}\cdot\vect{x}_j} +\order{k^4}}^2}  + \frac{2\phi^2}{\abs{\mathcal{V}_F}}\E{\abs{ \sum_{j={\mathcal{N}_m}+1}^\infty \fn{\tilde{m}}{\vect{k};\mathcal{R}_j}e^{-i\vect{k}\cdot\vect{x}_j} }^2}. \label{eq:chiv_multi_upper1}
\end{align}
Subsequently, we derive an upper bound on the second term in Eq. \eqref{eq:chiv_multi_upper1} by applying the triangle inequality: 
\begin{align}
& \E{\abs{\sum_{j={\mathcal{N}_m}+1}^\infty \fn{\tilde{m}}{\vect{k};\mathcal{R}_j}e^{-i\vect{k}\cdot\vect{x}_j} }^2} \leq  \E{\abs{\sum_{j={\mathcal{N}_m}+1}^\infty \abs{\fn{\tilde{m}}{\vect{k};\mathcal{R}_j}} }^2} \leq &\E{\abs{\sum_{j={\mathcal{N}_m}+1}^\infty \fn{v_1}{\mathcal{R}_j} }^2} =  \abs{\mathcal{V}_F}^2 (1-\eta_m)^2, \label{eq:residualSpecDens_crude}
\end{align}
where the inequality in Eq. \eqref{eq:residualSpecDens_crude} comes from the fact that $\abs{\fn{\tilde{m}}{\vect{k};R}} \leq \fn{v_1}{R} $; see Eq. \eqref{eq:FourierTransform_sphere}.
We note that the last term in Eq. \eqref{eq:residualSpecDens_crude} can be regarded as the largest volume-fraction fluctuations contributed from the uncovered gaps.
Combining Eqs. \eqref{eq:chiv_multi_upper1} and \eqref{eq:residualSpecDens_crude}, we obtain a rigorous bound as follows:
\begin{equation}\label{eq:chiv_upperbound_appendix}
\fn{\tilde{\chi}_{_V} ^{(m)}}{\vect{k}}  \leq 2\fn{F}{\vect{k};\phi}  + 2 \phi^2 \abs{\mathcal{V}_F} (1-\eta_m)^2,
\end{equation}
where $\fn{F}{\vect{k};\phi}$ is defined by Eq. \eqref{eq:chiv_multi_scaling}. We remark that this bound is derived without any prior knowledge of the cell-volume distribution.

\subsection{Derivation of Eq. \eqref{eq:chiv_multi_approx}}\label{sec:derivation_RandomPhaseApproximation}

We can approximate the spectral density given in Eq. \eqref{eq:chiv_multi_1-1} by assuming that its contributions from composite spheres [the first term in Eq. \eqref{eq:chiv_multi_1-3}] and uncovered gaps are uncorrelated, which effectively removes an ensemble average of their cross terms:
\begin{align}
\fn{\tilde{\chi}_{_V} ^{(m)}}{\vect{k}}&\approx \fn{F}{\vect{k};\phi}+\frac{\phi^2}{\abs{\mathcal{V}_F}} \E{\abs{\sum_{j={\mathcal{N}_m}+1}^\infty \fn{\tilde{m}}{\vect{k};\mathcal{R}_j}e^{-i\vect{k}\cdot\vect{x}_j} }^2}. \label{eq:chiv_multi_1-4} 
\end{align}
Here, the second term can be further simplified by assuming that the uncovered gaps are spatially uncorrelated:  
\begin{align}
&\frac{\phi^2}{\abs{\mathcal{V}_F}} \E{\abs{\sum_{j=\mathcal{N}_m+1}^\infty \fn{\tilde{m}}{\vect{k};\mathcal{R}_j}e^{-i\vect{k}\cdot\vect{x}_j} }^2} \approx \frac{\phi^2}{\abs{\mathcal{V}_F}} \E{\sum_{j=\mathcal{N}_m+1}^\infty \fn{\tilde{m}}{\vect{k};\mathcal{R}_j}^2} \approx \frac{\phi^2}{\abs{\mathcal{V}_F}} \E{\sum_{j=\mathcal{N}_m+1}^\infty \fn{v_1}{\mathcal{R}_j}^2 } , \label{eq:chiv_multi_1-5}
\end{align}
where the last approximation comes from $\abs{\fn{\tilde{m}}{k;R}}\approx \fn{v_1}{R} +\order{k^2} $ when $kR \ll 1$; see Eq.\eqref{eq:FourierTransform_sphere}.
\end{widetext}
}



%

\end{document}